\documentclass[apj,numberedappendix]{emulateapj}

\usepackage{bm}

\newcommand{\sao}{\altaffilmark{1}}
\newcommand{\mki}{\altaffilmark{2}}
\newcommand{\ngis}{\altaffilmark{3}}

\newcommand{\setC}{{\mathcal{C}}}
\newcommand{\setH}{{\mathcal{H}}}
\newcommand{\setM}{{\mathcal{M}}}
\newcommand{\setN}{{\mathcal{N}}}
\newcommand{\setP}{{\mathcal{P}}}
\newcommand{\setS}{{\mathcal{S}}}
\newcommand{\setU}{{\mathcal{U}}}
\newcommand{\1}{{\mathit{1}}}
\newcommand{\2}{{\mathit{2}}}
\newcommand{\3}{{\mathit{3}}}
\newcommand{\canon}[1]{\ensuremath{\langle #1 \rangle}}
\newcommand{\boldalpha}{{\boldsymbol{\alpha}}}
\newcommand{\boldxi}{{\boldsymbol{\xi}}}

\newcommand{\citetfap}{{F.~A. Primini et al.\ (2010, in preparation)}}
\newcommand{\citepfap}{{(F.~A. Primini et al.\ 2010, in preparation)}}
\newcommand{\citetkas}{{V.~Kashyap \& \ F.~A. Primini (2010, in preparation)}}

\shorttitle{{\it Chandra\/} Source Catalog}
\shortauthors{Evans et al.}
\submitted{Accepted May 22, 2010}

\begin{document}

\title{THE CHANDRA SOURCE CATALOG}

\author{Ian~N.~Evans,\sao\ Francis~A.~Primini,\sao\ Kenny~J.~Glotfelty,\sao\
  Craig~S.~Anderson,\sao\ Nina~R.~Bonaventura,\sao\ Judy~C.~Chen,\sao\
  John~E.~Davis,\mki\ Stephen~M.~Doe,\sao\ Janet~D.~Evans,\sao\
  Giuseppina~Fabbiano,\sao\ Elizabeth~C.~Galle,\sao\ Danny~G.~Gibbs~II,\sao\
  John~D.~Grier,\sao\ Roger~M.~Hain,\sao\ Diane~M.~Hall,\ngis\ Peter~N.~Harbo,\sao\
  Xiangqun~(Helen)~He,\sao\ John~C.~Houck,\mki\ Margarita~Karovska,\sao\
  Vinay~L.~Kashyap,\sao\ Jennifer~Lauer,\sao\ Michael~L.~McCollough,\sao\
  Jonathan~C.~McDowell,\sao\ Joseph~B.~Miller,\sao\ Arik~W.~Mitschang,\sao\
  Douglas~L.~Morgan,\sao\ Amy~E.~Mossman,\sao\ Joy~S.~Nichols,\sao\
  Michael~A.~Nowak,\mki\ David~A.~Plummer,\sao\ Brian~L.~Refsdal,\sao\
  Arnold~H.~Rots,\sao\ Aneta~Siemiginowska,\sao\ Beth~A.~Sundheim,\sao\
  Michael~S.~Tibbetts,\sao\ David~W.~Van~Stone,\sao\ Sherry~L.~Winkelman,\sao\
  and Panagoula~Zografou\sao}

\email{ievans@cfa.harvard.edu}

\altaffiltext{1}{Smithsonian Astrophysical Observatory, 60 Garden Street,
  Cambridge, MA 02138}

\altaffiltext{2}{MIT Kavli Institute for Astrophysics and Space Research, 77
  Massachusetts Avenue, Cambridge, MA 02139}

\altaffiltext{3}{Northrop Grumman, 60 Garden Street, Cambridge, MA 02138}

\begin{abstract}
The {\it Chandra\/} Source Catalog (CSC) is a general purpose virtual X-ray
astrophysics facility that provides access to a carefully selected set of
generally useful quantities for individual X-ray sources, and is designed to
satisfy the needs of a broad-based group of scientists, including those who
may be less familiar with astronomical data analysis in the X-ray regime.  The
first release of the CSC includes information about 94,676 distinct X-ray
sources detected in a subset of public ACIS imaging observations from roughly
the first eight years of the {\it Chandra\/} mission. This release of the
catalog includes point and compact sources with observed spatial extents
$\lesssim 30''$.  The catalog (1)~provides access to the best estimates of the
X-ray source properties for detected sources, with good scientific fidelity,
and directly supports scientific analysis using the individual source data;
(2)~facilitates analysis of a wide range of statistical properties for classes
of X-ray sources; and (3)~provides efficient access to calibrated observational
data and ancillary data products for individual X-ray sources, so that users can
perform detailed further analysis using existing tools.  The catalog includes
real X-ray sources detected with flux estimates that are at least 3 times their
estimated $1\,\sigma$ uncertainties in at least one energy band, while
maintaining the number of spurious sources at a level of $\lesssim 1$ false
source per field for a $100\rm\,ks$ observation.  For each detected source, the
CSC provides commonly tabulated quantities, including source position, extent,
multi-band fluxes, hardness ratios, and variability statistics, derived from the
observations in which the source is detected.  In addition to these traditional
catalog elements, for each X-ray source the CSC includes an extensive set of
file-based data products that can be manipulated interactively, including source
images, event lists, light curves, and spectra from each observation in which a
source is detected. 
\end{abstract}

\keywords{catalogs --- X-rays:~general}

\section{INTRODUCTION}

Ever since {\it Uhuru\/} \citep{gia71}, X-ray astronomy missions have had a
tradition of publishing catalogs of detected X-ray sources, and these catalogs
have provided the fundamental datasets used by numerous studies aimed at
characterizing the properties of the X-ray sky.  While source catalogs are the
primary data products from X-ray sky surveys \citep[e.g.,][]{gia72, for78,
elv92, vog93, vog99}, the {\it Einstein\/} IPC catalog \citep{har90}
demonstrated the utility of catalogs of {\it serendipitous\/} sources identified
in the fields of {\it pointed-observation\/} X-ray missions. More recent
serendipitous source catalogs \citep[e.g.,][]{gio90, whi94, ued05, wat08} have
further expanded the list of sources with X-ray data available for further
analysis by the astronomical community.

Source catalogs typically include a uniform reduction of the mission data. This
provides a significant advantage for the general scientific community because it
removes the need for end-users, who may be unfamiliar with the complexities of
the particular mission and its instruments, to perform detailed reductions for
each observation and detected source.

When compared to all previous and current X-ray missions, the {\it Chandra\/}
X-ray Observatory \citep[e.g.,][]{wei00, wei02} breaks the resolution barrier
with a sub-arcsecond on-axis point spread function (PSF). Launched in 1999, {\it
Chandra\/} continues to provide a unique high spatial resolution view of the
X-ray sky in the energy range from $0.1$ to $10\,\rm keV$, over a
$\sim\!60$--$250$ square arcminute field of view. The combination of excellent
spatial resolution, a reasonable field of view, and low instrumental background
translate into a high detectable-source density, with low confusion and good
astrometry. {\it Chandra\/} includes two instruments that record images of the
X-ray sky. The Advanced CCD Imaging Spectrometer \citep[ACIS;][]{bau98, gar03}
instrument incorporates ten $1024\times1024$ pixel CCD detectors (any six of
which can be active at one time) with an effective pixel size of $\sim\!0.5''$
on the sky, an energy resolution of order $110\,\rm eV$ at the Al-K edge
($1.49\,\rm keV$), and a typical time resolution of $\sim\!3.2\,\rm s$. The High
Resolution Camera \citep[HRC;][]{mur00} instrument consists of a pair of large
format micro-channel plate detectors with a pixel size $\sim\!0.13''$ on the sky
and a time resolution of $\sim\!15.6\ \mu\rm s$, but with minimal energy
resolution.  The wealth of information that can be extracted from identified
serendipitous sources included in {\it Chandra\/} observations is a powerful and
valuable resource for astronomy.

\begin{figure}
\epsscale{1.0}
\plotone{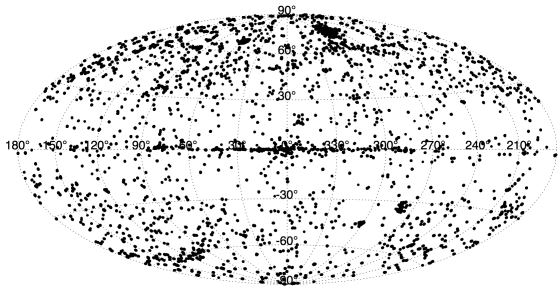}
\caption{\label{fig:sources}
Distribution of CSC release 1.0 master sources on the sky, in Galactic
coordinates.}
\end{figure}
 
The aim of the \dataset [ADS/Sa.CXO#CSC] {{\it Chandra\/} Source Catalog} (CSC)
is to disseminate this wealth of information by characterizing the X-ray sky as
seen by {\it Chandra\/}.  While numerous other catalogs of X-ray sources
detected by {\it Chandra\/} may be found in the literature
\citep[e.g.,][]{zez06, bra08, rom08, luo08, mun09, elv09}, the region of the sky
or set of observations that comprise these catalogs is restricted, and they are
typically aimed at maximizing specific scientific goals.  In contrast, the CSC
is intended to be an all-inclusive, uniformly processed dataset that can be
utilized to address a wide range of scientific questions.  The CSC is intended
ultimately to comprise a definitive catalog of X-ray sources detected by {\it
Chandra\/}, and is being made available to the astronomical community in a
series of increments with increasing capability over the next several years.

The first release of the CSC was published in 2009 March. This release includes
information about 135,914 source detections, corresponding to 94,676 distinct
X-ray sources on the sky, extracted from a {\it subset\/} of public imaging
observations obtained using the ACIS instrument during the first eight years of
the {\it Chandra\/} mission. The distribution of release~1 sources on the sky is
presented in Figure~1.

We expect that the CSC will be a highly valuable tool for many diverse
scientific investigations.  However, the catalog is constructed from pointed
observations obtained using the {\it Chandra\/} X-ray Observatory, and is
neither all-sky nor uniform in depth.  The first release of the catalog includes
only point and compact sources, with observed extents $\lesssim 30''$.  Because
of the difficulties inherent in detecting highly extended sources and point and
compact sources that lie close to them, and quantifying in a consistent and
robust way the properties of such sources, we have chosen to exclude entire
fields (or in some cases, individual ACIS CCDs) containing such sources from the
first release of the CSC, as described in \S~3.1.  Therefore, the
catalog does not include sources near some of the most famous {\it Chandra\/}
targets, and there may be selection effects that restrict the source content of
the catalog and which therefore may limit scientific studies that require
unbiased source samples.

The minimum flux significance threshold for a source to be included in the first
release of the CSC is set conservatively, and corresponds typically to
$\sim\!10$ detected source photons (on-axis) in the broad energy band integrated
over the total exposure time.  This conservative threshold was chosen to
maintain the spurious source rate at an acceptable level over the wide variety
of {\it Chandra\/} observations that are included in this release of the
catalog.  We expect to relax this criterion in future releases based on
experience gained constructing the current release.

A number of other {\it Chandra\/} catalogs do include sources with fewer net
counts than the CSC\null.  Such fainter thresholds are attainable typically
either because of specific attributes of the observations included in those
catalogs, or because of the assumptions made when constructing the catalog.

As an example of the former category, the XBootes survey catalog \citep{ken05}
includes sources that are roughly a factor of two fainter than the CSC flux
significance threshold.  That survey is constructed from short ($5\rm\,ks$)
observations obtained in an area with low line-of-sight absorption.  This
results in a negligible background level that substantially simplifies source
detection and enables identification of sources with very few counts.  Some {\it
Chandra\/} catalogs derived from observations with the range of exposures
comparable to those that comprise the CSC \citep[e.g.,][]{elv09, lai09, mun09}
also include fainter sources.  However, in these cases the additional source
fractions are in general not large, typically adding $\lesssim 10\%$ more
sources below the CSC threshold, as described in detail in \S~3.7.1.

For other {\it Chandra\/} catalogs, visual review and validation at the source
level is a planned part of the processing thread \citep[e.g.,][]{kim07, mun09}.
In some cases \citep[e.g.,][]{bro07}, visual review may be used to adjust
processing parameters for individual sources.  Such manual steps are
time-consuming, but enable lower significance levels to be achieved while
maintaining an acceptable spurious source rate.  In contrast, the CSC catalog
construction process requires that the processing pipelines run on a wide range
of observations with a minimum of manual intervention.  The scope of the CSC is
simply too large to require manual handling at the source level.  We do not
manually inspect individual source detections, nor do we adjust source detection
or processing parameters based on manual evaluation.  Instead, the CSC uses
a largely automated quality assurance approach, as described in
\S~3.14.

The sky coverage of the first catalog release (Fig.~2) totals
$\sim\!320$ square degrees, with coverage of $\sim\!310$ square degrees brighter
than a $0.5$--$7.0\rm\,keV$ flux limit of
$1.0\times10^{-13}\rm\,erg\,cm^{-2}\,s^{-1}$, decreasing to $\sim\!135$ square
degrees brighter than $1.0\times10^{-14}\rm\,erg\,cm^{-2}\,s^{-1}$, and
$\sim\!6$ square degrees brighter than
$1.0\times10^{-15}\rm\,erg\,cm^{-2}\,s^{-1}$. These numbers will continue to
grow as the {\it Chandra\/} mission continues, with a 15 year prediction of the
eventual sky coverage of the CSC of order 500 square degrees, or a little over
1\% of the sky.

\begin{figure}
\epsscale{1.0}
\plotone{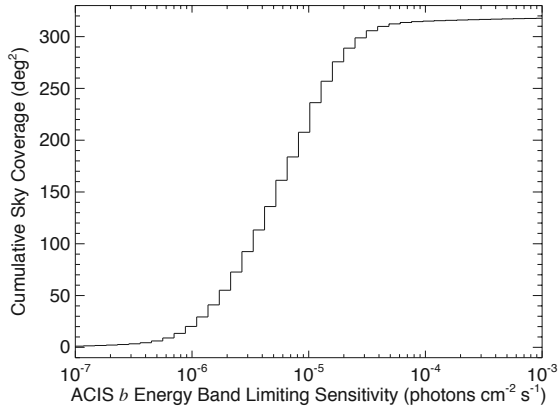}
\caption{\label{fig:skycov}
Sky coverage of release 1.0 of the CSC, in the ACIS broad energy band.  
The ordinate value is the total sky area included in the CSC that is sensitive
to point sources with fluxes at least as large as the corresponding value on the
abscissa.} 
\end{figure}
 
In this paper we describe in detail the content and construction of release~1 of
the CSC\null.  However, where appropriate we also discuss in addition the steps
required to process HRC instrument data used to construct release~1.1 of the
catalog, since the differences in the algorithms are small.  Release~1.1 of the
catalog is scheduled for spring 2010.  This paper is organized into 5 sections,
including the introduction.  In \S~2, we present a description of the catalog.
This includes the catalog design goals, an outline of the general
characteristics of {\it Chandra\/} data that are relevant to the catalog design,
the organization of the data within the catalog, approaches to data access, and
an outline of the data content of the catalog.  Section~3, which comprises the
bulk of the paper, describes in detail the methods used to extract the various
source properties that are included in the catalog, with particular detail
provided when the algorithms are new or have been adapted for use with {\it
Chandra\/} data.  A brief description of the principal statistical properties
of the catalog sources is presented in \S~4; this topic is treated
comprehensively by \citetfap.  Conclusions are presented in \S~5.  Finally,
Appendix~A contains details of the algorithm used to match source detection from
multiple overlapping observations, as well as the mathematical derivation of the
multivariate optimal weighting formalism used for combining source position and
positional uncertainty estimates from multiple observations.

\section{CATALOG DESCRIPTION}

\subsection{Design Goals}

The CSC is intended to be a general purpose virtual science facility, and
provides simple access to a carefully selected set of generally useful
quantities for individual sources or sets of sources matching user-specified
search criteria. The catalog is designed to satisfy the needs of a broad-based
group of scientists, including those who may be less familiar with astronomical
data analysis in the X-ray regime, while at the same time providing more
advanced data products suitable for use by astronomers familiar with {\it
Chandra\/} data.

The primary design goals for the CSC are to (1)~allow simple and quick access to
the best estimates of the X-ray source properties for detected sources, with
good scientific fidelity, and directly support scientific analysis using the
individual source data; (2)~facilitate analysis of a wide range of statistical
properties for classes of X-ray sources; (3)~provide efficient access to
calibrated observational data and ancillary data products for individual X-ray
sources, so that users can perform detailed further analysis using existing
tools such as those included in the {\it Chandra\/} Interactive Analysis of
Observations (CIAO; \citealp{fru06}) portable data analysis package; and
(4)~include all real X-ray sources detected down to a predefined threshold level
in all of the public {\it Chandra\/} datasets used to populate the catalog,
while maintaining the number of spurious sources at an acceptable level.

To achieve these goals, for each detected X-ray source the catalog records the
source position and a detailed set of source properties, including commonly used
quantities such as multi-band aperture fluxes, cross-band hardness ratios,
spectra, temporal variability information, and source extent estimates. In
addition to these traditional elements, the catalog includes file-based data
products that can be manipulated interactively by the user.  The primary data
products are photon event lists \citep[e.g.,][]{con92}, which record measures of
the location, time of arrival, and energy of each detected photon event in a
tabular format.  Additional data products derived from the photon event list
include images, light curves, and spectra for each source individually from each
observation in which a source is detected. The catalog release process is
carefully controlled, and a detailed characterization of the statistical
properties of the catalog to a well defined, high level of reliability
accompanies each release. Key properties evaluated as part of the statistical
characterization include limiting sensitivity, completeness, false source rates,
astrometric and photometric accuracy, and variability information.

\subsection{Data Characteristics}

Both ACIS and HRC cameras operate in a photon counting mode, and register
individual X-ray photon events. For each photon event, the two dimensional
position of the event on the detector is recorded, together with the time of
arrival and a measure of the energy of the event. In most operating modes, lists
of detected events are recorded over the duration of an observation, typically
between $1\,\rm ks$ and $160\,\rm ks$, and are then telemetered to the ground
for subsequent processing. 

To minimize the effect of bad detector pixels, and to avoid possible burn-in
degradation of the camera by bright X-ray sources, the pointing direction of the
telescope is normally constantly dithered in a Lissajous pattern, with a typical
scale length of about $20''$ on the sky and a period of order $1\,\rm ks$, while
taking data.  The motion of the telescope is recorded via an ``aspect camera''
\citep{ald00} that tracks the motion of a set of (usually 5) guide stars as a
function of time during the observation. The coordinate transformation needed to
remove the motion from the event (photon) positions is computed from the aspect
camera data and applied during data processing.

Breaking down the 4-dimensional X-ray data hypercube into spatial, spectral, and
temporal axes provides a natural focus on the properties that may be of interest
to the general user, but also identifies some of the complexities inherent in
{\it Chandra\/} data that must be addressed by catalog construction and data
analysis algorithms.

\begin{figure}
\epsscale{1.0}
\plotone{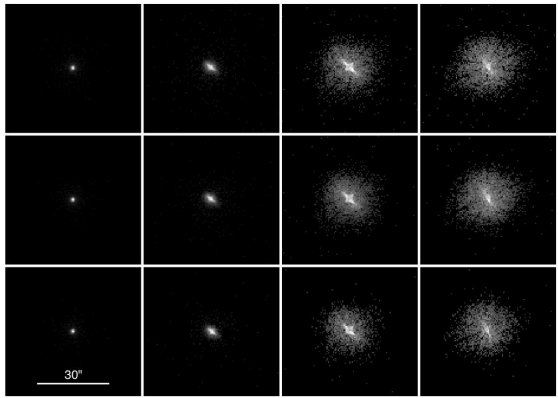}
\caption{\label{fig:psf}
Sample local {\it Chandra\/} model PSFs projected onto the ACIS detector
pixel plane extracted from the CSC.  The upper, middle, and lower sets of 4 
images correspond to PSF models computed at the monochromatic effective 
energies of the ACIS hard, medium, and soft energy bands, respectively.  From
left to right, the images correspond to PSFs determined at off-axis angles 
$\theta = 0'$, $5'$, $10'$, and $15'$, respectively. The orientation and details
of the PSF substructure varies with azimuthal angle, $\phi$.  The image
intensity scaling is proportional to the square root of the pixel flux.}
\end{figure}

Spatially, the {\it Chandra\/} PSF varies significantly with off-axis and
azimuthal angle (with the former variation dominating), as well as with incident
photon energy (Fig.~3). Close to the optical axis of the telescope,
the PSF is approximately symmetric with a 50\% enclosed energy fraction radius
of order $0.3''$ over a wide range of energies, but at $15'$ off-axis the PSF is
strongly energy-dependent, asymmetric, and significantly extended, with a 50\%
enclosed energy fraction radius of order $13''$ at $1.5\,\rm keV$.

For the widely used ACIS detector, the instrumental spectral energy resolution
is of order $100$--$200\,\rm eV$, and depends on incident photon energy and
location on the detector. Because the energy resolution is significantly lower
than the typical energy width of the features and absorption edges that define
the effective area of the telescope optics (and therefore the quantum efficiency
of the telescope plus detector system), a full matrix formulation that considers
the redistribution of source X-ray flux into the set of instrumental pulse
height analyzer bins must be used when performing spectral analyses. This is in
contrast to the more familiar scenario from many other wavebands, where the
instrumental resolution is often much higher than the spectral variation of
quantum efficiency, enabling the commonly used implicit assumption that the flux
redistribution matrix is diagonal (and is therefore not considered explicitly).

We note in passing that {\it Chandra\/} is equipped with a pair of transmission
gratings that can be inserted into the optical path, and is therefore capable of
performing high spectral resolution (slitless spectroscopy) observations.
However, such observations are not included in the current release of the
CSC\null.

\begin{figure}
\epsscale{1.0}
\plotone{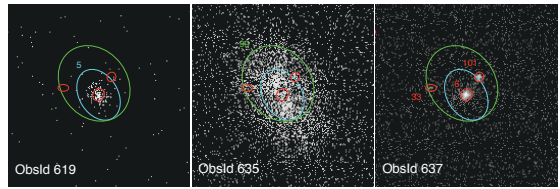}
\dataset [ADS/Sa.CXO#obs/00619] {}
\dataset [ADS/Sa.CXO#obs/00635] {}
\dataset [ADS/Sa.CXO#obs/00637] {}
\caption{\label{fig:srcobs}
Three separate observations that include the area surrounding the bright X-ray
source CXO~J162624.0-242448 are shown.  In each panel source detections from
observations 00619, 00635, and 00637 are identified in cyan, green, and red, 
respectively.  {\it Left:\/} Observation 00619 ($4.1\rm\,ks$ exposure).  In 
this short exposure, only the bright source visible at an off-axis angle of 
$\sim\!7.7'$.  The PSF is somewhat extended.  {\it Center:\/} Observation 00635 
($100.7\rm\,ks$ exposure).  In this deep exposure, the bright X-ray source is 
located $\sim\!15.6'$ off-axis in this deep exposure.  The extended PSF is 
clearly visible.  {\it Right:\/} Observation 00637 ($96.4\rm\,ks$ exposure).  
The bright source is located $\sim3.0'$ off-axis, and the combination of the 
compact PSF and long exposure resolves the region into 3 distinct source 
detections.}
\end{figure}

Time domain analyses must consider the impact of spacecraft dither within an
observation. Strong false variability signatures at the dither frequency can
arise because of variations of the quantum efficiency over the detector, or
because the source or background region dithers off the detector edge or across
a gap between adjacent ACIS CCDs. Corrections for these effects, as well as for
cosmic X-ray background flares that can be highly variable over periods of a few
kiloseconds, must be applied when computing light-curves. The extremely low
photon event rates common for many faint X-ray sources typically require time
domain statistics to be evaluated using event arrival-time formulations instead
of rate-based approaches.

An additional level of complexity occurs because many astronomical sources of
interest that will be included in the catalog are extremely faint.  Rigorous 
application of Poisson counting statistics is required when deriving source 
properties and associated errors, separating X-ray analyses from many other 
wavebands where Gaussian statistics are typically assumed.

\subsection{Data Organization\label{sec:org}}

The tabulated properties included in the CSC are organized conceptually into two
separate tables, the {\it Source Observations Table\/} and the {\it Master
Sources Table\/}. Distinguishing between source detections (as identified
within a single observation) and X-ray sources physically present on the sky is
necessary because many sources are detected in multiple observations and at
different off-axis angles (and therefore have different PSF extents).

Each record included in the Source Observations Table tabulates properties
derived from a source detection in a single observation. These entries also
include pointers to the associated file-based data products that are included in
the catalog, which are all observation-specific in the first catalog release.
Each record in the Source Observations Table is further split internally into a
set of source-specific data and a set of observation-specific, but
source-independent, data. The latter are recorded once to avoid duplication.  A
description of the data columns recorded in the Source Observations Table for
each source detection is provided in Table~1.

\begin{figure*}
\epsscale{1.4}
\vglue -0.75truein
\hglue -0.75truein
\plotone{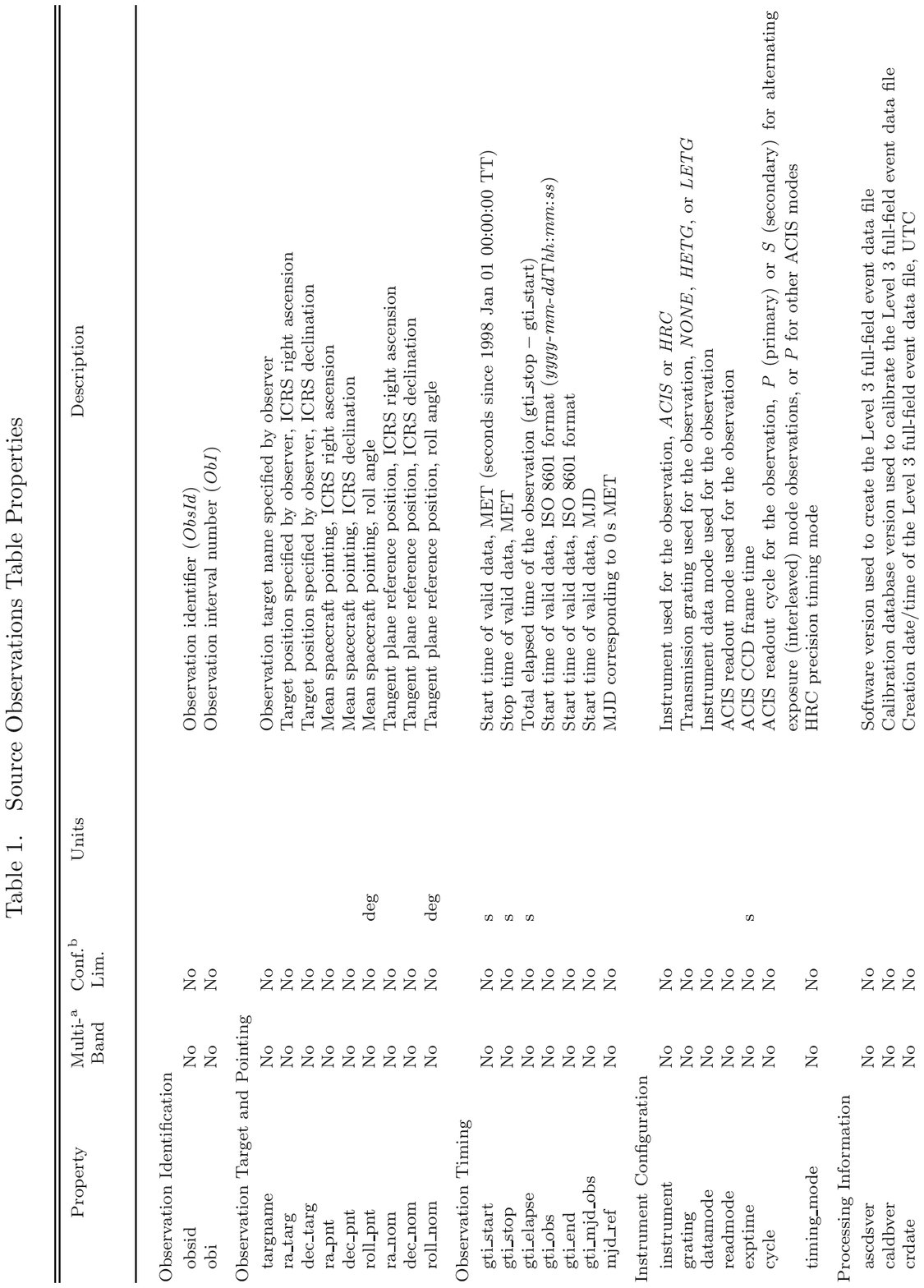}
\end{figure*}

\begin{figure*}
\epsscale{1.4}
\vglue -0.75truein
\hglue -0.75truein
\plotone{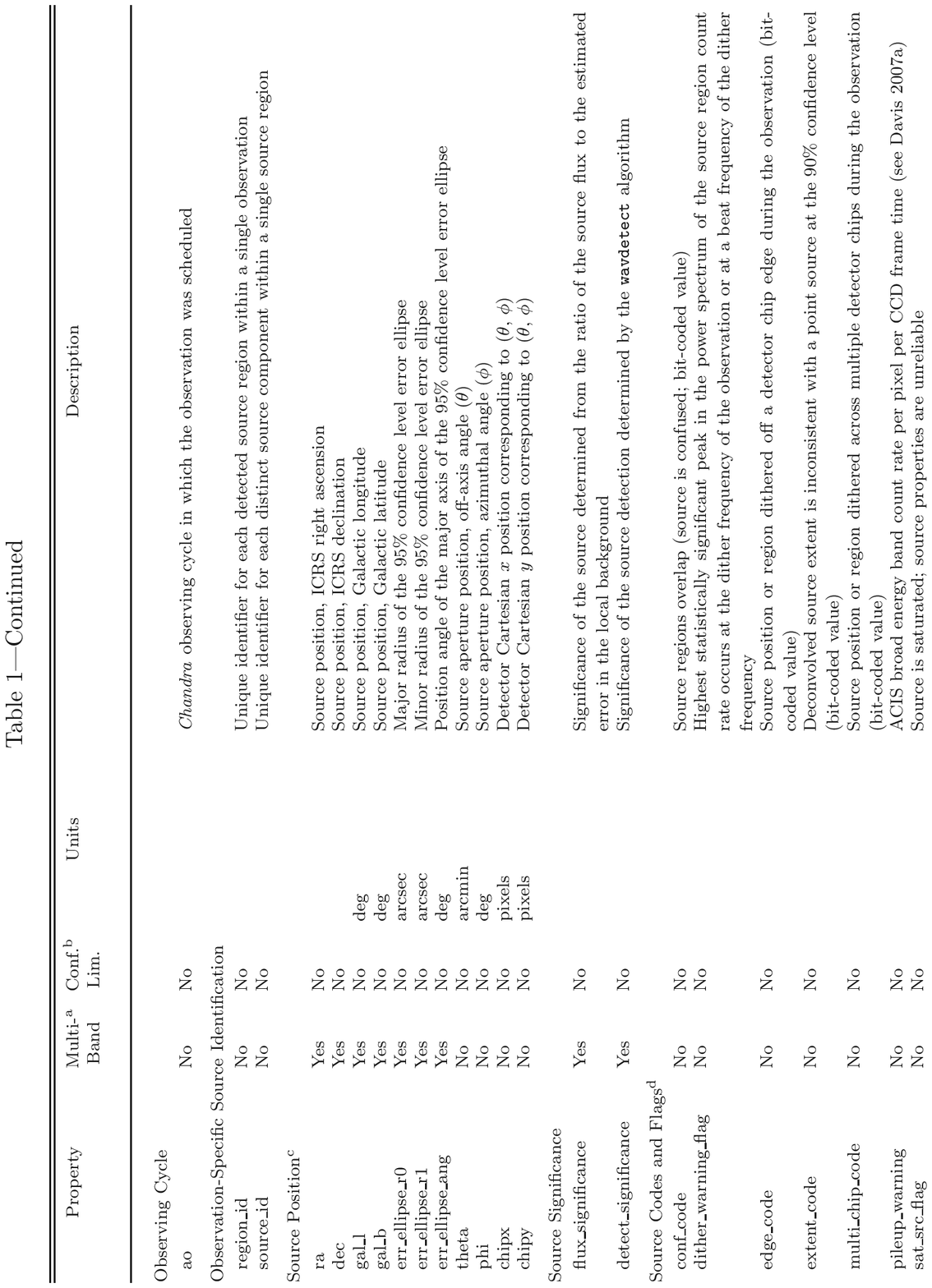}
\end{figure*}

\begin{figure*}
\epsscale{1.4}
\vglue -0.75truein
\hglue -0.75truein
\plotone{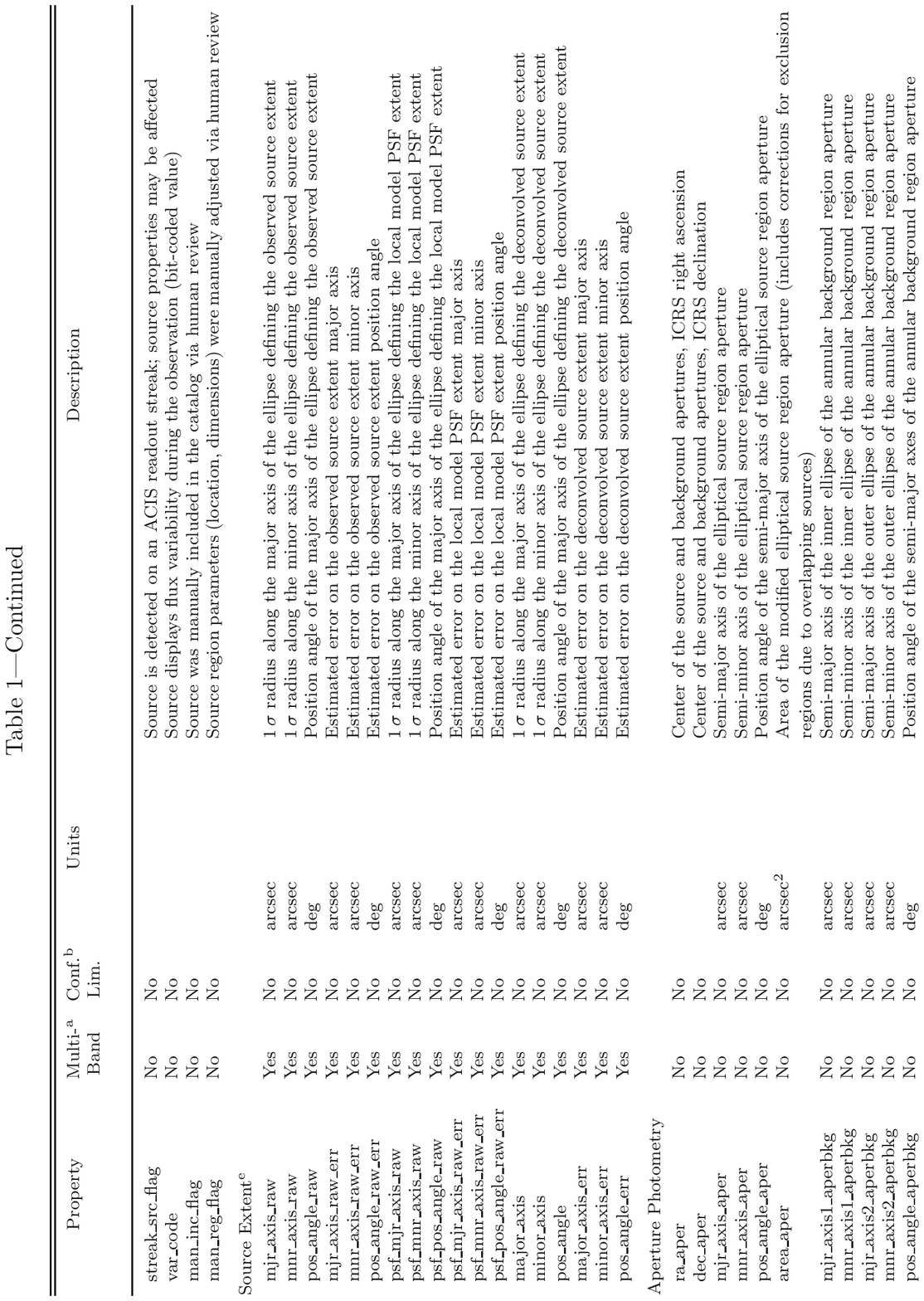}
\end{figure*}

\begin{figure*}
\epsscale{1.4}
\vglue -0.75truein
\hglue -0.75truein
\plotone{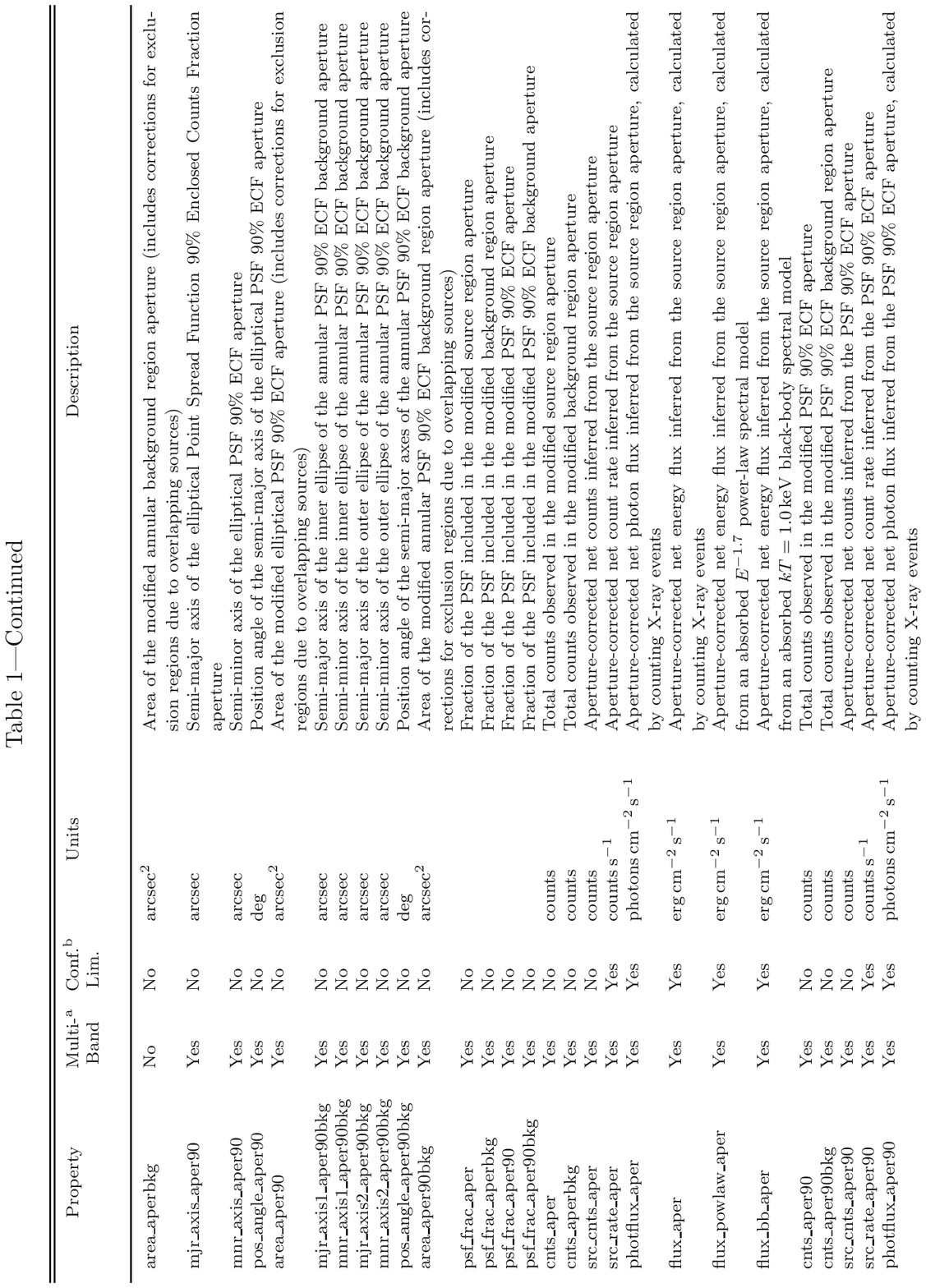}
\end{figure*}

\begin{figure*}
\epsscale{1.4}
\vglue -0.75truein
\hglue -0.75truein
\plotone{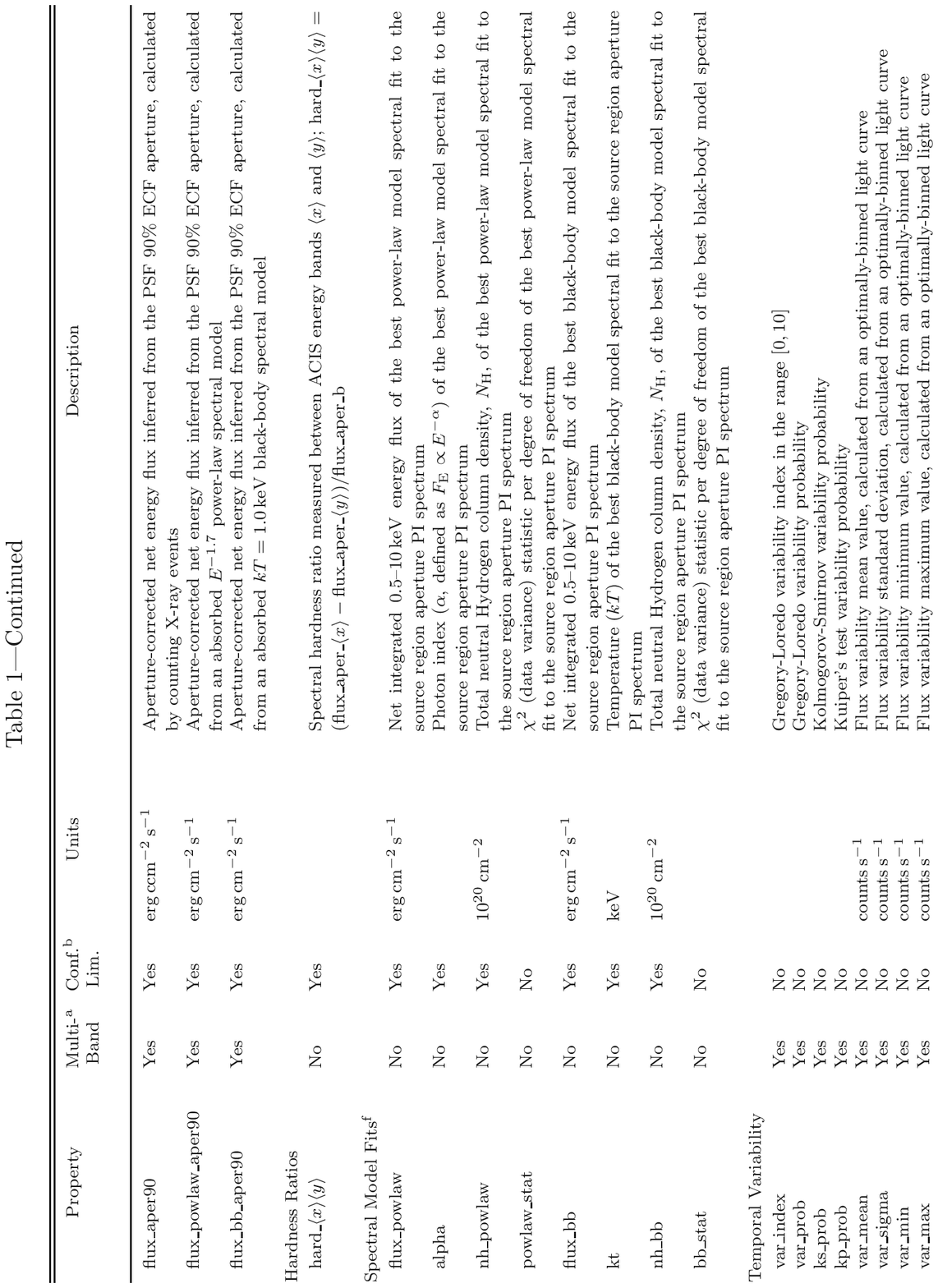}
\end{figure*}

\begin{figure*}
\epsscale{1.4}
\vglue -0.75truein
\hglue -0.75truein
\plotone{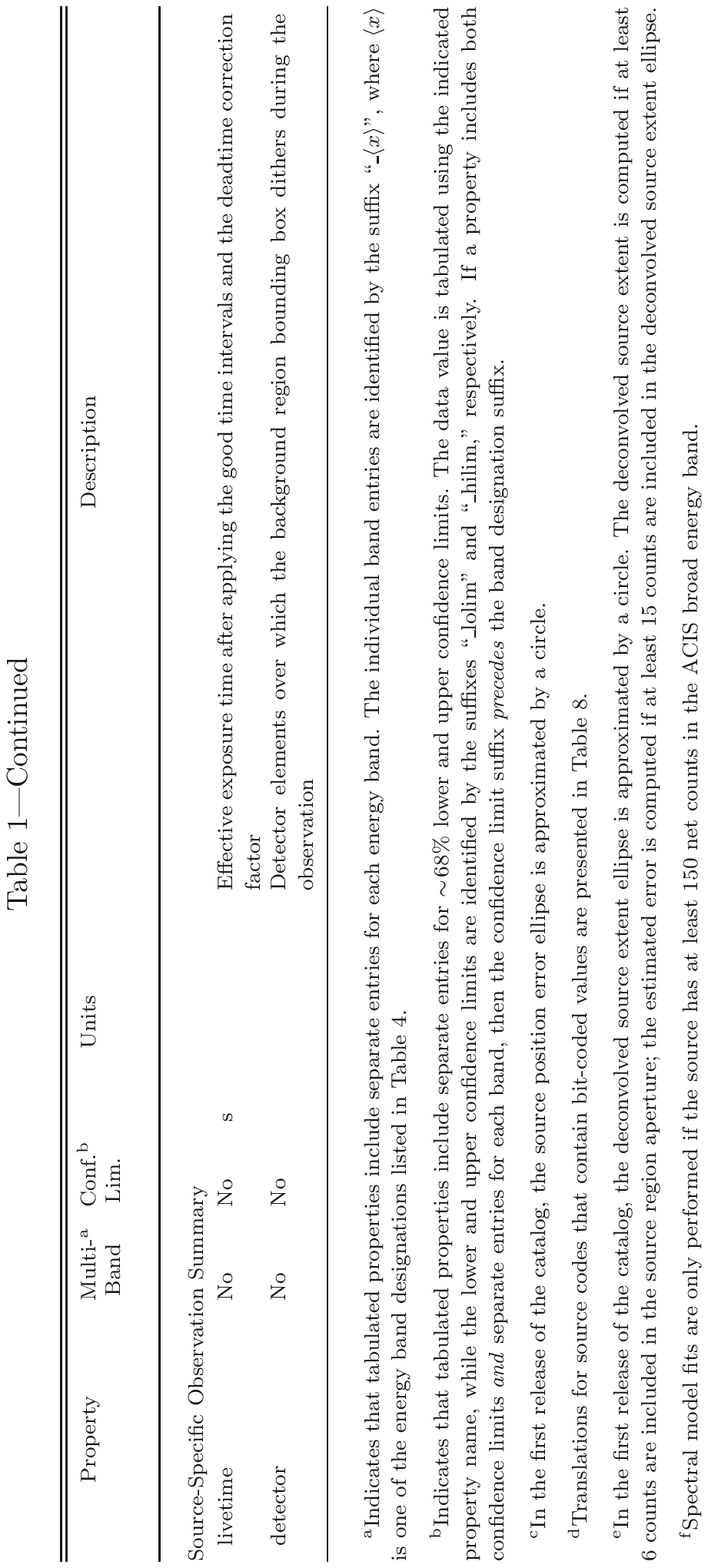}
\end{figure*}

\begin{figure*}
\epsscale{1.4}
\vglue -0.75truein
\hglue -0.75truein
\plotone{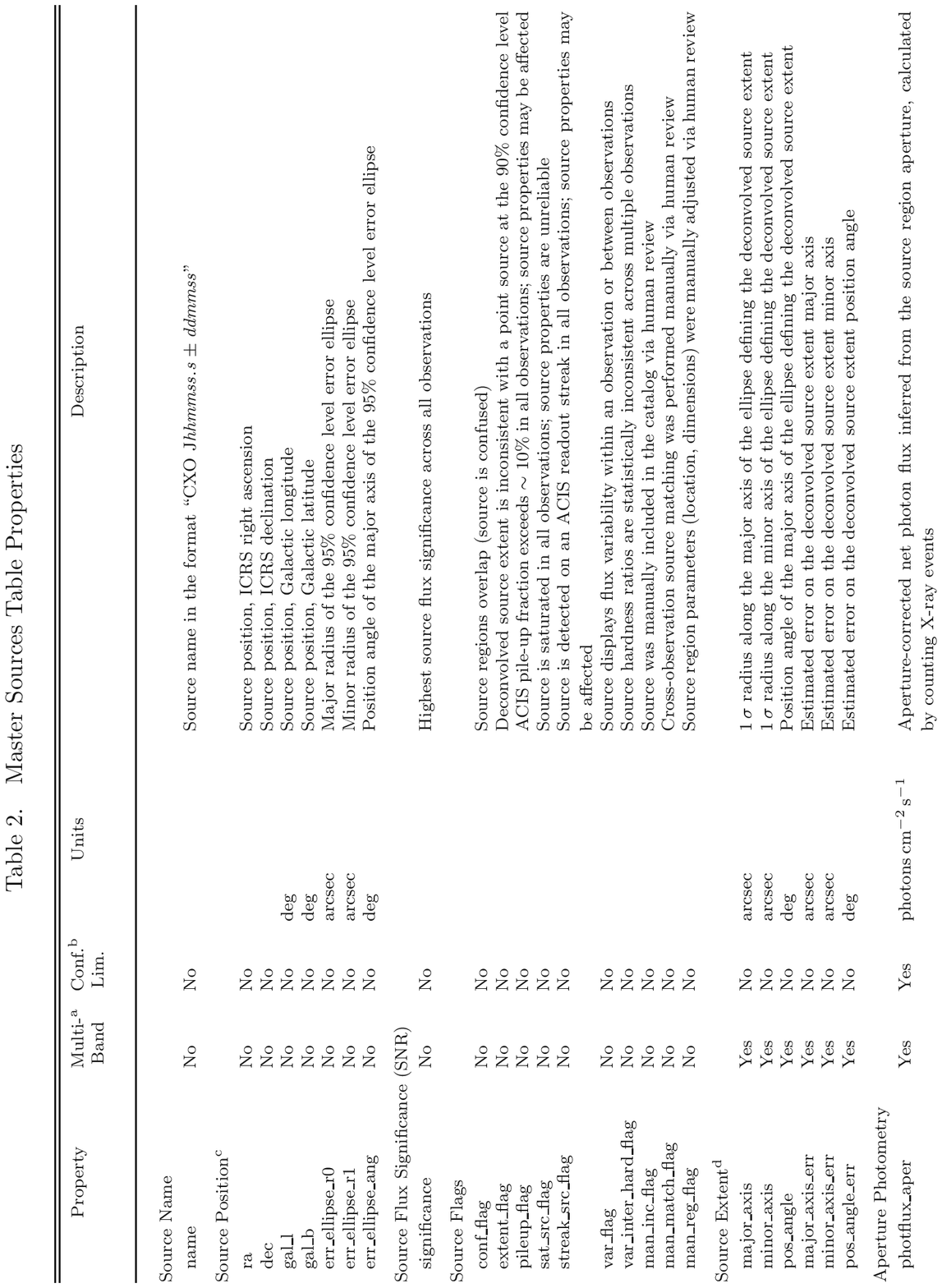}
\end{figure*}

\begin{figure*}
\epsscale{1.4}
\vglue -0.75truein
\hglue -0.75truein
\plotone{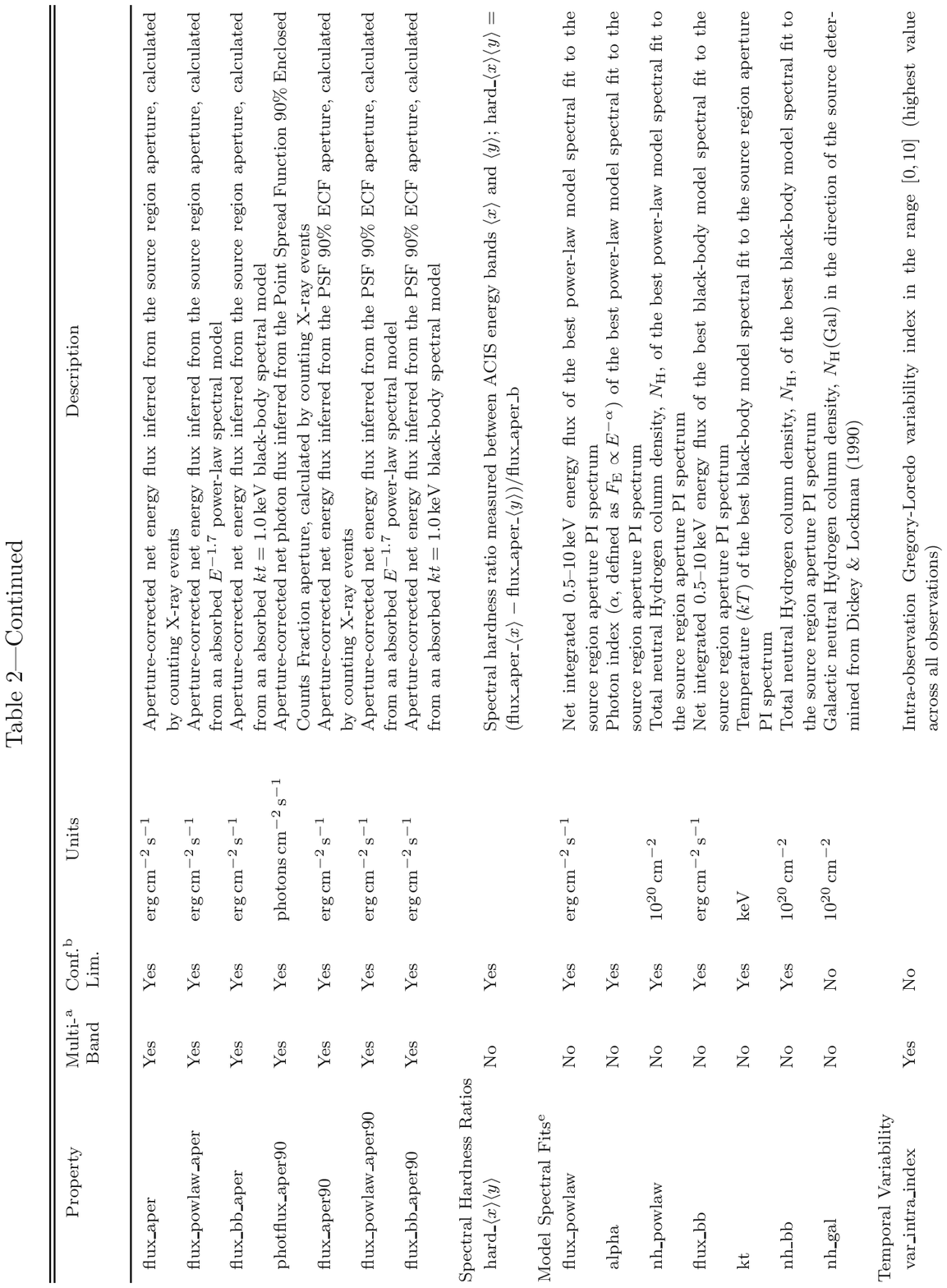}
\end{figure*}

\begin{figure*}
\epsscale{1.4}
\vglue -0.75truein
\hglue -0.75truein
\plotone{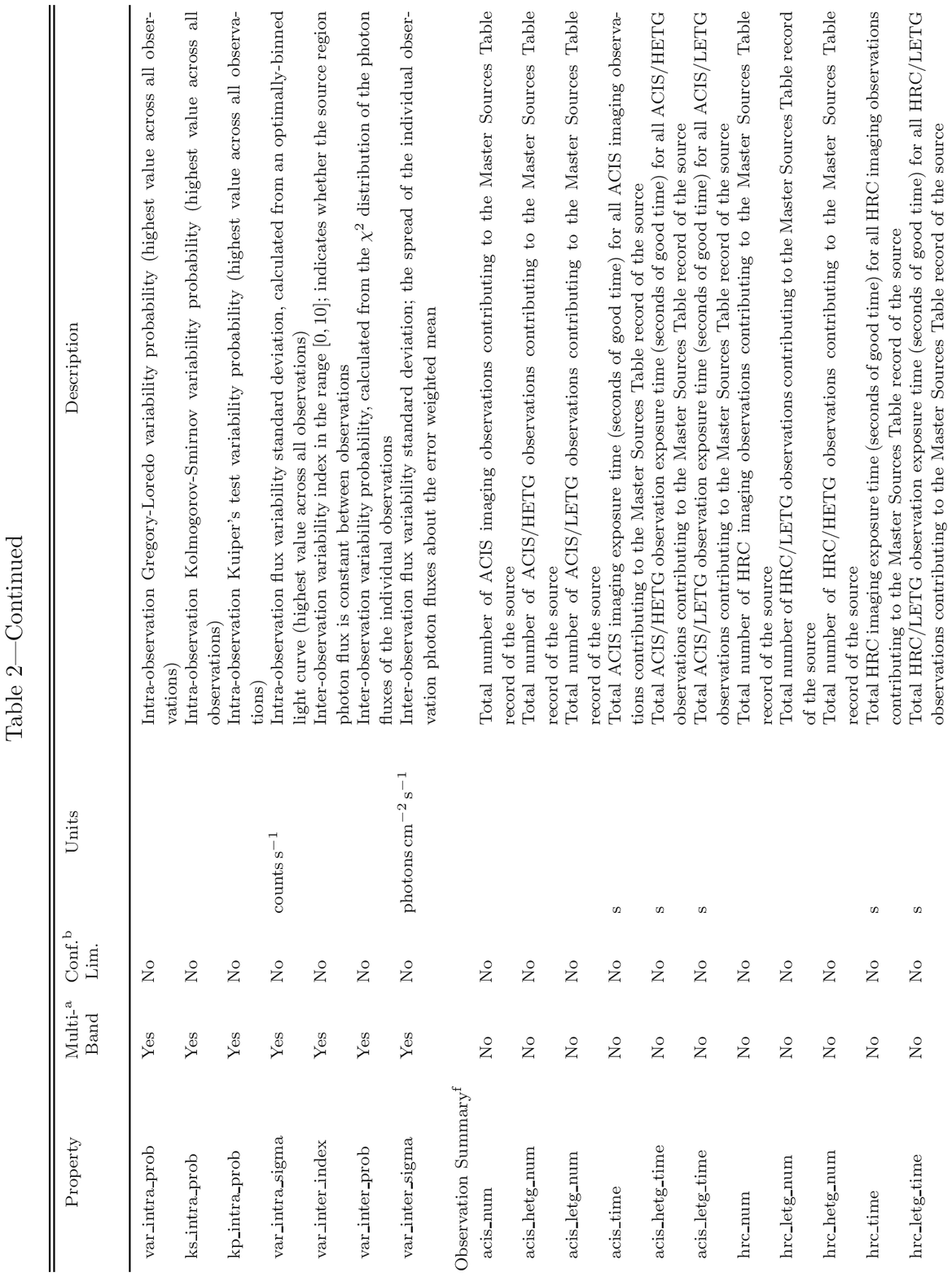}
\end{figure*}

\begin{figure*}
\epsscale{1.4}
\vglue -0.75truein
\hglue -0.75truein
\plotone{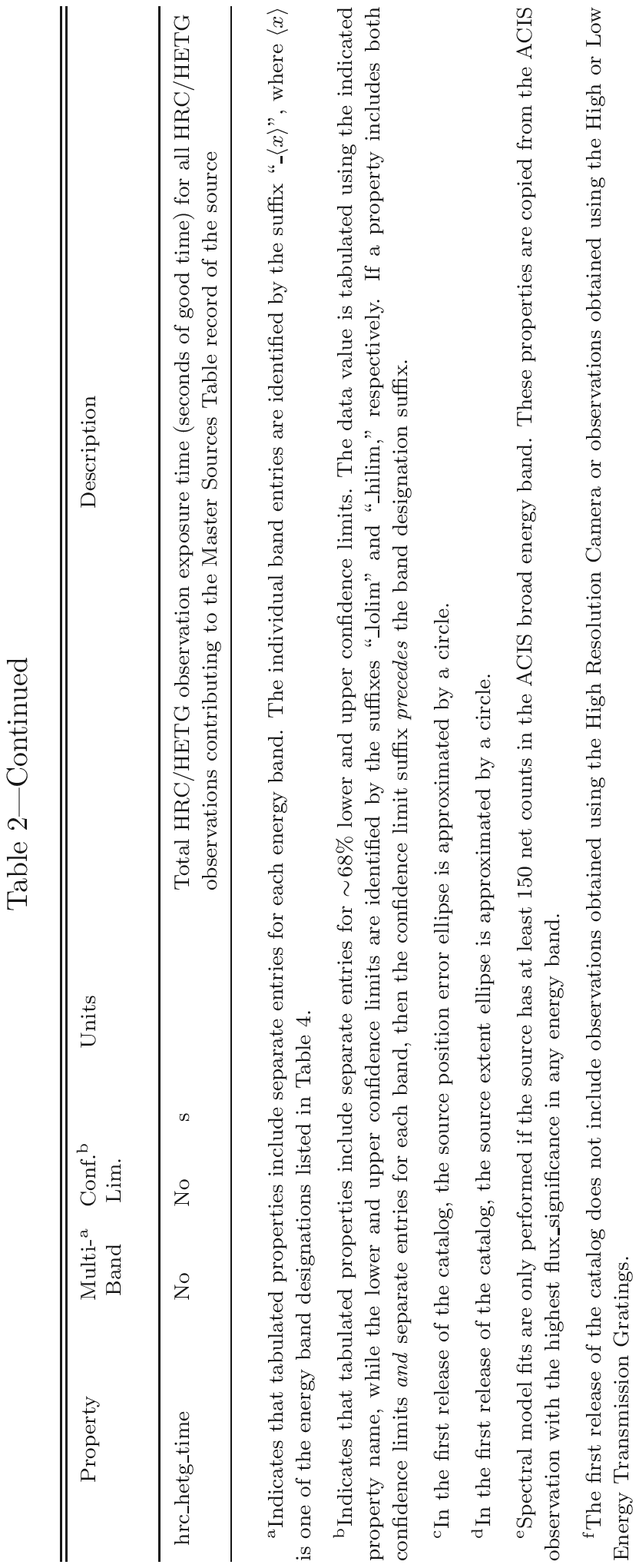}
\end{figure*}

\begin{figure*}
\epsscale{1.4}
\vglue -0.75truein
\hglue -0.75truein
\plotone{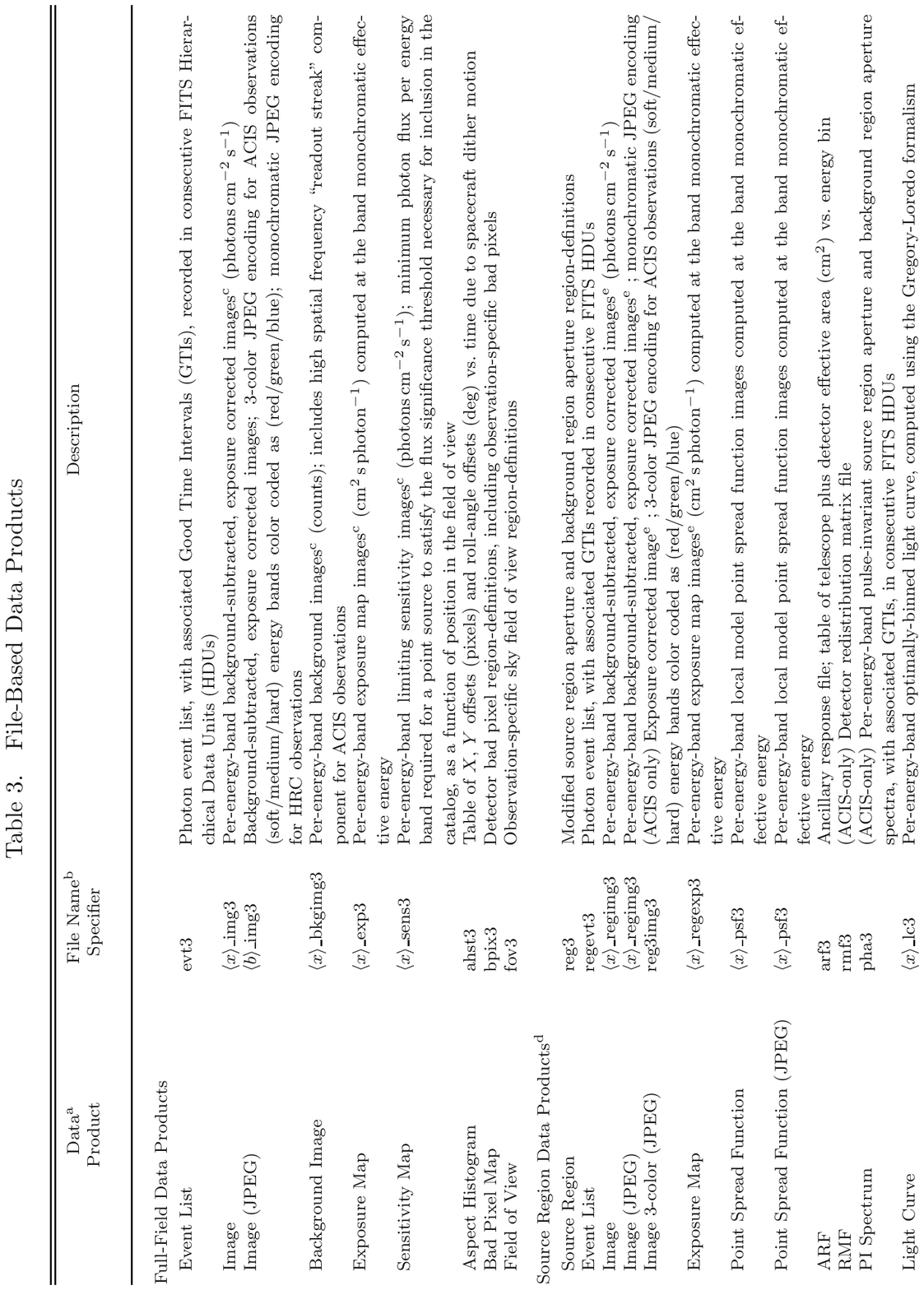}
\end{figure*}

\begin{figure*}
\epsscale{1.4}
\vglue -0.75truein
\hglue -0.75truein
\plotone{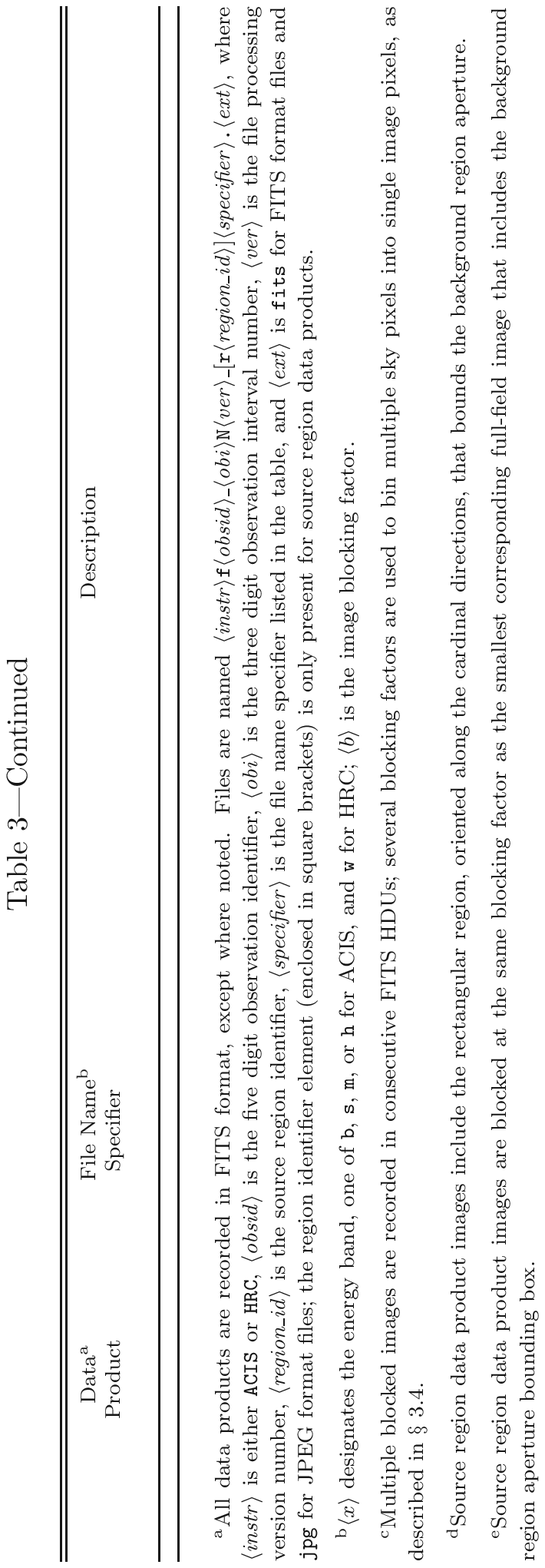}
\end{figure*}

\setcounter{table}{3}

Because of the dependence of the PSF extent with off-axis angle, multiple
distinct sources detected on-axis in one observation may be detected as a single
source if located far off-axis in a different observation
(Fig.~4). During catalog processing, source detections from all
observations that overlap the same region of the sky are spatially matched to
identify distinct X-ray sources. Estimates of the tabulated properties for each
distinct X-ray source are derived by combining the data extracted from all
source detections and observations that can be uniquely associated, according to
the algorithms described in \S~3. The best estimates of the
source properties for each distinct X-ray source are recorded in the Master
Sources Table. A description of the data columns recorded in the Master Sources
Table for each source is provided in Table~2.

Each distinct X-ray source is thus conceptually represented in the catalog by a
single entry in the Master Sources Table, and one or more associated entries in
the Source Observations Table (one for each observation in which the source was
detected).

All of the tabulated properties included in both the Master Sources Table and
the Source Observations Table can be queried by the user. Bi-directional links
between the entries in the two tables are managed transparently by the database,
so that the user can access all observation data for a single source seamlessly.

If a source detection included in the Source Observations Table can be related
unambiguously to a single X-ray source in the Master Sources Table, then the
corresponding table entries will be associated by ``unique'' linkages.  Source
detections included in the Source Observations Table that cannot be related
uniquely to a single X-ray source in the Master Source Table will have their
entries associated by ``ambiguous'' linkages (Fig.~5).

\begin{figure}
\epsscale{1.0}
\plotone{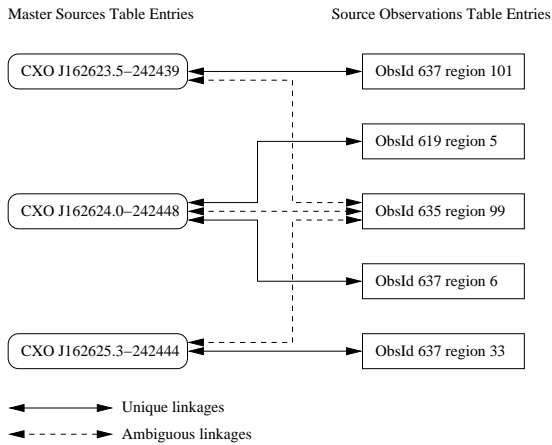}
\caption{\label{fig:linkages}
Linkages between the Master Sources Table and the Source Observations Table
entries for the source detections from Fig.~4 are depicted.
The 3 source detections in observation 00637 are uniquely identified with
distinct X-ray sources on the sky, and will be associated with the
corresponding master sources through ``unique'' linkages.  Similarly, the
single source detection (region 5) in observation 00619 is an unambiguous
match to region 6 in observation 00637, and so will also be associated with
the same master source via a unique linkage.  The confused detection, region
99 in observation 00635 overlaps the 3 source detections in observation 00637,
and so is associated with the corresponding master sources with ``ambiguous''
linkages.}
\end{figure}

The data from ambiguous source detections are not used when computing the best
estimates of the source properties included in the Master Sources Table.  In the
case of ACIS observations, source detections for which the estimated photon
pile-up fraction \citep{jed07a} exceeds $\sim\!10\%$ will not be used if source
detections in other ACIS observations do not exceed this threshold.

Using the linkages between the entries in the two tables, the user will
nevertheless be able to identify all of the X-ray sources in the catalog that
could be associated with a specific detection in a single observation, and
vice-versa. These linkages may be important, for example, when identifying
candidate targets for follow-up studies based on a data signature that is only
visible in the observation data for a confused source.

\subsection{Data Access}

The primary user tool for querying the CSC is the CSCview web-browser interface
\citep{zog08}, which can be accessed from the public catalog
web-site\footnote{\url{http://cxc.cfa.harvard.edu/csc/}}.  The user can directly
query any of the tabulated properties included in either the Master Sources
Table or the Source Observations Table, display the contents of an arbitrary set
of properties for matching sources, and retrieve any of the associated
file-based data products for further analysis. CSCview provides a form-based
data-mining interface, but also allows users to enter queries written using the
Astronomical Data Query Language (ADQL; \citealp{adql}) standard directly.
Query results can be viewed directly on the screen, or saved to a data file in
multiple formats, including tab-delimited ASCII (which can be read directly by
several commonly used astronomical applications) and International Virtual
Observatory Alliance\footnote{\url{http://www.ivoa.net/}} (IVOA) standard
formats such as VOTable \citep{VOT}.

Automated access to query the catalog from data analysis applications and
scripts running on the user's home platform was identified as being needed for
several science use cases.  VO standard interfaces, including Simple Cone Search
\citep{VOcone} and Simple Image Access \citep{SIAP}, provide limited query and
data access capabilities, while more sophisticated interactions are possible
through a direct URL connection. Support for VO workflows using applicable
standards will be added in the future as these standards stabilize. An interface
that integrates catalog access with a visual sky browser provides a simple
mechanism for visualizing the regions of the sky included in the catalog, and
may also be particularly beneficial for education and public outreach purposes.

Since {\it Chandra\/} is an ongoing mission, the CSC includes a mechanism to
permit newly released observations to be added to the catalog and be made
visible to end users, while at the same time providing stable, well-defined, and
statistically well-characterized released catalog versions to the community.
This is achieved by maintaining a revision history for each database table
record, together with flags that establish whether catalog quality assurance and
catalog inclusion criteria are met, and using distinct views of the catalog
databases that utilize these metadata.

``Catalog release views'' provide access to each released version of the
catalog, with the latest released version being the default. Catalog releases
will be infrequent (no more than of order 1 per year) because of the controls
built in to the release process, and because of the requirement that each
release be accompanied by a detailed statistical characterization of the
included source properties. Once data are included in a catalog release view,
then they are frozen in that view, even if the source properties are revised or
the source is deleted in a later catalog release. A source may be deleted 
if the detection is subsequently determined to be an artifact of the data or
processing, but the most likely reason that a source is deleted from a later 
catalog release is that additional observations included in the later release 
resolve the former detection into multiple distinct sources.

\begin{deluxetable*}{ccccccc}
\tabletypesize{\small}
\tablecolumns{7}
\tablewidth{0pt}
\tablecaption{CSC Energy Bands\label{tab:ebands}}
\tablehead{
  \multicolumn{2}{c}{Band} 		& \colhead{Energy\tablenotemark{a}}	& \colhead{Monochromatic\tablenotemark{a}} & \multicolumn{3}{c}{Integrated Effective Area\tablenotemark{b}} \\
\colhead{Name}	& \colhead{Designation}	& \colhead{Range}	& \colhead{Energy}  & \colhead{ACIS-I} & \colhead{ACIS-S} & \colhead{HRC-I}
}
\startdata
\cutinhead{ACIS Energy Bands}
Ultra-soft	& {\it u\/}	& 0.2--0.5	& 0.4\phn & 7.36--2.24 & 68.7--23.0 & \nodata \\
Soft		& {\it s\/}	& 0.5--1.2	& 0.92    & 216--155   & 411--274   & \nodata \\
Medium		& {\it m\/}	& 1.2--2.0	& 1.56    & 438--401   & 539--493   & \nodata \\
Hard		& {\it h\/}	& 2.0--7.0	& 3.8\phn & 1590--1580 & 1680--1670 & \nodata \\
Broad		& {\it b\/}	& 0.5--7.0	& 2.3\phn & 2240--2140 & 2630--2440 & \nodata \\
\cutinhead{HRC Energy Band}
Wide		& {\it w\/}	& 0.1--10	& 1.5\phn & \nodata    & \nodata    & 605 \\
\enddata
\tablenotetext{a}{$\rm keV$.}
\tablenotetext{b}{$\rm keV\,cm^2$, computed at the ACIS-I, ACIS-S, and HRC-I aimpoints.  For ACIS
energy bands, the pair of values are the integrated effective area with zero focal plane contamination 
(first number) and with the late 2009 level of focal plane contamination (second number).}
\end{deluxetable*}

``Database views'' provide access to the catalog database, including any new
content that may not be present in an existing catalog release. Because on-going
processing is continually modifying the catalog database, tabulated data and
file-based data products in a database view may be superseded at any time, and
the statistical properties of the data are not guaranteed.

We anticipate that users who require a stable, well-characterized dataset will
choose primarily to access the catalog through the latest catalog release view.
On the other hand, users who are interested in searching the latest data to
identify sources with specific signatures for further study will likely use the
latest database view.

\subsection{Data Content\label{sec:dcontent}}

The first release of the CSC includes detected sources whose flux estimates are
at least 3 times their estimated $1\,\sigma$ uncertainties, which typically
corresponds to about 10 net (source) counts on-axis and roughly 20--30 net
counts off-axis, in at least one energy band.  In this release, multiple
observations of the same field are {\it not\/} combined prior to source
detection, so the flux significance criterion applies to each observation
separately.  

For each source detected in an observation, the catalog includes approximately
120 tabulated properties. Most values have associated lower and upper confidence
limits, and many are recorded in multiple energy bands. The total number of
columns included in the Source Observations Table (including all values and
associated confidence limits for all energy bands) is 599.
 
Roughly 60 master properties are tabulated for each distinct X-ray source on the
sky, generated by combining measurements from multiple observations that include
the source. Combining all values and associated confidence limits for all energy
bands yields a total of 287 columns included in the Master Sources Table.

The tabulated source properties fall mostly into the following broad categories:
source name, source positions and position errors, estimates of the raw
(measured) extents of the source and the local point spread function, and the
deconvolved source extents, aperture photometry fluxes and confidence intervals
measured or inferred in several ways, spectral hardness ratios, power-law and
thermal black-body spectral fits for bright ($>150$ net counts) sources, and
several source variability measures (Gregory-Loredo, Kolmogorov-Smirnov, and
Kuiper tests).

Also included in the CSC are a number of file-based data products in formats
suitable for further analysis in CIAO\null. These products, described in
Table~3, include both full-field data products for each
observation, and products specific to each detected observation-specific source 
region.

The full-field data products include a ``white-light'' full-field photon event
list, and multi-band exposure maps, background images, exposure-corrected and
background-subtracted images, and limiting sensitivity maps.

Source-specific data products include a white-light photon event list, the
source and background region definitions, a weighted ancillary response file
(the time-averaged product of the combined telescope/instrument effective area
and the detector quantum efficiency), multi-band exposure maps, images, model
ray-trace PSF images, and optimally binned light-curves. Observations obtained
using the ACIS instrument additionally include low-resolution ($E/{\Delta
E}\sim\!10$--$40$, depending on incident photon energy and location on the
array) source and background spectra and a weighted detector redistribution
matrix file (the probability matrix that maps photon energy to detector pulse
height).

\subsubsection{Energy Bands}

The energy bands used to derive many CSC properties are defined in
Table~4.  The energy bands are chosen to optimize the
detectability of X-ray sources while simultaneously maximizing the
discrimination between different spectral shapes on X-ray color-color diagrams.

The effective area of the telescope (including both the {\it Chandra\/} High
Resolution Mirror Assembly [HRMA] and the detectors) is shown in
Figure~6 as a function of energy, together with the average ACIS
quiescent backgrounds derived from blank sky observations \citep{mar01a}.  The
effective area is measured at the locations of the nominal ``ACIS-S'' aimpoint
on the ACIS S3 CCD, and the nominal ``ACIS-I'' aimpoint on the ACIS I3 CCD\null.

\begin{figure}
\epsscale{1.0}
\plotone{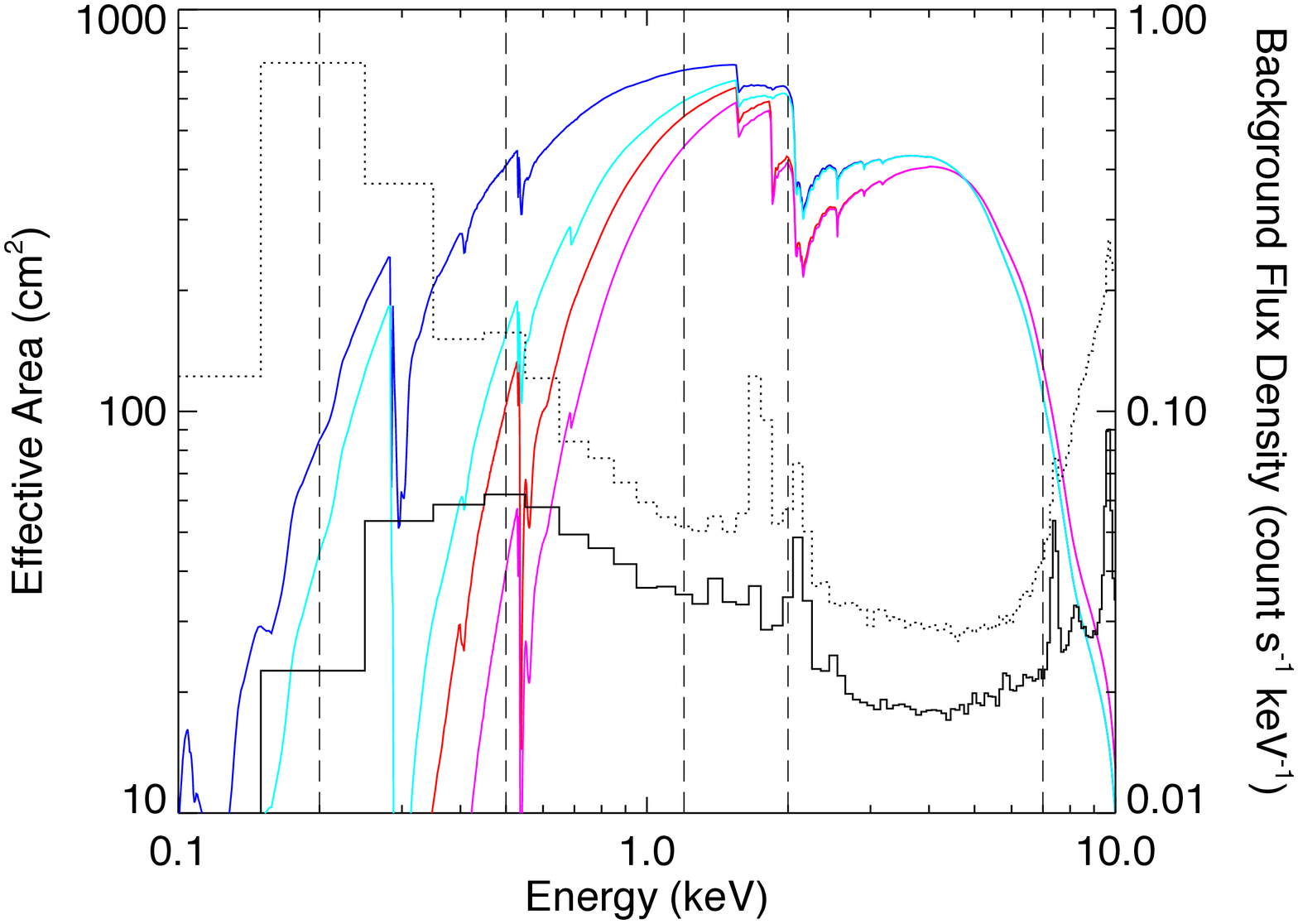}
\caption{\label{fig:eabkg}
{\it Chandra\/} effective area and average ACIS quiescent background as a 
function of energy.  The blue and cyan curves present the combined HRMA plus
ACIS effective area at the ACIS-S aimpoint, with zero and the late 2009 level
of focal plane contamination, respectively.  The red and cyan curves show the
effective area at the ACIS-I aimpoint, again with zero and late 2009 
contamination, respectively.  The dotted black line shows the quiescent
background flux density on the ACIS S3 CCD, while the solid black line 
represents the ACIS I3 CCD background.  The energies corresponding to the
edges of the CSC energy bands are shown by vertical dashed lines.}
\end{figure}
 
Where possible, the energy bands are chosen to avoid large changes of effective
area within the central region of the band, since such variations degrade the
accuracy of the monochromatic effective energy approximation described below.
For example, the M-edge of the Iridium coating on the HRMA has significant
structure in the $\sim\!2.0$--$2.5\rm\,keV$ energy range that provides a natural
breakpoint between the ACIS medium and hard energy bands.  Note however, that
large effective area variations are unavoidable within the ACIS broad and soft
energy bands and the HRC wide energy band.

Weighting the effective area by the source spectral shape and integrating over
the bandpass provides an indication of the relative detectability of a source in
the energy band.  Selecting energy band boundaries so that source detectability
is roughly the same in different energy bands more uniformly distributes Poisson
errors across the bands, and so enhances detectability in the various bands.

Several different source types were simulated when selecting energy bands.
These included absorbed non-thermal (power-law) models with photon index values
ranging from 1 to~4, absorbed black-body models with temperature varying from
$20\rm\,eV$ to $2.0\rm\,keV$, and absorbed, hot, optically-thin thermal plasma
models \citep{ray77} with $kT=0.25$--$4.0\rm\,keV$.  In all cases, the Hydrogen
absorbing column was varied over the range
$1.0\times10^{20}$--$1.0\times10^{22}\rm\,cm^{-2}$.  Detected X-ray spectra were
simulated using PIMMS \citep{pimms}, and then folded through the bandpasses to
construct synthetic X-ray color-color diagrams (see Fig.~7 for
example color-color diagrams based on the final band parameters).  Energy bands
chosen to fill the color-color diagrams maximally provide the best
discrimination between different spectral shapes.  For detailed X-ray
spectral-line modeling, the \citet{ray77} models have been superseded by more
recent X-ray plasma models \citep[e.g.,][]{MEKAL, APEC}.  However, since the
radiated power of the newer models as a function of temperature is not
significantly different from the 1993 versions of the \citet{ray77} models used
here, the latter are entirely adequate for the purpose of evaluating coverage of
the X-ray color-color diagrams and the task is greatly simplified because of
their availability in PIMMS\footnote{The newer Mekal and APEC models are
included in PIMMS v4.0.}\null.

\citet{gri09} compared broad band X-ray photometry with accurate ACIS spectral
fits and found that model-independent fluxes could be derived from the
photometry measurements to an accuracy of about 50\% or better for a broad range
of plausible spectra.  They used similar but not identical energy bands to those
adopted for the CSC, but did not use the method of deriving fluxes from
individual photon energies employed herein.

Combining all of these considerations \citep{mcc07} yields the following
selection of energy bands for the CSC\null.

The ACIS soft ({\it s\/}) energy band spans the energy range 0.5--$1.2\rm\,keV$.
The lower bound is a compromise that is set by several considerations.  ACIS
calibration uncertainties increase rapidly below $0.5\rm\,keV$, so this
establishes a fairly hard lower limit to avoid degrading source measurements in
the energy band.  As shown in Figure~8, below about
$0.6\rm\,keV$ the background count rate begins to increase rapidly, while the
integrated effective area rises very slowly resulting in few additional source
counts.  While pushing the band edge to higher energy will result in a lower
background, the integrated effective area drops rapidly if the lower bound is
raised above $\sim\!0.8\rm\,keV$, reducing the number of source counts collected
in the band.  We choose to set the lower bound equal to $0.5\rm\,keV$ since
doing so enhances the detectability of super-soft sources, while not noticeably
impacting measurements of other sources.  The upper cutoff for the soft energy
band is set equal to $1.2\,\rm keV$, which balances the preference for uniform
integrated effective areas amongst the energy bands with the desire to maximize
the area of X-ray color-color plot parameter space spanned by the simulations.

The lower bound of the ACIS medium ({\it m\/}) energy band matches the upper
bound of the soft energy band.  We locate the upper band cutoff at $2.0\rm\,keV$
since this value tends to maximize the coverage of the X-ray color-color
diagram.  This value also moves the Iridium M-edge out of the sensitive medium
band, and instead placing it immediately above the lower boundary of the ACIS
hard ({\it h\/}) energy band.

The high energy boundary of the latter band is set to $7.0\rm\,keV$.  This
cutoff provides a good compromise between maximizing integrated effective area
and minimizing total background counts (Fig.~8).  Above
$7.0\rm\,keV$, the background rate increases rapidly at the ACIS-S, while below
this energy the integrated effective area decreases rapidly at the ACIS-I
aimpoint.  Placing the hard energy band cutoff at $7.0\rm\,keV$ also has the
advantage that the $\rm Fe\,K\alpha$ line is included in the band, allowing
intense Fe line sources to be detected without compromising the measurement
quality for typical catalog sources.

\begin{figure*}
\epsscale{1.0}
\plotone{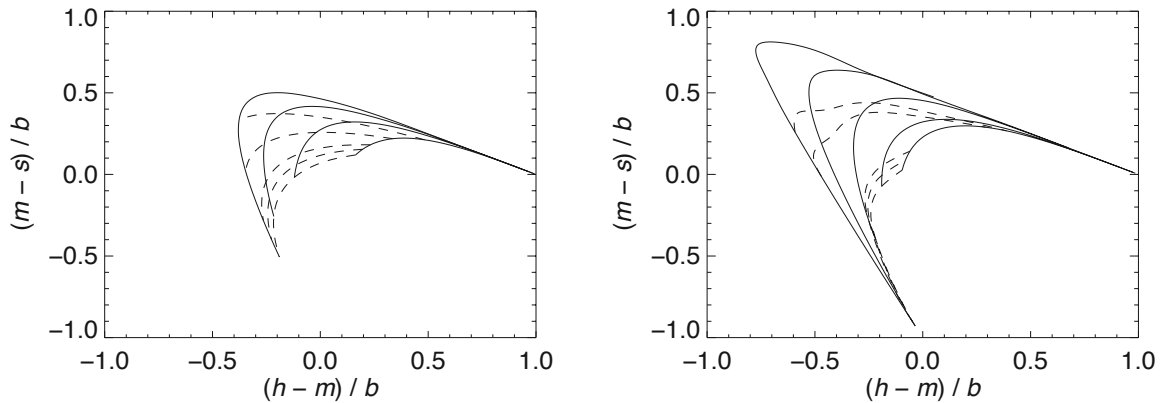}
\caption{\label{fig:ebandcc}
Synthetic color-color diagrams computed for the ACIS hard ($h$), medium ($m$),
soft ($s$), and broad ($b = h + m + s$) energy bands.  {\it Left:\/} Absorbed
power-law models.  The solid lines are lines of constant photon index $\Gamma
= 1.0$, $2.0$, $3.0$, and $4.0$ (from right to left).  The dashed lines are
lines of constant neutral Hydrogen column densities $N_{\rm H} = 1.0\times
10^{20}$, $1.0\times 10^{21}$, $2.0\times 10^{21}$, $5.0\times 10^{21}$, and
$1.0\times 10^{22}\rm\,cm^{-2}$ (from bottom to top).  {\it Right:\/} Hot,
optically thin thermal plasma models \citep{ray77}.  The solid lines are lines 
of constant temperature $kT = 0.25$, $0.5$, $1.0$, $2.0$, and $4.0\rm\,keV$ 
(from left to right).  The dashed lines are lines of constant neutral Hydrogen
column densities $N_{\rm H} = 1.0\times 10^{20}$, $1.0\times 10^{21}$,
$2.0\times 10^{21}$, $1.4\times 10^{22}$, and $1.75\times 10^{22}\rm\,cm^{-2}$
(from bottom to top). Energy bands were chosen to optimize the ability to
estimate spectral parameters from color-color diagrams.}
\end{figure*}
 
The ACIS broad ({\it b\/}) band covers the same energy range as the combined
soft, medium, and hard bands, and therefore spans the energy range 
0.5--$7.0\rm\,keV$.  

Simulations indicate that an additional energy band extending below $0.5\,\rm
keV$ is beneficial for discriminating super-soft X-ray sources in color-color
plots. The ACIS front-illuminated CCDs have minimal quantum efficiency below
$0.3\,\rm keV$, while the response of the back-illuminated CCDs extends down to
$\sim 0.1\,\rm keV$. Hydrocarbon contamination is present on both the HRMA
optics \citep{jer05} and the ACIS optical blocking filter \citep{mar04}. The
latter reduces the effective area at low energies, and enhances the depth of the
Carbon K-edge. An ACIS ultra-soft ({\it u\/}) band covering $0.2$--$0.5\,\rm
keV$ is added to provide better discrimination of super-soft sources. Source
detection is {\it not\/} performed in this energy band, because of the typical
lower overall signal-to-noise ratio (SNR) and the resulting enhanced
false-source rate.

Finally, since the HRC (particularly HRC-I) has minimal spectral resolution, a
single wide ({\it w\/}) band that includes essentially the entire pulse height
spectrum (specifically, PI values $0:254$), roughly equivalent to 0.1-10 keV,
is used for HRC observations.

While bands in these general energy ranges give the best balance of count rate
and spectral discrimination, our simulations indicate that the exact choice of
band boundary energies is not critical at the 10\% level.

\subsubsection{Band Effective Energies\label{sec:effeng}}

In principle, the variations of HRMA effective area, detector quantum
efficiency, and (for ACIS) focal plane contamination, with energy imply that
energy-dependent data products such as exposure maps or PSFs should be
constructed by integrating the source spectrum over the energy band. This
approach would be both extremely time-consuming, and require knowledge of the
source spectrum that is typically not available {\it a priori\/}.  In practice,
a monochromatic effective energy is chosen for each energy band to be used to
construct energy dependent data products \citep{mcc07}.

The monochromatic effective energy for each band is determined using the
relation
\begin{equation}
\label{eqn:eeff}
E_{\rm eff} = \frac{\int dE \, E A(E) Q(E) C(E) S(E)}{\int dE \, A(E) Q(E) C(E) S(E)},
\end{equation}
where $E$ is the energy, $A$ is the effective area of the HRMA, $Q$ is the
detector quantum efficiency, $C$ is the reduction in transmission due to focal
plane contamination, $S$ is a power-law spectral weighting function of the form
$(E/E_0)^{-\alpha}$, and the integral is performed over the energy band.

\begin{figure*}
\epsscale{1.0}
\plotone{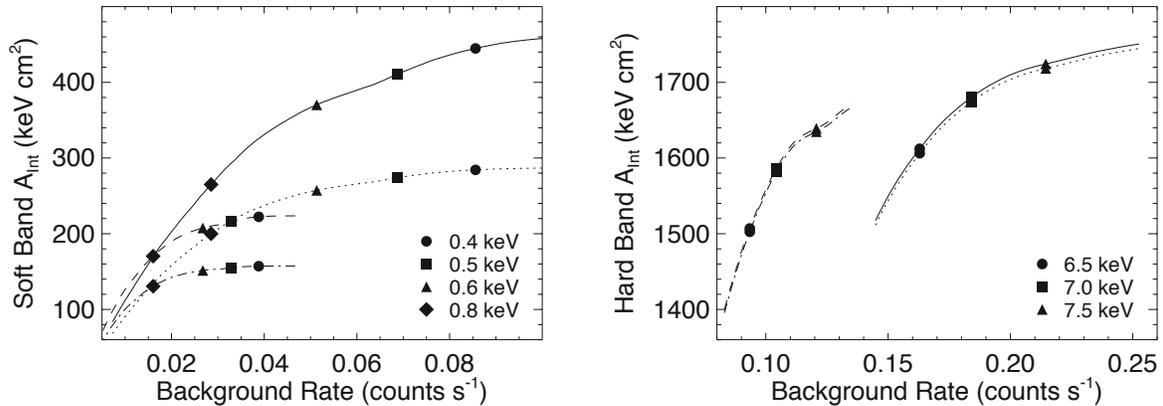}
\caption{\label{fig:bandedge}
{\it Left:\/} Plot shows how the ACIS soft ({\it s\/}) energy band integrated 
effective area and background count rate per CCD vary with the choice of lower
bound for the energy band.  Markers for different lower bounds are shown.  The 
individual curves show the relationship at the ACIS-S and ACIS-I aimpoints, and 
with zero and the late 2009 level of focal plane contamination, as follows.  
Solid line: ACIS-S aimpoint, no contamination; dotted line: ACIS-S aimpoint, 
late 2009 contamination; dashed line: ACIS-I aimpoint, no contamination; 
dash-dotted line: ACIS-I aimpoint, late 2009 contamination.  
{\it Right:\/} Plot shows how the ACIS hard ({\it h\/}) energy band integrated 
effective area and background count rate per CCD vary with the choice of upper
bound for the energy band.  Markers and line styles are the same as in the left
panel.}
\end{figure*}
 
The monochromatic effective energies for each energy band were calculated for
sources located at the ACIS-I and ACIS-S aimpoints, and also for the nominal
aimpoint on the HRC-I detector.  Since the CSC is constructed from observations
acquired throughout the {\it Chandra\/} mission, ACIS focal plane contamination
models with both zero contamination (appropriate for observations obtained early
in the mission) and the contamination level current as of late 2009 were
employed.  Power-law spectral weighting functions with $\alpha$ varying from 0.0
to 2.0 were used.  Setting $\alpha=1$ gives a spectral weighting function that
approximates an absorbed $\Gamma = 1.7$ power-law spectrum, and the limits for
$\alpha$ were chosen to span the typical range of values determined from fits to
a canonical subset of {\it Chandra\/} datasets.  The remaining parameters in
equation~(\ref{eqn:eeff}) are extracted from the {\it Chandra\/} calibration
database \citep[CalDB;][]{geo05,caldb06}\null.  The monochromatic effective
energies for ACIS were chosen to be the approximate arithmetic means of the
$\alpha = 1$ values derived for the ACIS-I and ACIS-S aimpoints, with zero and
late 2009 focal plane contamination.  For ACIS energy bands other than the the
broad band, the monochromatic effective energies computed for a single value of
$\alpha$ all agree within $\lesssim 0.1\rm\,keV$.  The dependence on $\alpha$ is
similarly small, except for the hard energy band, where varying $\alpha$ from
0.0 to 2.0 changes the monochromatic effective energy from $\sim\!4.2\rm\,keV$
to $\sim\!3.4\rm\,keV$.  For the ACIS broad energy band, the agreement between
the different models for a single value of $\alpha$ is $\sim\!\pm 0.3\rm\,keV$.
However, for this band the dependence on $\alpha$ is more significant, varying
from $\sim\!3.3\rm\,keV$ for $\alpha=0.0$ to $\sim\!1.6\rm\,keV$ for
$\alpha=2.0$.  The monochromatic effective energies used to construct the CSC
are reported in Table~4.

Although the use of a single monochromatic effective energy for each energy band
simplifies data analysis by removing the dependence on the source spectrum, some
error will be introduced for sources that have either {\it extremely\/} soft or
{\it extremely\/} hard spectra compared to the canonical $\alpha=1.0$ power-law 
spectral weighting function.  Knowledge of the expected magnitude of the error 
that may be introduced is helpful when evaluating catalog properties.

For both the ACIS medium and hard energy bands, neither extremely soft nor
extremely hard source spectra induce variations in exposure map levels that are
greater than $\sim\!10\%$, so photometric errors due to source spectral shape
should not exceed this value.  In the ACIS soft energy band, very soft spectra
may produce deviations of order 5--20\%, with the largest excursions expected
for the front-illuminated CCDs.  These differences increase to $\sim\!15$--35\% 
for the ACIS ultra-soft energy band, with the largest values once again
associated with the front-illuminated CCDs.  For all of the ACIS narrow energy
bands, the errors induced by extremely hard spectra are much smaller than those
caused by extremely soft spectra.  The presence of the Iridium edge and the large
energy ranges included in the ACIS broad and HRC wide energy bands may produce
significantly larger variations for extreme spectral shapes.  Very soft spectra
can alter exposure map values by $\sim\!65$--90\% in the ACIS broad energy band, 
although there is little impact in the HRC wide energy band.  Conversely, 
extremely hard spectra may induce changes up to $\sim\!70\%$ in the HRC wide 
energy band, and $\sim\!25$--30\% in the ACIS broad energy band.  As described 
in \S~4.4, model-based statistical characterization of CSC source 
fluxes \citepfap\ produces results that are generally consistent with these 
expectation, with the exception that flux errors in the ACIS broad energy band
appear to be $\sim\!10\%$ for most sources.

When computing fluxes for point sources, an aperture correction is applied to
compensate for the fraction of the PSF that is not included in the aperture.
Since the extent of the {\it Chandra\/} PSF varies with energy, using a 
monochromatic effective energy can introduce a flux error because the energy
dependence of the PSF fraction is not considered.  This error can be bounded by
a {\it post facto\/} comparison of PSF fractions for catalog source detections
in the 5 ACIS energy bands.  The majority of variations {\it between\/} energy 
bands fall in the range 4--8\%, with 90\% of source detections showing $<10\%$ 
differences.  These values represent an upper bound on the error introduced 
{\it within\/} an energy band by the use of a monochromatic effective energy.

\subsubsection{Coordinate Systems and Image Binning\label{sec:imgbin}}

As described previously, X-ray photon event data are recorded in the form of a
photon event list.  The pixel position on the detector where a photon was
detected is recorded in the ``chip'' pixel coordinate system.  Event positions
are remapped to celestial coordinates through a series of transforms, as
described by \citet{mcd01}.  The first step in this process remaps chip
coordinates to a uniform {\it real-valued\/} virtual ``detector'' pixel space by
applying corrections for the measured detector geometry, and instrumental and
telescope optical system distortions recorded in the CalDB\null.  Subsequent
application of the time-dependent aspect solution removes the spacecraft dither
motion, and maps the event positions to a uniform virtual ``sky'' pixel plane.
The latter has the same pixel scale as the original instrumental pixels, but is
oriented with North up ($+Y$ direction) and is centered at the celestial
coordinates of the tangent plane position for the observation.  As an aid to
users, the location of each event in each coordinate system is recorded in the
calibrated photon event list.  A simple unrotated world coordinate system
transform maps sky positions to ICRS right ascension and declination by applying
the plate scale calibration to the difference between the position of the source
and a fiducial point, which is typically the optical axis of the telescope.  The
celestial coordinates of the fiducial point are determined from the aspect
solution.

\begin{figure*}
\epsscale{0.9}
\plotone{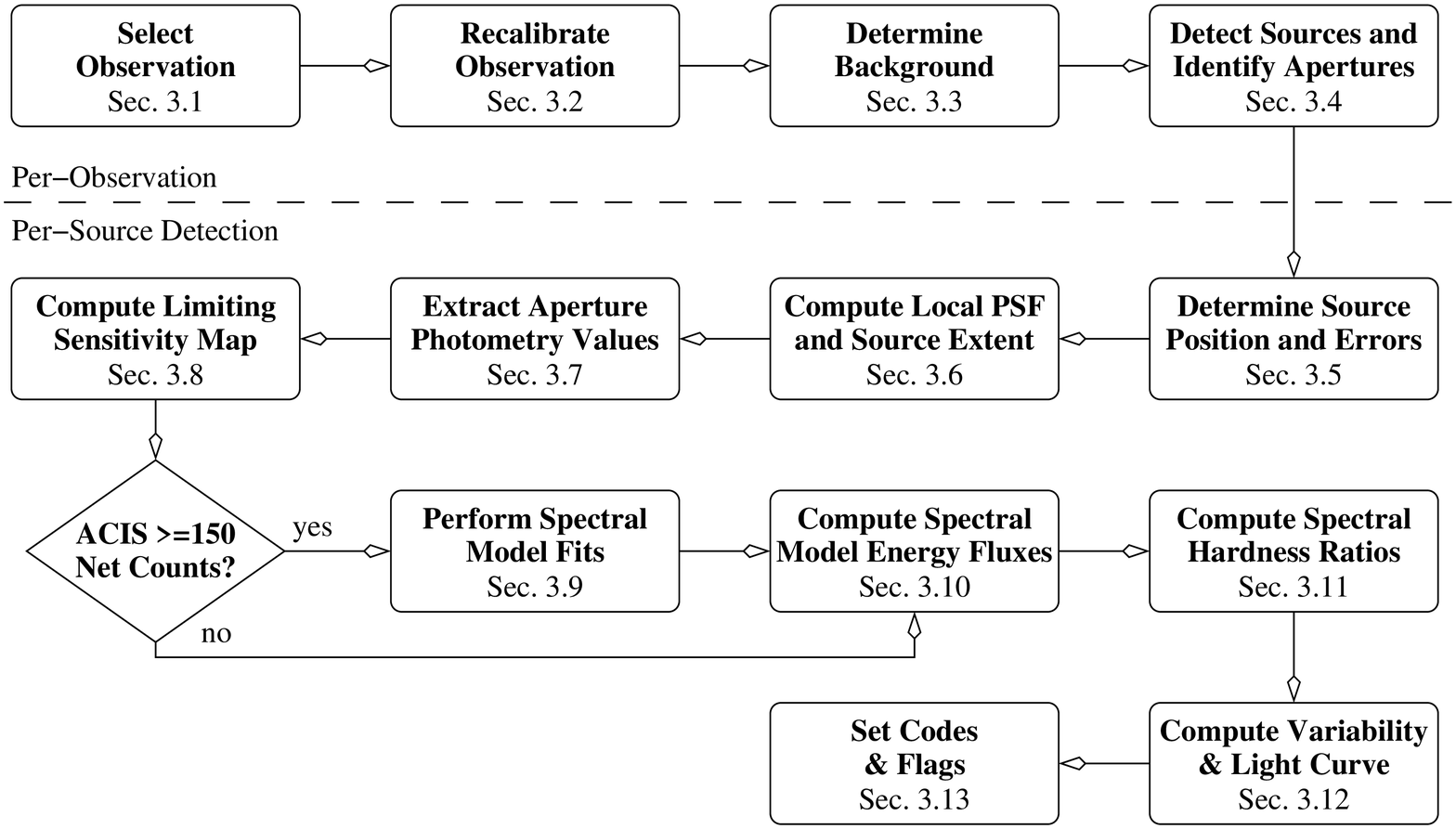}
\caption{\label{fig:blocko}
High-level flow diagram depicting the steps used to process each observation's
full field-of-view, detect sources, and extract the physical properties for each
detected source.  The references identify the relevant sections of the text that
describe in detail the methods used.}
\end{figure*}
 
Sky images are constructed from the calibrated photon event lists by binning
photon positions in sky coordinates into a regular, rectangular image pixel
grid.  A consequence of constructing images by binning in sky coordinates is
that {\it Chandra\/} images are always oriented with North up.  The choice of
image blocking factor determines the number of sky pixels that are binned into a
single image pixel.  Full field image products associated with ACIS observations
are constructed by binning the area covered by the inner $2048\times 2048$ sky
pixels at single pixel resolution, then binning the inner $4096\times 4096$ sky
pixels at block 2, and finally binning the entire $8192\times 8192$ sky pixel
field at block 4.  The corresponding blocking factors for HRC-I observations are
2, 5, and 12.  Using a constant blocked image size of $2048\times 2048$ pixels
reduces overall data volume, while preserving resolution in the outer areas of
the field of view where the PSF size is significantly larger than a single
pixel.

\section{CATALOG GENERATION\label{sec:catgen}}

In this section, we describe in detail the methods used to derive the X-ray
source properties that are included in the CSC, with particular detail provided
in cases where the algorithms are new or have been newly adapted for use with
{\it Chandra\/} data.

The principal steps necessary to generate the catalog consist of processing the
data for each observation's full field-of-view, detecting X-ray sources included
within that field of view, and then extracting the spatial, photometric,
spectral, and temporal properties of each detected source.
Figure~9 is a depiction of the high-level flow used to perform
these steps. In the figure, each block references the section of the text that
describes in detail the methods used. The physical properties associated with
each source detection are recorded as a separate row in the catalog Source
Observations Table. 

Once the source detections from each observation have been evaluated, they are
correlated with source detections from all other spatially overlapping
observations to identify distinct X-ray sources on the sky. The steps required
to perform the source cross-matching, and then combine the data from multiple
observations of a single source to evaluate the source's properties, follow a
similar flow to the one presented in Figure~10. Many of the
elements that comprise the second flow are built on the foundations developed
for the related steps from the first flow. For convenience and continuity of
notation, the former are described in the same text sections as the latter. The
properties for each distinct X-ray source are included as a separate row in the
catalog Master Sources Table.

Data processing for release 1 of the CSC was performed using versions 3.0--3.0.7
of the {\it Chandra\/} X-ray Center data system \citep[CXCDS;][]{eva06a, eva06b}
catalog processing system (``CAT''), with calibration data extracted from CalDB
version 3.5.0\null.  The observation recalibration steps included in CAT3.0
correspond {\it approximately\/} with those included CIAO 4.0\null.  In several
cases, programs developed for CAT3.0 to evaluate source properties have been
repackaged with new interfaces for interactive use in subsequent CIAO releases
(see Table~5).

\subsection{Observation Selection\label{sec:obsel}}

While the CSC ultimately aims to be a comprehensive catalog of X-ray sources
detected by {\it Chandra\/}, all of the functionality required to achieve that
goal are not included in the release 1 processing system. A set of pre-filters
is used to limit the data content to the set of observations that the catalog
processing system is capable of handling.

\begin{figure*}
\epsscale{0.9}
\plotone{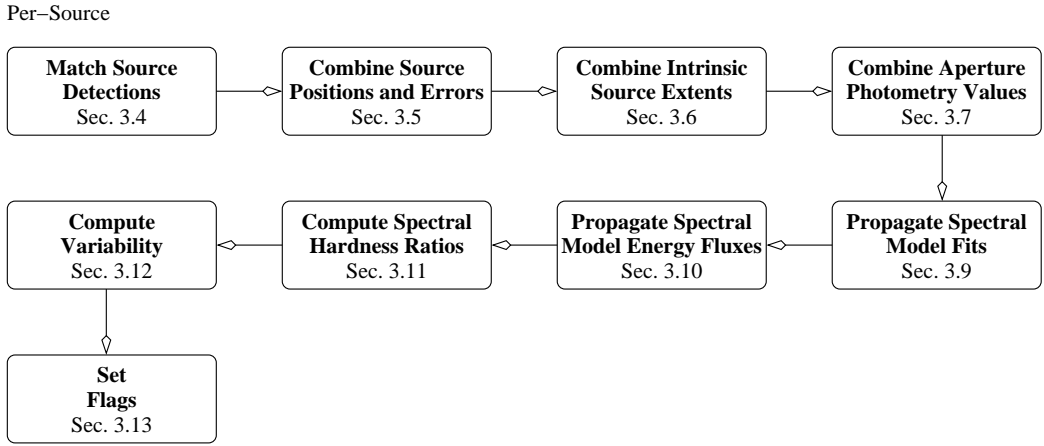}
\caption{\label{fig:blockm}
As Fig.~9, except that the steps used to cross-match source
detections, and combine data from multiple observations to evaluate source
properties, are shown.}
\end{figure*}
 
For release 1, only public ACIS ``timed-exposure'' readout mode imaging
observations obtained using either the ``faint,'' ``very faint,'' or ``faint
with bias'' datamodes are included. ACIS observations that are obtained using
CCD subarrays with $\leq 128$ rows are also excluded, because there are too few
rows to ensure that source-free regions can be identified reliably when
constructing the high spatial frequency background map. HRC-I imaging mode
observations are not included in release 1 of the catalog, but are included in
incremental release 1.1. HRC-S observations are excluded because of the presence
of background features associated with the edges of ``T''-shaped
energy-suppression filter regions that form part of the UV/ion-shield.
Observations of solar system objects are not included in the CSC.

All observations included in the CSC must have been processed using the standard
data processing pipelines included in version 7.6.7, or later, of the
CXCDS\null.  This version of the data system was used to perform the most recent
bulk reprocessing of {\it Chandra\/} data, and includes revisions to the
pipelines that compute the aspect solution that is used to correct for the
spacecraft dither motion and register the source events on the sky. Observations
must have successfully passed the ``validation and verification'' (quality
assurance) checks that are performed upon completion of standard data
processing.

The largest scale lengths used to detect sources to be included in the CSC have
angular extents $\sim\!30''$.  Sources with apparent sizes greater than this are
either not detected, or may be detected incorrectly as multiple close sources.
Prior to catalog construction, all observations are inspected visually for the
presence of extended sources that may be detected incorrectly, and such
observations are excluded from catalog processing.  For ACIS observations, if
the presence of any spatially extended emission is restricted to a single CCD
only, then the data from that CCD are dropped, and sources detected on any
remaining CCDs are typically included in the catalog.  The latter rule allows
many sources surrounding bright, extended cores of galaxies to be included in
the catalog, rather than having the entire observation rejected outright.

While the visual inspection and rejection process is inherently subjective in
nature, an attempt was made to calibrate the method by constructing a ``training
set'' of several hundred observations that were processed through a test version
of the catalog pipelines.  The training set observations included a wide variety
of point, compact, and extended sources, with differing exposures and SNR, which
were classified as accept/reject based on the actual results of running the
pipeline source detection and source property extraction steps.  These
observations and classifications were then used to train the personnel who
performed the visual inspection process.

\subsection{Observation Recalibration\label{sec:recal}}

Although all observations included in the CSC have been processed through the
CXCDS standard data processing pipelines, we nevertheless re-run the
instrument-specific calibration steps as the first step in catalog construction.
One reason for reapplying the instrumental calibrations is that they are subject
to continuous improvements, and may have been revised since the last time the
observations were processed or reprocessed.  A second reason is to ensure that a
single set of calibrations are applied to all datasets, so that the resulting
catalog will be calibrated as homogeneously as possible.

For ACIS, the principal instrument-specific calibrations that are re-applied are
the (time-dependent) gain calibration and the correction for CCD charge transfer
inefficiency (CTI).  The former calibration maps the measured pulse height for
each detected X-ray event into a measurement of the energy of the corresponding
incident X-ray photon.  CTI correction attempts to account for charge lost to
traps in the CCD substrate when the charge is being read out.  This effect is
considerably larger than anticipated prior to launch because of damage to the
ACIS CCDs caused by the spacecraft's radiation environment.  Additionally,
observation-specific bad pixels and hot pixels are flagged for removal, as are
``streak'' events on CCD S4 (ACIS-8).  The latter apparently result from a flaw
in the serial readout electronics \citep{hou00}.  Pixel afterglow events, which
arise because of energy deposited into the CCD substrate by cosmic ray charged
particles, are removed using the {\tt acis\_run\_hotpix} tool that is also
included in CIAO\null.  Although this program can miss some real faint
afterglows, such events are very unlikely to exceed the flux significance
threshold required for inclusion in the catalog.  The default 0.5 pixel event
position randomization in chip coordinates is used when the calibrations are
reapplied.

\begin{deluxetable*}{lcl}
\tablecolumns{3}
\tablewidth{0pt}
\tablecaption{CSC-Related CIAO Tools\label{tab:ciaotools}}
\tablehead{
\colhead{Tool Name}	& \colhead{CIAO Version}	& \colhead{Description}
}
\startdata
{\tt aprates}		& 4.1	& Calculate source aperture photometry properties \\
{\tt dmellipse}		& 4.1	& Calculate ellipse including specified encircled fraction \\
{\tt eff2evt}		& 4.1	& Calculate energy flux from event energies \\
{\tt lim\_sens}		& 4.1	& Create a limiting sensitivity map \\
{\tt mkpsfmap}		& 4.1	& Look up PSF size for each pixel in an image \\
{\tt acis\_streak\_map}	& 4.1.2	& Create a high spatial frequency background map \\
{\tt dither\_region}	& 4.1.2	& Calculate region on detector covered by a sky region \\
{\tt evalpos}		& 4.1.2	& Get image values at specified world coordinates \\
{\tt glvary}		& 4.1.2	& Search for variability using Gregory-Loredo algorithm \\
{\tt pileup\_map}	& 4.1.2	& Create image that gives indication of pileup \\
{\tt modelflux}		& 4.1.2	& Calculate spectral model energy flux \\
{\tt srcextent}		& 4.1.2	& Compute source extent  \\
{\tt create\_bkg\_map}	& 4.2	& Create a background map from event data \\
{\tt dmimgpm}		& 4.2	& Create a low spatial frequency background map \\
\enddata
\end{deluxetable*}

The main instrument specific calibrations for HRC data relate to the
``degapping'' correction that is applied to the raw X-ray event positions to
compensate for distortions introduced by the HRC detector readout hardware.
Several additional calibrations compensate for effects introduced by amplifier
range switching and ringing in the HRC electronics, and a number of validity
tests are performed to flag X-ray event positions that cannot be properly
corrected due to amplifier saturation and other effects.

Since data are recorded continuously during an observation, a ``Mission Time
Line'' is constructed during standard data processing that records the values
of key spacecraft and instrument parameters as a function of time.  These 
parameters are compared with a set of criteria that define acceptable values, 
and ``Good Time Intervals'' (GTIs) that include scientifically valid data are 
computed for the observation.  The GTI filter from standard data processing is
reapplied without change as part of the recalibration process.

Background event screening performed as part of catalog data recalibration is
somewhat more aggressive than that performed as part of standard data
processing, typically reducing the non-X-ray background.  For a $10\rm\,ks$
observation, the median catalog background rate is roughly 80\% of the nominal
field background rates \citep{POG}, although there is considerable scatter.
\citetfap\ include a detailed statistical analysis of the improvements to the
non-X-ray background afforded by this screening.

The reduction of the background event rate is achieved by removing time
intervals containing strong background flares.  These time intervals are
identified separately for each chip.  First, the background regions of the image
are identified by constructing a histogram of the event data, determining the
mean and standard deviations of the histogram values, and rejecting all pixels
that have values more than 3 standard deviations above the mean.  An
optimally-binned light curve of the background pixels is then created using the
Gregory-Loredo algorithm (see \S~3.12.1).  Time bins for which the
count rate exceeds $10\times$ the minimum light curve value are identified.  The
corresponding intervals are considered to be background flares, and the GTIs are
revised to exclude those periods.

We emphasize that the objective of this procedure is to remove only the most
intense background flares, which occur relatively infrequently.  Time intervals
that include moderately enhanced background rates are not rejected by this
process, since their contributions increase the overall SNR\null.  The aggregate
loss of good exposure time exceeds 25\% for less than 1.5\% of the observations
included in the catalog; the loss is greater than 10\% for 3\% of the
observations, and greater than 5\% for 5\% of the observations.

For each observation included in the CSC, the recalibrated photon event list is
archived, together with several additional full-field data products.  These
include multi-resolution exposure maps computed at the monochromatic effective
energies of each energy band and the associated ancillary data products (aspect
histogram, bad pixel map, and field of view region definition), used to
construct them (see Table~3).

\subsection{Background Map Creation}

For the first release of the CSC, background maps are used for automated source
detection.  They are created directly from each individual observation with the
necessary accuracy.  The general observation background is assumed to vary
smoothly with position, and is modeled using a single low spatial frequency
component.  Although this assumption is in general satisfied across the fields
of view included in this catalog release, there may be localized regions where
the background intensity has a strong spatial dependence, and therefore where
the detectability of sources may be reduced.  Several different approaches were
considered for constructing the low spatial frequency background component,
including spatial transforms, low pass filters, and data smoothing.  However,
the most effective and physically meaningful technique is a modified form of a
Poisson mean.  This method, described below, estimates the local background from
the peak of the Poisson count distribution included in a defined sampling area.
The dimensions of the sampling area act effectively as a spatial low pass filter
that determines the minimum angular size that contributes to the background.

\begin{figure*}
\epsscale{1.0}
\plotone{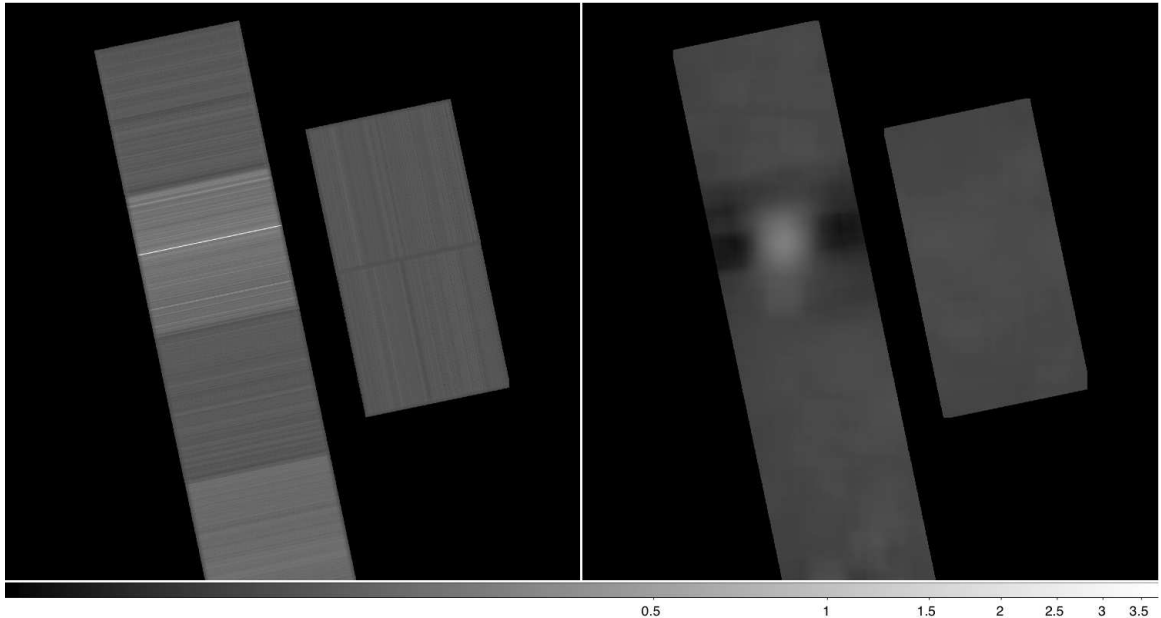}
\caption{\label{fig:bkg}
ACIS broad-band high and low spatial frequency background maps for observation
00735, as used for catalog source detection (Fig.~12, {\it
Right\/}). 
{\it Left:\/} ACIS high spatial frequency background map component.  Each image
pixel represents $2\times 2$ blocked sky pixels.  Intensities have an an offset
of $+0.1\,\hbox{\rm count}\,\hbox{\rm (image pixel)}^{-1}$ added, and the result
is scaled logarithmically over the range 0.0375--3.75. The readout streak
associated with the bright source is clearly visible.  {\it Right:\/} ACIS low
spatial frequency background map component.  The Poisson mean includes a
residual image of the bright source, at a peak level of $\sim\!0.15\,\hbox{\rm
count}\,\hbox{\rm (image pixel)}^{-1}$.}
\end{figure*}
 
High spatial frequency linear features, commonly referred to as ``readout
streaks,'' result when bright X-ray sources are observed with ACIS.  These
streaks arise from source photons that are detected during the CCD readout frame
transfer interval ($\sim\!40\,\mu\rm s$ per row) following each exposure
($\sim\!3.2\,\rm s$ per exposure for a typical observation).  All pixels along a
given readout column are effectively exposed to all points on the sky that lie
along that column during the frame transfer interval, so that columns including
bright X-ray sources have enhanced count rates along their length.  Unless
accounted for by the source detection step, the increased counts in the bright
readout streak are detected as multiple sources.  Although readout streaks are
comprised of mis-located source photons, we choose to model them as a background
component.

Background maps computed for ACIS observations include contributions from both
components, while HRC background maps include only the low spatial frequency
component.

The reader should note that background maps are {\it not\/} used when deriving
source properties such as aperture photometry.  Instead, a local background
value determined in an annular aperture surrounding the source is used, as
described in \S~3.4.1.  Significant spatial variations of the
observed X-ray flux on the scale of the background aperture will increase the
background local variance, thus reducing the significance of the source
detection, perhaps below the threshold required for inclusion of the source in
the catalog.  This effect is seen in some galaxy cores, where the unresolved
emission contributes X-ray flux to the annular background apertures surrounding
each source.

\subsubsection{ACIS High Spatial Frequency Background}

The algorithm described here is a refinement of method used by \citet{mcc05} to
address the impact of readout streaks on source detection.  The streak map is
computed at single pixel resolution independently for each ACIS CCD and energy
band.  The first step is to identify the bright-source-free regions on the
detector. For ease of computation the orientation of the $X$-axis is defined to
be along the chip rows (perpendicular to the readout direction) and the $Y$-axis
is defined to be along the direction of the readout columns. To identify the
source-free regions, the photon event totals, $X_{\rm sum}$ summed along the
$X$-axis are constructed, and the median ($\tilde{X}$), mode ($\hat{X}$), and
standard deviation ($\sigma_X$) of the distribution of the $X_{\rm sum}$ values
are computed. These values provide a basic characterization of the background. 
From an examination of many data histograms, the maximum value of $X_{\rm sum}$ 
which can still considered background dominated is given by
\begin{displaymath}
X_{\rm sum} ({\rm max}) = \min[\tilde{X} + n \, \sigma_X, 2 \, \hat{X}],
\end{displaymath}
where $n$ is set to 1\null. Rows for which $X_{\rm sum} \gg X_{\rm sum}({\rm
max})$ include a substantial bright source contribution. All rows with $X_{\rm
sum} \leq X_{\rm sum} (\rm max)$ (excluding off-chip and dither regions) are
considered to comprise the source-free regions and are used to calculate the
streak map.

The average number of events per pixel is calculated separately for each readout
column ($Y$-axis direction) from all of the rows in the source-free regions.
These values are replicated across each CCD row to create an image that includes
the sum of the readout streak contribution and the mean one-dimensional low
spatial frequency background component. The latter must be accounted for when
combining the high spatial frequency readout streak map with the two-dimensional
low spatial frequency background map.

\begin{figure*}
\epsscale{1.0}
\plotone{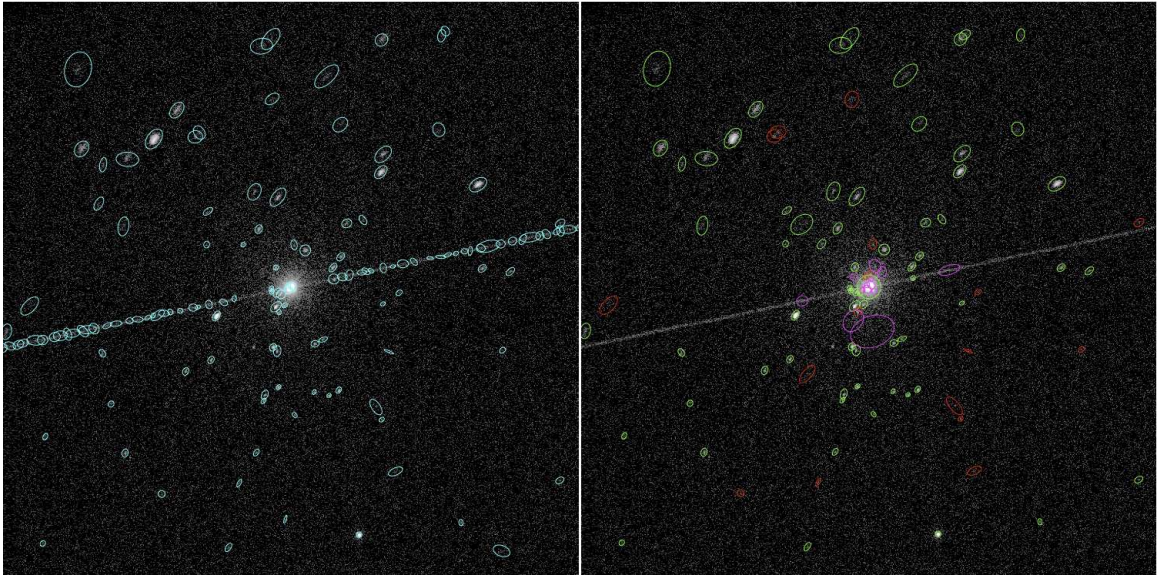}
\caption{\label{fig:detstrk}
ACIS broad-band image of the central region of the field of observation 00735
(M81), which includes an extremely bright source that produces a very bright
readout streak.  Because photon pile-up has eroded the central peak of the
bright source, the source is detected incorrectly as multiple close sources that
must be rejected manually.  
{\it Left:\/} Numerous false sources are detected along the length of the
readout streak if the latter is not modeled as part of the background.  Source
detections are shown in cyan.  No quality assurance processing has been applied
to these detections.
{\it Right:\/} When the background map described in the text is used, the
false sources are suppressed.  Source detections in green are included in the
catalog; sources in red do not meet the minimum flux significance criteria for
inclusion in the catalog; sources in magenta have been rejected manually during
quality assurance processing.}
\end{figure*}
 
For the algorithm to obtain a good measure of the background, of order 100
bright-source-free rows are required.  This condition is satisfied for most
observations.  Observations with too few source-free rows poorly sample the
background.  This can lead to erroneously low intensities for bright readout
streaks in the resulting background map, which may enhance the false source rate
along these streaks.  Faint sources that fall in the source-free rows will be
considered to be part of the background, which can lead to similar results.
Nevertheless, the algorithm is remarkably effective, even in crowded regions
such as the Orion complex and the Galactic center fields.  An example broad-band
ACIS streak map, created for \dataset [ADS/Sa.CXO#obs/00735] {observation 00735}
(M81), is shown in Figure~11, {\it Left\/}.

\subsubsection{Low Spatial Frequency Background}

For each observation, a low spatial frequency background map is constructed
separately for each energy band and image blocking factor (see
\S~3.4).  \citet{mcc08} provide an initial discussion of this
algorithm and general background map creation.

As described above, for ACIS observations the high spatial frequency background
map includes a component that represents the one-dimensional average of the low
frequency background over the rows used to create the streak map.  This
component, as well as the high spatial frequency background, are removed by
subtracting the streak map from the original image from which it was created.
For each image blocking factor, the difference image is constructed by
subtracting the appropriately regridded streak map from the corresponding
blocked original image.

For each pixel in the resulting difference image, a centered sampling region
with dimensions $n \times n$ pixels is defined.  Spatial scales smaller than
$\sim\! n$ pixels are attenuated.  The sampling regions are truncated at the
edges of the images, and so some higher frequency information may propagate into
the background map.  However this effect has not been found to have any
significant impact on the utility of the resulting map.

A histogram of the count distribution is constructed from the pixels included in
the sampling region associated with each image pixel.  The first histogram bin
will typically span the count range from $-0.5$ to $+0.5$ for ACIS observations,
since the readout streak map has been subtracted and there will be some negative
pixels.  The low spatial frequency background at this image pixel location is
computed using a modified form of a Poisson mean
\begin{displaymath}
b_{\rm lf} = \mbox{\rm mean}[h(a) \cup h(b) \cup h(c)],
\end{displaymath}
where $h(x)$ is the number of counts in histogram bin $x$, $a$ is the bin with
the maximum number of histogram counts, and $b$ and $c$ are the lower and higher
bins immediately adjacent to $a$. The low spatial frequency background map is
formed by computing $b_{\rm lf}$ for each pixel location in the image. For ACIS
observations, $n = 129$ pixels, corresponding to a spatial scale of order $1'$
for images blocked at single pixel resolution.  Figure~11, {\it
Right\/} displays the ACIS broad-band low spatial frequency map for observation
00735 (M81) that corresponds to the streak map shown in the left hand panel of
the figure.

\subsubsection{Total Background Map}
 
The first step in creating the total background map is to correct the readout
streak map (for ACIS observations only) for the effects of reduced exposure
near the edges of the observation that arise due to the spacecraft dither, by
dividing by the appropriate band-specific normalized exposure map.  Similarly,
the low spatial frequency background map is corrected by dividing by the
smoothed, band-specific normalized exposure map.  The smoothing that is applied
to the normalized exposure map in the latter case matches the smoothing applied
when constructing the low spatial frequency background map.  Finally, the two
background components for each energy band are summed to produce the total
exposure-normalized background map that is required for source detection.

Figure~12 displays the central region of the broad-band ACIS
image of M81 (observation 00735), with source detections overlayed.  The source
detections shown in the left-hand panel are those that result if the background
is modeled internally by {\tt wavdetect} (see \S~3.4, below); the
panel on the right shows the source detections resulting from using the total
background from Fig.~11.  Using the background map has eliminated the
false sources detected on the readout streak.

The total background maps for each energy band are also archived and accessible
through the catalog.  These maps differ from those used for source detection in
that they have been multiplied by the normalized band-specific exposure map, and
are therefore recorded in units of counts.  For the convenience of the user, we
also store multi-resolution photon-flux images for the full field of each
observation, created by filtering the photon event list by energy band, binning
to the appropriate image resolution, subtracting the total background map
appropriate to the energy band, and dividing by the corresponding exposure map.

\subsection{Source Detection\label{sec:srcdet}}

Candidate sources for inclusion in the CSC are identified using the CIAO {\tt
wavdetect} wavelet-based source detection algorithm \citep{fre02}.  {\tt
wavdetect} has been used successfully with {\it Chandra\/} data by a number of
authors \citep[e.g.,][]{bra01, gia02, leh05, kim07, mun09}, and its capabilities
and limitations are well known \citep[e.g.,][]{val01}.

Early in the catalog processing pipeline development cycle, several different
methods for detecting sources were evaluated.  In addition to {\tt wavdetect},
these included the CIAO implementations of the sliding cell \citep{har84, cal01}
and Voronoi tessellation and percolation \citep{ebe93} algorithms, and a version
of the {\tt SExtractor} package \citep{ber96} modified locally to use Poisson
errors in the low count regime.

The Voronoi tessellation and percolation algorithm was quickly discarded because
of the significant computational requirements and complexities for automated
use.  A series of simulations was used to compare the performance of the
remaining methods with respect to source detection efficiency for isolated point
sources, the efficiency with which close, equally-bright pairs of point sources
with $2''$ and $4''$ separations are resolved, and false source detection rate
\citep[A.~Dobrzycki, private communication;][]{hai04}.  The first two properties
were evaluated for point sources containing 10, 30, 100, and 2000 counts, with
off-axis angles 0--$10'$ with $1'$ spacing, and nominal background rates for
exposure times of 3, 10, 30, and $100\rm\,ks$.  The false source rate was
evaluated as a function of off-axis angle for the same exposure times.  

All three detection algorithms performed reliably for bright, isolated sources
located close to the optical axis.  Compared to the remaining methods, {\tt
wavdetect} had better source detection efficiency for faint sources located
several arcminutes off-axis, and was able to resolve close pairs of sources more
reliably than the sliding cell technique. The locally modified version of {\tt
SExtractor} provided inconsistent results, in some cases detecting large
numbers of spurious sources.

These simulations were performed early in the catalog processing pipeline
development cycle, as an aid in selecting the source detection algorithm to be
used for catalog construction.  They did not make use of the background maps
described in the previous section.  The actual performance of the source
detection process used to construct the CSC is established from more detailed
and robust simulations, as described in \S~4 and references
therein.

Based on the results of the simulations, {\tt wavdetect} was selected as the
source detection method of choice for the CSC\null.

The {\tt wavdetect} algorithm does not require a uniform PSF over the field of
view, and is effective in detecting compact sources in moderately crowded fields
with variable exposure and Poisson background statistics.  To detect candidate
sources in a two-dimensional image $D$, {\tt wavdetect} repeatedly constructs
the two-dimensional correlation integral 
\begin{equation}
\label{eqn:corrint}
C(x, y; \boldalpha) = \int\!\int dx^\prime\,dy^\prime\, W(x-x^\prime, y-y^\prime; \boldalpha)\,
D(x^\prime, y^\prime)
\end{equation}
for a set of Marr (``Mexican Hat'') wavelet functions, $W$, with scale sizes
that are appropriate to the source dimensions to be detected. The elliptical
form of the Marr wavelet may be written in the dimensionless form
\begin{equation}
\label{eqn:mexhat}
W(x, y; \boldalpha) = (2 - \rho^2)\exp(-\rho^2/2),
\end{equation}
where
\begin{displaymath}
\rho^2 = \frac{1}{a_1^2}(x\cos\phi + y\sin\phi)^2 + \frac{1}{a_2^2}(-x\sin\phi + y\cos\phi)^2
\end{displaymath}
and the parameters $\boldalpha = (a_1, a_2, \phi)$ define the semi-major
and semi-minor radii and rotation angle of the Mexican Hat.

A localized clump of counts in the image $D$ will produce a local maximum of $C$
if the scale sizes defined by $\boldalpha$ are approximately the same as,
or larger than, the dimension of the clump. To determine whether a local maximum
of $C$ is due to the presence of a {\it source\/}, the detection significance, 
$S_{i, j}$, in each image pixel $(i, j)$ is determined from
\begin{displaymath}
S_{i, j} = \int_{C_{i, j}}^\infty dC\,p(C|n_{B, i, j}),
\end{displaymath}
where $n_{B, i, j}$ is the number of background counts within the limited
spatial extent of $W$, and $p(C|n_{B, i, j})$ is the probability of $C$ given
the background $B$. If $S_{i, j} \leq S_0$, where $S_0$ is a defined limiting
significance level, then pixel $(i, j)$ is identified as a source pixel. 

\begin{figure*}
\epsscale{0.6}
\plotone{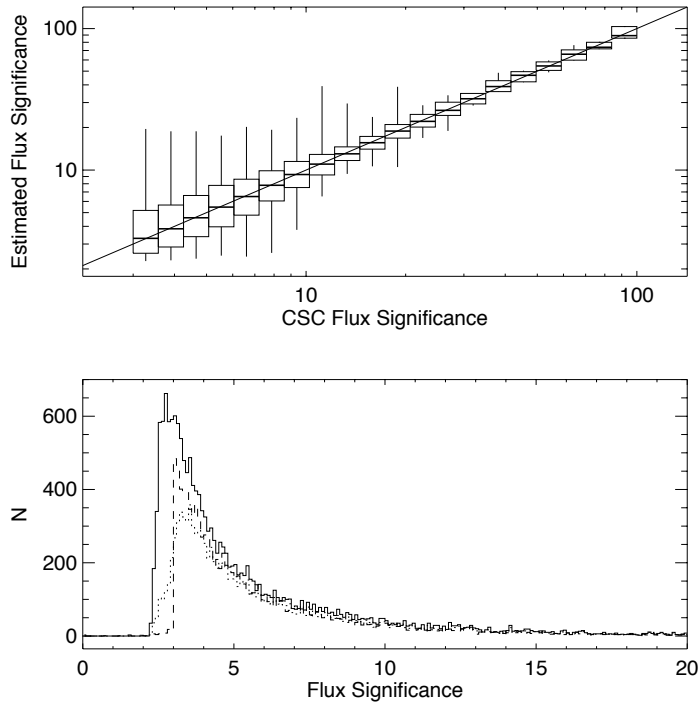}
\caption{\label{fig:fluxsig}
{\it Top:\/} Estimated flux significance versus catalog flux significance for 
$\sim\!11,000$ sources detected in the ACIS broad energy band in a pre-release 
test version of the CSC\null.  The ``estimated'' flux significance is defined 
as the ratio ${\tt net\_counts}/{\tt net\_counts\_err}$, as reported by {\tt 
wavdetect}, and correlates well with the actual flux significance used the 
determine catalog inclusion.  Horizontal lines indicate the median of the points 
in each bin, and the vertical lines identify the extreme points.  Boxes include 
90\% of the points in each bin.
{\it Bottom:\/} Distribution of estimated flux significances for all detected 
sources (solid line), including those which fell below the flux significance 
threshold for the test catalog.  The distribution of estimated significances for 
sources included in the catalog is shown by the dotted line; the dashed line is 
the distribution of actual flux significances for the same sources.  The flux 
significances for all detected sources extends well below the distributions for 
sources included in the catalog.}
\end{figure*}
 
The limiting significance level used to generate the CSC is set to $S_0 =
2.5\times10^{-7}$.  This formally corresponds to $\sim\!1$ false source due to
random fluctuations per $2048\times2048$ pixel image, although due to the
heuristics of the algorithm, the actual number of false sources may be lower.
The situation is further complicated in our case because the final candidate
source list output from the CSC source detection pipeline is a combination of
several {\tt wavdetect} runs in different energy bands (see below).  We note
that reliable quantitative estimates of the false source rates and detection
efficiency can only be provided through simulations, as discussed in
\S~4\null.  As described in \S~2.5, we impose an
additional restriction on the flux significance of a source.  To ensure that the
flux significance requirement is the defining criterion for a source to be
included in the catalog, we have verified that the flux significances of sources
that pass our {\tt wavdetect} threshold extend well below that required to
satisfy the flux significance rule (see Figure~13).  We estimate
that roughly $\sim\!1/3$ of all the sources detected by {\tt wavdetect} fall
below this threshold.

Source detection is performed recursively by applying {\tt wavdetect} to
multiply-blocked sky images constructed as described in \S~2.5.3.
The use of a constant blocked image size maintains algorithm efficiency while
not compromising detection efficiency in the outer areas of the field of view
where the PSF size is significantly larger than a single pixel.

Applied to the CSC, wavelets with scales $a_i = 1$, $2$, $4$, $8$, and $16$
(blocked) pixels are computed for each image blocking factor and each energy
band except for the ACIS ultra-soft band. This combination of wavelet scales and
image blocking factors provides good sensitivity for detection of sources with
observed angular extents $\lesssim 30''$.  Some point sources with extreme
off-axis angles, $\theta > 20'$, may not be detected because the size of the
local PSF exceeds the largest wavelet scale/blocking factor combination.
\citetfap\ calibrate this effect statistically.

Source detection is not performed in the ACIS ultra-soft energy band.  This band
is impacted heavily both by increased background and by decreased effective area
because of ACIS focal plane contamination (the ratio of integrated background to
effective area is 1--2 orders of magnitude larger for the {\it u\/} band when
compared to the other ACIS energy bands).  Under these circumstances we are
limited by the accuracy of the background map determination; small errors in the
background map result in an unacceptable fraction of spurious source detections.

The {\tt wavdetect} algorithm incorporates steps to compare nearby correlation
maxima identified at multiple wavelet scales to ensure that each source is
counted only once. After duplicates are eliminated, a source cell that includes
the pixels containing the majority of the source flux is constructed.  Although
a source cell may have an arbitrary shape, for simplicity an elliptical
representation of the source region is used throughout the CSC\null. The lengths
of the semi-axes of this source region ellipse are set equal to the $3\,\sigma$
orthogonal deviations of the distribution of the counts in the source cell.

\begin{figure}
\epsscale{1.0}
\plotone{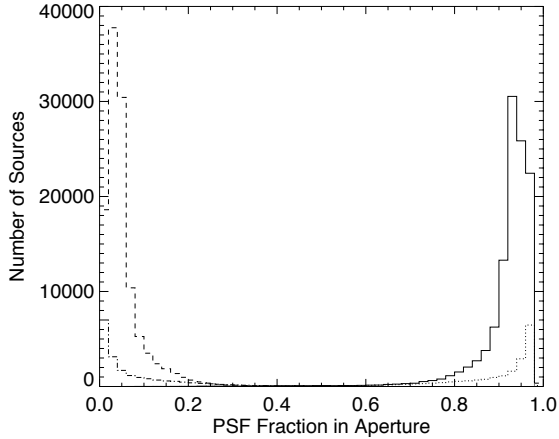}
\caption{\label{fig:psffrac}
Histogram of detected source PSF fractions.  For sources with off-axis angles 
$\theta \leq 10'$, the PSF fraction included in the source region aperture is 
shown by the solid line, while the dashed line displays the PSF fraction 
included in the background region aperture.  For sources with $\theta > 10'$,
the dotted line represents the PSF fraction included in the source region 
aperture and the dash-dotted line indicates the PSF fraction included in the 
background region aperture.}
\end{figure}
 
Source region ellipses for candidate sources detected within a single
observation from images with different blocking factors or in different energy
bands are combined outside of {\tt wavdetect} to produce a single merged source
list. This step rejects any detections that have RMS radii smaller than the 50\%
enclosed counts fraction radius of the local PSF, calculated at the
monochromatic effective energy of the band in which the source is detected.
Such detections are likely artifacts arising from cosmic ray impacts. Candidate
source detections whose centroids are closer than the local PSF radius, or that
are closer than $3/4$ of the mean detected source ellipse radii, are deemed to
be duplicates. If any duplicates are identified, then the detection from the
image with the smallest blocking factor is kept, and if the image blocking
factors are equal, then the detection with the highest significance is
used. This approach ensures that data from the highest spatial resolution
blocked image will be used to detect point and compact sources.  However, knotty
emission that is located on top of extended structures will tend to be
identified as distinct compact sources, while the extended emission is not
recorded.

\subsubsection{Source Apertures\label{sec:srcaper}}

Numerous source-specific catalog properties are evaluated within defined
apertures.  We define the ``PSF 90\% ECF (enclosed counts fraction) aperture''
for each source to be the ellipse that encloses 90\% of the total counts in a
model PSF centered on the source position. Because the size of the PSF is energy
dependent, the dimensions of the PSF 90\% ECF aperture vary with energy band.

We define the ``source region aperture'' for each source to be equal to the
corresponding $3\,\sigma$ source region ellipse included in the merged source
list, scaled by a factor of $1.5\times$.  Like the PSF 90\% ECF aperture, the
source region aperture is also centered on the source position, but the
dimensions of the aperture are {\it independent\/} of energy band.  Evaluation
of model PSFs with off-axis angles $\lesssim 10'$ demonstrates that the
dimensions of the source region aperture correspond {\it approximately\/} to the
dimensions of the PSF 90\% ECF ellipses for the ACIS broad energy band.  This is
confirmed {\it a posteriori\/} by examining the distribution of PSF aperture
fractions in source and background (see below) region apertures of all
individual catalog sources with ACIS broad band flux significance ${}\geq 3.0$.
Figure~14 demonstrates that the source region apertures typically
include $\sim\!90$--95\% of the PSF, while the background region apertures
contain $\lesssim 5$--10\%.  We emphasize that while these fractions are
typical, the actual PSF fractions, determined by integrating the model PSF over
the source and background region apertures and excluding regions from
contaminating sources, are used for the actual determination of source fluxes
(see \S~3.7).

Comparison of the source fluxes within the PSF 90\% ECF aperture and the
source region aperture provide a crude indication whether a source is
extended. If the flux in the source region aperture is significantly greater
than the flux in the PSF 90\% ECF aperture, then the source region determined by
{\tt wavdetect} is considerably larger than the local PSF, and the source is
likely extended.

Both the PSF 90\% ECF aperture and the source region aperture are surrounded by
corresponding background region annular apertures. In both cases, the inner edge
of the annulus is set equal to the outer edge of the corresponding source
aperture, while the radius of the outer edge of the annulus is set equal to
$5\times$ the inner radius of the source region aperture.  Although the
background region apertures defined in this manner include $\sim\!5$--10\% of 
the X-rays from the source, this contamination is accounted for explicitly when 
computing aperture photometry fluxes.

Overlapping sources could contaminate any measurements obtained through the
source and background apertures. To avoid this, both types of apertures are
modified to exclude areas that are included in any overlapping source region
apertures, or that fall off the detector.  Areas surrounding ACIS readout
streaks are also excluded from the modified background apertures. {\it
Aperture-specific catalog quantities are derived from the event data in the
appropriate modified aperture.\/} The fractions of the local model PSF counts
that are included in the modified apertures are recorded in the catalog for each
source, and are used to apply aperture corrections when computing fluxes, under
the assumption that the source is well modeled by the PSF.

The modified source region and background region aperture definitions are
recorded as FITS files using the spatial region file convention \citep{rot08}.
CIAO \citep{fru06} can be used to apply these regions as spatial filters to
extract the photon event data for the source (or background) from the archived
photon event list.  To simplify access to file-based data products (see
Table~3) for individual sources, we also separately store the
source region photon event list, per-band exposure maps, and per-band source
region images.  These products include data from the rectangular region of the
sky that is oriented North--South/East--West and that bounds the background
region.

\subsubsection{Matching Source Detections from Multiple Observations\label{sec:mrg}}

Each source record in the CSC Master Sources Table is constructed by combining
source detections included in the Source Observations Table from one or more
observations. A necessary first step in this process requires matching the
source detections from all of the observations that include the same region of
the sky. Cross-matching algorithms \citep[e.g.,][]{dev05, gra06} are
often focussed on efficiently matching large catalogs, and typically use
criteria on the position difference distribution, or cross-correlation approach
techniques, for identifying matches. In many cases, these approaches assume
(often implicitly) that the source PSF is at least approximately spatially
uniform across the field of view, and comparable between the datasets being
matched.

\begin{figure*}
\epsscale{1.0}
\plotone{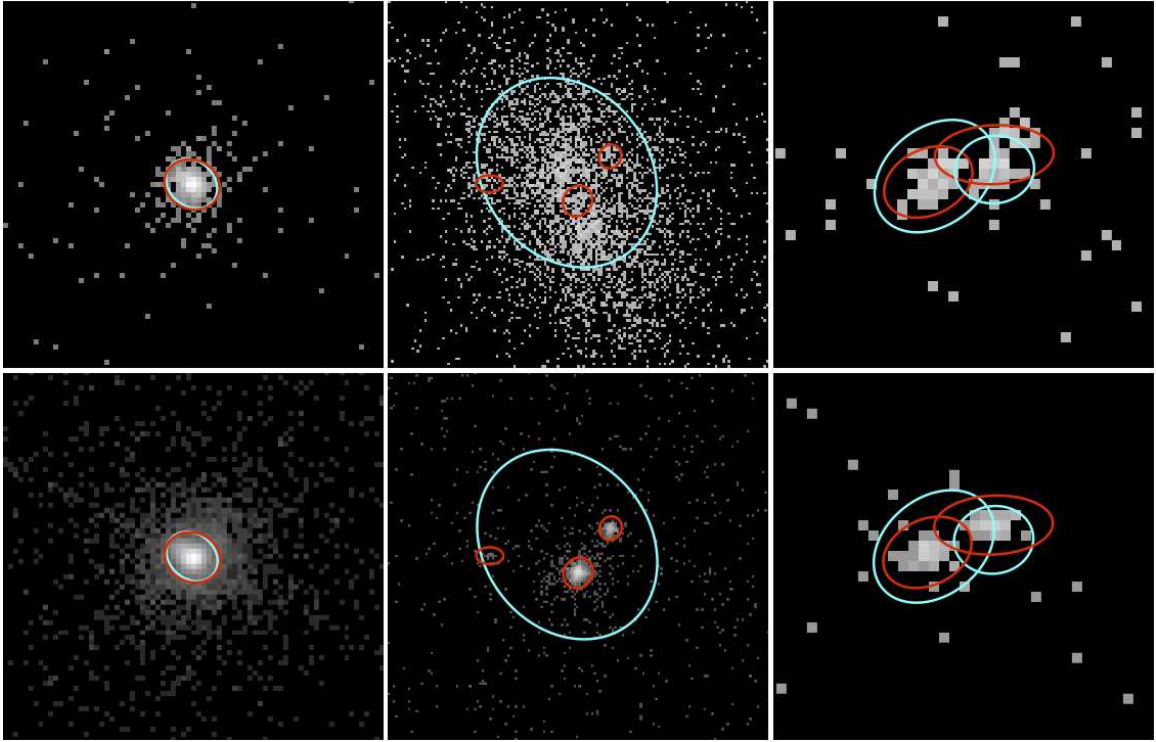}
\caption{\label{fig:mrgcases}
{\it Left:\/} Upper and lower images illustrate the common source matching
case where the source detections from the individual observations all uniquely
match a single source on the sky.  The source region aperture determined from
the upper image is shown in cyan, while the source region aperture determined
from the lower image is shown in red.
{\it Center:\/} In this case, the off-axis source region aperture computed from
the source detection in the upper image, shown in cyan, overlaps multiple source
region apertures from the observation in the lower image, shown in red.  The
cyan source detection is confused, and will be connected to the master sources
associated with the red source detections using ``ambiguous'' linkages.
{\it Right:\/} The sources detected in these observations for a confused ``pair
of pairs.''  The fractional overlaps between the pair of cyan source region
apertures and the pair of red source region apertures is sufficiently large that
these detections are assigned to be resolved by human review.}
\end{figure*}
 
However, when matching source detections across multiple {\it Chandra\/}
observations, the strong dependence of the PSF size with off-axis angle must be
considered explicitly, since source detections that are well off-axis in one
observation are often resolved into multiple sources close to the optical axis
in other observations. Under these circumstances the source positions determined
by {\tt wavdetect} are {\it not\/} comparable, and cannot be used for source
matching.  Instead, the source matching approach used for the CSC is based on
the overlaps between the PSF 90\% ECF apertures of the source detections from
the individual observations. Although empirical in nature, this algorithm works
well for matching compact source detections between {\it Chandra\/}
observations.

The detailed algorithm is described in Appendix~A.  The method
identifies the overlap fractions between the PSF 90\% ECF apertures of
overlapping source detections from the observations, and separates them into
three different categories.

The first category is the simplest, where the source detections from the various
observations have apertures that all mutually overlap (Fig.~15,
{\it Left\/}). This is the most common situation, and corresponds to the case
where the source detections all uniquely match a single source on the sky.
Roughly 90\% of the $\sim\!18,000$ sources in the Master Sources Table that are
linked to more than one source detection in the Source Observations Table fall
into this category. Each of the matching entries in the latter table will be
associated with the corresponding Master Sources Table entry with a ``unique''
linkage, as described in \S~2.3.

In the second category, the aperture associated with a source detection in one
observation overlaps the apertures associated with multiple distinct source
detections from other observations. This circumstance typically arises because
source detections from a single observation are always assumed to be distinct;
this assumption can fail very far off-axis ($\theta \gtrsim 20'$), where the PSF
size exceeds the maximum {\tt wavdetect} wavelet scale/image blocking factor
combination. This category is illustrated in Figure~15, {\it
  Center\/}, and arises most often because a source detection in one observation
is resolved into multiple sources by one or more of the overlapping
observations. The unresolved source detection in the Source Observations Table
will be connected to all Master Sources Table entries associated with the
matching resolved source detections via ``ambiguous'' linkages, and the
detection will be flagged as confused. The X-ray photon events associated with
the unresolved detection cannot be distributed across the matching resolved
sources. In release 1 of the CSC, source properties derived from the detection
will not be used to compute the source properties included in the Master Sources
Table. Upper limits for photometric quantities could in principle be extracted
from the unresolved source detection, and these would be quite valuable for
variability studies. This capability will be included in a future release of
the CSC\null.

In a few cases, a set of aperture overlaps cannot be resolved automatically
using the current algorithm. This third category typically occurs when there are
multiple overlapping, confused source detections. In this case, the source
detections are flagged for review by a human, who is then responsible for
resolving the matches. Only 415 (out of 94,676) master sources include source
detections that required manual review, and the majority of these were readily
resolved as confused ``pairs of pairs.'' The latter case, which is illustrated
in Figure~15, {\it Right\/} commonly occurs because a pair of
source detections included in a single observation both overlap a pair of source
detections in another observation. Each of the source detections in the first
observation overlaps both source detections in the second observation, making
the detections confused, and vice-versa. If a manual review is required to
complete a match for a specific source, then a flag is set in the catalog to
indicate this fact.

The actual processing required to perform the matches is complex. Whenever new
overlapping source detections are identified during catalog processing, the set
of matches is recomputed using all of the observations processed so far. The
algorithm then queries the prior state of the catalog, and determines the set of
updates that are necessary to migrate from that state to the newly determined
state. This procedure works regardless of the order in which observations
processed, and also permits an already-processed observation to be reprocessed
should an error have occurred.

\subsubsection{Source Naming}

Each distinct X-ray source included in the Master Sources Table is assigned a
name that is derived from the source's location on the sky.  Catalog sources are
designated ``CXO J$\hbox{\it HHMMSS.s\/}\pm \hbox{\it DDMMSS\/}$,'' where
$\hbox{\it HHMMSS.s\/}$ and $\pm\hbox{\it DDMMSS\/}$ are the ICRS right
ascension and declination, respectively, of the source position, {\it
truncated\/} to the indicated precision. This format complies with the
International Astronomical Union (IAU) Recommendations for
Nomenclature\footnote{\url{http://cdsweb.u-strasbg.fr/cgi-bin/Dic/iau-spec.htx}}.
The ``CXO'' prefix is registered with Commission~5 of the IAU for exclusive
use in source designations issued by the {\it Chandra\/} X-ray Center. 

The name assigned to a source is determined from the combined source position
once the detections of the source are merged according to \S~3.4.2.  If
the source has never been included in a released version of the CSC, then the
source name may be revised if a source detection in a subsequently processed
observation modifies the combined source position.  Therefore, the name assigned
to a source that is visible in a database view {\it may\/} change as new
observations are processed, if the source has never been included in a catalog
release.

Once a source is included in a released version of the CSC, then the name of
that source is frozen. The name will not be changed in either future catalog
release views (i.e., subsequent catalog releases) or database views, even if
additional observations refine the source position. Therefore minor
discrepancies can arise between the latter and the source designation. However,
if an observation included in a later catalog release resolves an apparently
single source included in an earlier version of the catalog into multiple
distinct sources, then the previous source designation is retired and new names
are assigned to the resolved sources.

\subsection{Source Position Determination}

\begin{figure}
\epsscale{1.0}
\plotone{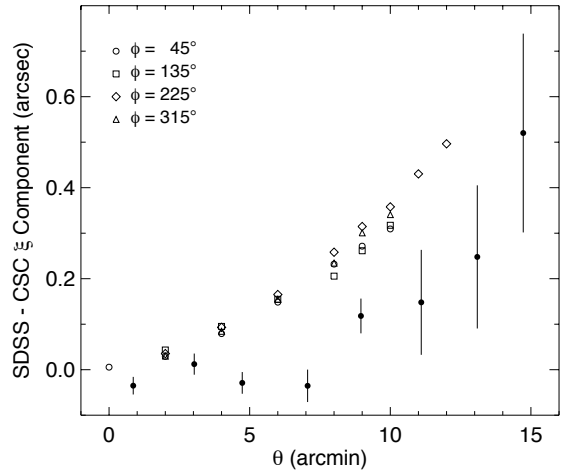}
\caption{\label{fig:centerr}
Mean (signed) coordinate differences between SDSS and CSC source positions
measured along the $\boldxi$ vector (defined in the text) as a function of
off-axis angle, $\theta$, for individual observations of CSC sources with at 
least 500 net counts are shown as filled circles.  Open symbols indicate expected
values computed from high SNR ray-trace simulations at the indicated values of
$\theta$ and $\phi$\null.  Note that the measured mean coordinate differences 
are consistent with zero offset for $\theta\lesssim 8'$, and are a factor 
$\sim\!2$ times smaller than the model predictions for larger values of $\theta$.}
\end{figure}
 
Within a single observation, the detected source positions are those assigned by
the {\tt wavdetect} algorithm.  Their accuracy can be estimated by evaluating
the mean (signed) coordinate differences between the {\tt wavdetect} positions
and the positions of matching sources in the seventh data release (DR7) of the
Sloan Digital Sky Survey \citep[SDSS;][]{aba09}.  The latter are extracted from
the CSC/SDSS Cross-Match
Catalog\footnote{\url{http://cxc.cfa.harvard.edu/cgi-gen/cda/CSC-SDSSxmatch .html}}.
As shown in Figure~16, the mean coordinate differences
demonstrate good position agreement between the CSC and SDSS for sources with
$\theta\lesssim 8'$ (where we have restricted the comparison to include only
individual observations of CSC sources with at least 500 net counts to minimize
statistical errors).

For larger off-axis angles, Figure~16 suggests that there may be
a systematic offset between the {\tt wavdetect} source positions and the SDSS
source positions.  The measured mean position difference is $\lesssim 0\farcs
3$ for $\theta\lesssim 15'$, but appears to increase with off-axis angle.  The
exact cause of this offset is uncertain.  Some authors \citep[e.g.,][]{ale03,
leh05, luo08} have reported a similar effect, which they attribute to
centroiding errors introduced by of the asymmetric nature of the {\it Chandra\/}
PSF at large off-axis angles.

High quality PSF simulations generated using the SAOTrace (formerly SAOsac)
ray-trace code \citep{jer95, jer04} confirm that the asymmetry can displace the
measured centroid from the requested location of the simulated PSF on a uniform
pixel grid.  We designate the vector orientation from the measured centroid
position to the requested location of the PSF as $\boldxi$\null.  As shown in
Figure~16, the expected centroid displacement along $\boldxi$
computed from the simulations is not a good measure of the actual mean
coordinate difference, which is a factor of order 2 times smaller than predicted
by the models.  One possible reason for the disagreement between the model and
actual measurements is that the {\it Chandra\/} plate scale calibration was
derived from observations of NGC~2516 and LMC~X-1 using source centroid
measurements that were {\it not\/} corrected for the asymmetry of the PSF
\citep{mar01b}.

As described in \S~2.5.3, the celestial coordinates of a source
observed by {\it Chandra\/} are computed by applying a series of transforms to
the measured position of the source on the detector.  The final step in this
process applies the measured plate scale calibration to the difference between
the position of the source on the virtual sky pixel plane and a known fiducial
point, which is typically the telescope optical axis.  If all of the star-star
baselines used to calibrate the plate scale were oriented parallel to $\boldxi$,
then the lack of correction for the PSF asymmetry would to first order
compensate for the linear component of the systematic position offset when
measuring real sources whose locations are fixed in world coordinates rather
than virtual pixel plane coordinates.  Since not all star-star baselines were so
aligned, some residual systematic position offset may be expected, but at an
undetermined level that is less than predicted from the PSF simulations.

Systematic position offsets can also arise from uncertainties in the detector
geometry.  As an example, consider imaging observations that use the nominal 
ACIS-I aimpoint.  Sources with $\theta\lesssim 8'$ will be located on the same 
CCD array as the aimpoint.  As $\theta$ increases, an increasing fractions of 
sources will instead be positioned on ACIS-S array CCDs, until for $\theta 
\gtrsim 11'$ all sources will be located on the ACIS-S array.  Uncertainties in 
the relative positions and tilts of the ACIS-I and ACIS-S arrays would therefore 
introduce systematic position offsets for sources with large off-axis angles, 
while not impacting sources that fall on the same CCD array as the aimpoint.  
This signature is consistent with the absence of mean position differences
measured for $\theta\lesssim 8'$.

Because of the small magnitude of the mean position differences measured for
$\theta\lesssim 15'$, we have chosen not to apply an uncertain correction to
source positions in release 1 of the catalog.  We plan to investigate in detail
the cause of the systematic position offsets at large off-axis angles, and adjust
source positions in future catalog releases if appropriate.

\subsubsection{Source Position Uncertainty\label{sec:posunc}}

In addition to reporting measured source positions, {\tt wavdetect} also reports
positional errors associated with each source detection. The reported errors are
based on a statistical moments analysis, and do not consider instrumental
effects such as pixelization, aspect-induced blur, or asymmetrical PSF structure
that may contribute to the total positional uncertainties. Simulations that
compare the reference positions of artificially generated sources with their
positions determined by {\tt wavdetect} indicate that the positional
uncertainties computed by {\tt wavdetect} are underestimated for sources with
large off-axis angles. The simulation results, which quantify the dependence of
positional uncertainties of simulated sources on off-axis angle, were found to
be consistent with the more extensive simulations uses to construct the {\it
Chandra\/} Multiwavelength Project (ChaMP) X-ray point source catalog
\citep{kim07}.

In the first release of the CSC, source position error ellipses are substituted by
error circles computed using the ChaMP positional uncertainty relations 
\begin{equation}
\label{eqn:PU}
\log P = \left\{ 
  \begin{array}{l}
    0.1145\theta - 0.4957 \log S_w + 0.1932 \\
    \quad\quad\quad\quad\quad\quad 0.0000 < \log S_w \leq 2.1393 \\
    0.0968\theta - 0.2064 \log S_w - 0.4260 \\
    \quad\quad\quad\quad\quad\quad 2.1393 < \log S_w \leq 3.3000.
 \end{array}
\right.
\end{equation}
In these equations, $P$ is the positional uncertainty in arcseconds, $\theta$ is
the off-axis angle in arcminutes, and $S_w$ is the source net counts reported by
{\tt wavdetect}. These relations were derived to characterize the positional
uncertainties at the 95\% confidence level of X-ray point sources in the ChaMP
X-ray point source catalog, which includes $\sim\!6,800$ X-ray sources detected
in 149 {\it Chandra\/} observations.  The values of $\log P$ computed using
equations~(\ref{eqn:PU}) are not equal at the boundary where $\log S_w = 2.1393$
(roughly 138 net counts). However, this error is negligible for
$\theta\lesssim\!10'$.

Although HRC observations are not included in the first release of the CSC, we 
have used a series of simulations to derive an improved positional uncertainty 
relation that is appropriate for sources detected in HRC-I observations.  The 
simulations include $\sim\!6,000$ point sources spanning $0< \theta < 22'$ and 
$9 < S_w < 3600$.  The best fit surface for the 95\% position uncertainty 
quantile is
\begin{eqnarray}
\label{eqn:HPU}
\log P &=& 0.752569 + 0.216985\theta + 0.000242\theta^2\nonumber\\
&&- 1.142476 \log S_w + 0.172132\log^2 S_w \\
&&- 0.040549\theta\log S_w\nonumber.
\end{eqnarray}
The simulations do not sample the region with $\theta > 20'$ and $S_w < \sim\!50$,
and so we impose an upper bound of $\log P = 2.128393$ on this relation, to 
cap the uncertainty in this regime.

In release 1.1 of the CSC, positional uncertainties for sources detected in ACIS
observations are computed using equation~(\ref{eqn:PU}) while positional
uncertainties for sources detected in HRC-I observations are computed using
equation~(\ref{eqn:HPU}).

The positional uncertainties from equations~(\ref{eqn:PU}) and~(\ref{eqn:HPU})
provide a good measure of the statistical uncertainty of the location of the
source in the frame of the observation, but do not consider potential sources of
error that are external to the observation.  These include the error in the mean
aspect solution for the observation, the astrometric errors in the AXAF ({\it
Chandra\/}) Guide and Acquisition Star Catalog \citep{sch03}, and the
calibration of the geometry of the spacecraft and focal plane.  As described in
\S~4.3, \citet{rot09} has recently used the CSC/SDSS Cross-Match 
Catalog to calibrate the combined external error by analyzing the statistical 
distribution of the measured separations of CSC point source detections from 
individual observations with their counterparts in SDSS DR7 \citep{aba09}.

The resulting external astrometric error is $0\farcs16\pm0\farcs0.01$ 
($1\,\sigma$), or $0\farcs39$ (95\% confidence).  The latter must be added in 
quadrature to the position uncertainties from equations~(\ref{eqn:PU}) 
and~(\ref{eqn:HPU}) to compute the absolute position error for CSC sources.  
In release~1 of the catalog, the positional error reported in the catalog 
tables is taken directly from equation~(\ref{eqn:PU}), so the quadrature 
addition of the external astrometric error component {\it must be performed 
by the user.\/}  Release~1.1 of the CSC will {\it include\/} this error 
component directly in the tabulated values.

\begin{figure}
\epsscale{1.0}
\plotone{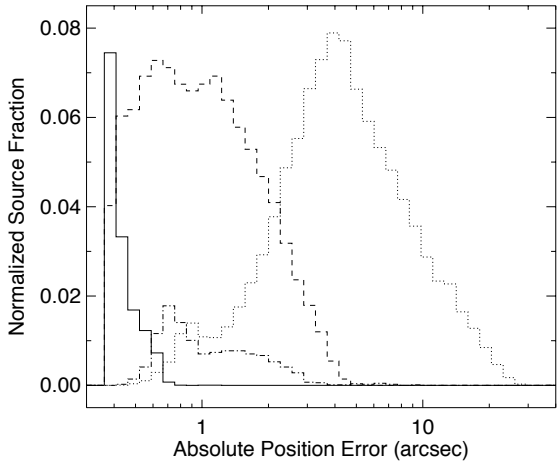}
\caption{\label{fig:poserr} 
Histograms of the total absolute position error for source detections included 
in release~1 of the CSC\null.  The dashed and dotted lines include all source 
detections with off-axis angles $\theta < 8'$ and $8' \leq \theta < 15'$, 
respectively.  Each histogram is normalized by the total number of source 
detections within the relevant $\theta$ range.  Source detections with at least 
500 net counts are shown with the solid line ($\theta < 8'$) and dash-dotted 
line ($8' \leq \theta < 15'$).  For the latter two histograms, the normalized
source fraction has been scaled by a factor of $4\times$.}
\end{figure}
 
A {\it post facto\/} histogram of the 95\% confidence positional uncertainties,
including the external astrometric error, for source detections included in
release~1 of the CSC is presented in Figure~17.  The figure
demonstrates that statistical errors due to lack of net source counts does
dominate the positional uncertainty for all but the brightest sources.  However,
for bright sources detected at small off-axis angles, the external astrometric
error limits the accuracy of the derived source positions.  Any error introduced
by using uncorrected source centroid positions from {\tt wavdetect} is
negligible for the overwhelming majority of source detections.

\subsubsection{Combining Source Positions from Multiple Observations\label{sec:mow}}

Improved estimates of the position and positional uncertainty of each X-ray
source are determined from the statistically independent source detections
included in the set of individual observations using a multivariate optimal
weighting formalism.  This technique is effective in cases where simple
averaging fails, for example when the area defining the source position varies
significantly from observation to observation. We express the uncertainties of
the estimates in the form of error ellipses centered upon the estimated source
positions.  An equivalent approach has been used for weapons targeting
\citep{ore96}.  To our knowledge the usage here is the first documented
application to astrophysical data.

In the multivariate optimal weighting formalism, given a set of estimates,
$X_i$, of the mean of some two-dimensional quantity, and the $2 \times 2$
covariance matrices, $\sigma_i^2$, associated with these estimates, an
improved estimate, $X$, of the mean, and the associated covariance matrix,
$\sigma^2$, are \citep[e.g.,][]{jed07b}
\begin{eqnarray}
\label{eqn:mow}
X = \sigma^2 \sum_i \frac{X_i}{\sigma_i^2} &;\quad& \sigma^2 = \left[ \sum_i
  \frac{1}{\sigma_i^2} \right]^{-1}.
\end{eqnarray}

For the application described here, we take $X_i$ to be the $i$th estimate of
the source position, projected onto a common tangent plane (which is constructed
at the mean position of the ellipse centers).  The corresponding covariance
matrix is 
\begin{equation}
\label{eqn:cov}
\begin{array}{l}
\!\!\sigma_i^2 =\\
\;\left(
\begin{array}{cc}
\sigma^{\prime 2}_{1,i}\cos^2 \vartheta_i + \sigma^{\prime 2}_{2,i} \sin^2
\vartheta_i & (\sigma^{\prime 2}_{2,i} - \sigma^{\prime 2}_{1,i}) \cos
\vartheta_i \sin \vartheta_i \\ 
(\sigma^{\prime 2}_{2,i} - \sigma^{\prime 2}_{1,i}) \cos \vartheta_i \sin
\vartheta_i & \sigma^{\prime 2}_{1,i}\sin^2 \vartheta_i + \sigma^{\prime
  2}_{2,i} \cos^2 \vartheta_i 
\end{array}
\right),\end{array}\nonumber
\end{equation}
where $\sigma^{\prime}_{1,i}$ and $\sigma^{\prime}_{2,i}$ are the lengths of the
semi-minor and semi-major axes, respectively, of the $i$th error ellipse
projected onto the common tangent plane, and $\vartheta_i$ is the angle that the
major axis of the $i$th error ellipse makes with respect to the tangent plane
$y$ axis.  The derivation of equation~(\ref{eqn:cov}) is presented in
Appendix~B. 

Once the covariance matrices corresponding to the error ellipses for each
individual source observation, equation~(\ref{eqn:cov}), are computed, the error
ellipses are combined using equation~(\ref{eqn:mow}).  This yields the optimally
weighted source position and position error ellipse for the combined set of
observations, on the common tangent plane.  Mapping these back to the
celestial sphere provides the combined source position and error ellipse
estimates.

\subsection{Source Extent Estimates}

The observed spatial extent of a source is estimated using a rotated elliptical
Gaussian parameterization of the form
\begin{eqnarray}
\label{eqn:egaus}
S(x, y; \alpha) = \frac{s_0}{\sigma_1 \sigma_2} \exp\left[-\pi(\mathcal{A}\mathbf{x})^2\right],
\end{eqnarray}
where
\begin{eqnarray*}
\mathcal{A} = \left({\begin{array}{cc}
\sigma_1^{-1} & 0 \\
0 & \sigma_2^{-1}
\end{array}}\right)
\left({\begin{array}{cc}
\cos\phi_0 & \sin\phi_0 \\
-\sin\phi_0 & \cos\phi_0
\end{array}}\right) &;\quad& 
\mathbf{x} = \left({\begin{array}{cc}
x \\ y\end{array}}\right),
\end{eqnarray*}
where $(x, y)$ is the Cartesian center location of the Gaussian, and the
parameters $\alpha = (\sigma_1, \sigma_2, \phi_0)$ are the $1\,\sigma$ radii
along the major and minor ellipse axes, and the position angle of the major axis
of the ellipse, respectively.

The parameters of $S$ are determined using a wavelet-based approach that is
similar to that used for source detection.  The methods differ in their details
and assumptions, however.

For source detection, the choice of wavelet scales used by {\tt wavdetect} to
detect sources is determined {\it a priori\/}.  Because of the strong variation
of PSF size with off-axis angle, multiple wavelet scales and input image
blocking factors are required to search for sources at all off-axis angles, as
described in \S~3.4.  Stepping between the discrete wavelet scales
and image blocking factors as a function of off-axis angle introduces small but
systematic biases in the derived dimensions of the source region ellipses (and
therefore the source region apertures; see Figure~18).

Photometric, spectral, and temporal properties determined from the X-ray events
included in the aperture are not impacted by these effects, since the aperture
dimensions are sufficiently large that they typically enclose $\sim\!90\%$ of
the PSF counts, and a correction factor is applied for the fraction of the PSF
that falls outside of the aperture.  However, these biases render the source
region aperture dimensions unsuitable for use as a measure of the source extent.

\begin{figure*}
\epsscale{1.0}
\plotone{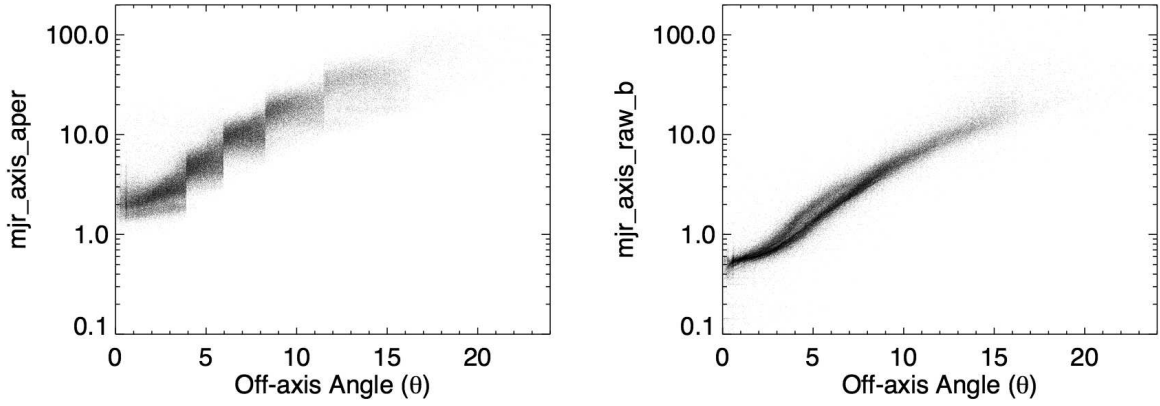}
\caption{\label{fig:apertheta}
Comparison of source region aperture dimensions with raw (undeconvolved) source
extent for the ACIS broad ({\it b\/}) energy band for all source observations 
included in the first release of the CSC\null.  {\it Left:\/} Semi-major axis 
of the source region aperture is plotted for each source observation versus source
off-axis angle, $\theta$\null.  The step-like structure visible in the plot
results from the stepping between discrete wavelet scales and image blocking
factors as a function of off-axis angle.  {\it Right:\/} Semi-major axis of the
raw (undeconvolved) ACIS broad ({\it b\/}) energy band source extent is plotted
versus $\theta$\null.  The source extent is derived using a scaleless wavelet 
approach, as described in the text, and varies smoothly with $\theta$.  Differing 
geometric tilts of the ACIS-I and ACIS-S CCD arrays relative to the focal plane
split the distribution into separate populations for $3'\lesssim\theta\lesssim 7'$.
The lower population corresponds to point sources located on ACIS-I array CCDs,
while point sources located on ACIS-S array CCDs comprise the upper population.
The difference in the vertical scales between the two plot panels results because 
the source region aperture scales {\it approximately\/} as the PSF 90\% ECF, 
whereas the source extent estimates the $1\,\sigma$ scale of an elliptical Gaussian 
parameterization of the source.}
\end{figure*}
 
The wavelet-based approach used for estimating the source extent determines the
optimal wavelet scale size directly from the data under the assumption that a
source exists at approximately the location determined by {\tt wavdetect\/}.
The raw source extent estimates derived using this approach vary smoothly with
off-axis angle (Fig.~18).

The two-dimensional correlation integral
\begin{equation}
\label{eqn:corrint2}
\begin{array}{l}
\!\!C(x, y; \boldalpha) =\\
\quad \int_{-X}^X\int_{-Y}^Y dx^\prime\,dy^\prime\, W(x-x^\prime, y-y^\prime; \boldalpha)\,S(x^\prime, y^\prime; \boldalpha),
\end{array}
\end{equation}
where the region of interest is $|x^\prime|\leq X$ and $|y^\prime|\leq Y$, and
$W$ is again specified by equation~(\ref{eqn:mexhat}), is computed first.

We choose a coordinate system in which the peak of $S$ is centered at the
origin. The quantity $\psi(x, y; \boldalpha) = C(x, y;
\boldalpha)/(a_1a_2)^{1/2}$ has a maximum value at the origin when
$a_i=\sigma_i\sqrt{3}$ and $\phi=\phi_0$ \citep{dam97}. The source parameters
are determined by maximizing $\psi$.

In practice, equation~(\ref{eqn:corrint2}) is evaluated as a discrete sum over
the pixels of the image. Although integration of equation~(\ref{eqn:mexhat})
does not yield a simple closed-form solution, and numerical integration is
computationally expensive, for the purpose of optimizing $\psi_0 = C(0, 0;
\boldalpha)/(a_1a_2)^{1/2}$, a rectangular approximation for the integral over
each pixel is sufficient. In this approximation,
\begin{displaymath}
W_{mn}(x_i, y_j; \boldalpha) \approx W(x_m - x_i, y_n - y_j; \boldalpha) \Delta x \Delta y,
\end{displaymath}
where $W(x, y; \boldalpha)$ is evaluated at the center of each pixel and the
pixel area is $\Delta x \Delta y$.

A small sub-image of the source is extracted centered on the source position
determined by {\tt wavdetect}.  The accuracy of this source position is refined
by searching the center of the sub-image for the coordinates $(x_0, y_0 )$ that
maximize $\psi_0(x, y; a, a, 0)$.  A new sub-image is then extracted using the
improved source position.

Finally, the size and orientation of the elliptical Gaussian source
parameterization are derived by maximizing $\psi(x_0, y_0; \boldalpha)$.
Choosing good initial values for $a_i$ helps to ensure that this optimization
step converges reliably.  We set the initial values $a_1 = a_2 = \max{[(d_1
  d_2)^{1/2}, a_{\rm grid}}]$, where $d_i$ are the ellipse semi-axes derived by
{\tt wavdetect}, and the value of $a_{\rm grid}$ is obtained by examining
$\psi_0(x_0, y_0; a, a, 0)$ on a grid of $a$ values spanning the half-width of
the source image; $a_{\rm grid}$ is usually the smallest $a$ that corresponds to
either a local maximum or an inflection point.  This choice is motivated by the
observation that when a single source is present, the location of a local
maximum provides a good estimate of the source size (see Figure~19).
Similarly, when the source of interest is blended with other nearby sources, an
inflection point where $\partial^2_a\psi_0 = 0$ often occurs near the ``edge''
of the central source.  When the first occurrence of $\partial_a\psi_0=0$ occurs
at a local minimum, $a_{\rm grid}$ is the smallest value of $a$ on the pixel
grid.

\subsubsection{PSF Extent\label{sec:psf}}

The spatial extent of the local PSF is determined for comparison with the
observed source extent, and as an aid to assessing the intrinsic extent of the
source. Since the size of the {\it Chandra\/} PSF is a strong function of
off-axis angle, a ray-trace model is constructed at the measured off-axis and
azimuthal angles $(\theta, \phi)$ separately for each detected source. Although
the shape of the PSF is energy-dependent, within each energy band the ray-trace
model is computed only at the monochromatic effective energy (see
\S~2.5.2) of the band. This approximation results in an error that is
dependent on the actual spectrum of the source, but that does not exceed
$\sim\!10\%$ for typical power-law or black-body source spectra.

The ray-trace model is computed using version 1.0.0 of the SAOTrace simulation
code \citep{jer95, jer04} with the latest HRMA optical coefficients\footnote{The
{$\tt orbit\_XRCF\mathtt{+}tilts\mathtt{+}ol\_01b$} calibration model
configuration.} derived from calibration observations.  A ray density of $\rm
0.2\,\hbox{rays}/\hbox{mm}^2$ is used for the ray-trace.  The rays are then
projected onto the detector focal plane, a Gaussian blur is applied to account
for the degradation due to de-dithering the telescope motion
($\sigma=0\farcs148$ for ACIS, $0\farcs2$ for HRC), and the image is resampled
onto the pixel plane of the detector.  Each PSF image is recorded in the CSC as
a file-based data product (see Table~3), which can be
retrieved by the catalog user and compared directly to the corresponding source
region image.

\begin{figure}
\epsscale{1.0}
\plotone{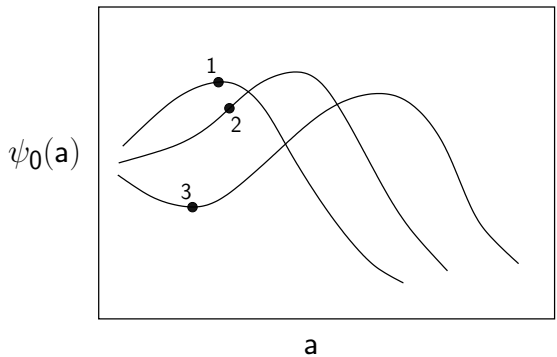}
\caption{\label{fig:psi}
Deriving a correlation scale length from the shape of $\psi_0(a)$.  Three curves
are shown, schematically illustrating the different shapes that $\psi_0(a)$ can
have.  An unblended source normally yields a distinct local maximum (point 1).
Depending on the source spacing, closely spaced sources may yield an inflection
point (point 2) or a local minimum (point 3).  The text describes how each of
these cases is treated.}
\end{figure}
 
The ray density used in the ray-trace models typically produces a total of
$\sim\!6$--$8\times10^3$ counts (with a full range of $\sim\!4$--$13\times10^3$
counts) in the resulting PSF model in the ACIS broad, hard, and soft energy
bands, where the combined HRMA/instrument effective area at the monochromatic
effective energies of the bands is $\sim\!300$--$400\rm\,cm^2$.  The number of
counts in the ACIS medium energy band PSF models is about 60\% higher, because
of the higher effective area at that band's monochromatic effective energy.  For
the ACIS ultra-soft energy band the total PSF counts may be only a few hundred
because of the poor quantum efficiency in the band.  Although the PSF models
computed here are sufficient for their intended purposes of providing basic
estimates of source extent and point source aperture corrections for aperture
photometry, they are not suitable for analyses such as image deconvolution that
require detailed PSF models.

Once the image of the local model PSF is constructed, the rotated elliptical
Gaussian parameterization, $p(x, y; b_1, b_2, \psi)$, of the spatial extent is
computed in the same way that the observed source extent, $S(x, y; \sigma_1,
\sigma_2, \phi_0)$, is computed from the source image.

\subsubsection{Intrinsic Source Extent}

Using the rotated elliptical Gaussian parameterizations derived above, the
observed source extent, $S(x, y; \sigma_1, \sigma_2, \phi_0)$, can be treated as
the convolution of the local PSF, $p(x, y; b_1, b_2, \psi)$ with the intrinsic
source extent, $s(x, y; a_1, a_2, \phi)$, where we parameterize the latter
similarly to equation~(\ref{eqn:egaus}).  In general, $\phi_0 \neq \phi$, since
the PSF-convolved ellipse need not have the same orientation as the intrinsic
source ellipse.

In principle, one can determine the parameters, $(a_1, a_2, \phi)$, of the
intrinsic source ellipse by solving a nonlinear system of equations involving
the PSF parameters, $(b_1, b_2, \psi)$, and the observed source parameters,
$(\sigma_1, \sigma_2, \phi_0)$. However, because these equations are based on
simple assumptions regarding the source and PSF profiles, and because the input
parameters are often uncertain, such an elaborate calculation seems unjustified.
 
A much simpler and more robust approach makes use of the identity 
\begin{displaymath}
\sigma_1^2 + \sigma_2^2 = a_1^2 + a_2^2 + b_1^2 + b_2^2,
\end{displaymath}
which applies to the convolution of two elliptical Gaussians having arbitrary
relative sizes and position angles.  Using this identity, one can define a
{\it root-sum-square\/} intrinsic source size,
\begin{eqnarray}
\label{eqn:arss}
a_{\rm rss} &=& \frac{1}{\sqrt{2}}\left(a_1^2+a_2^2\right)^{1/2} \nonumber \\
&=&\frac{1}{\sqrt{2}}\max \left[0, (\sigma_1^2+\sigma_2^2) -
  (b_1^2+b_2^2)\right]^{1/2}, 
\end{eqnarray}
that depends only on the sizes of the relevant ellipses and is independent of
their orientations. This expression is analogous to the well-known result for
convolution of one-dimensional Gaussians and for convolution of circular
Gaussians in two dimensions. The factors of $1/\sqrt{2}$ ensure that the
statistic value gives the radius of the source image when applied to circular
source images.

Using equation~(\ref{eqn:arss}), one can derive an analytic expression for the
uncertainty in $a_{\rm rss}$ in terms of the measurement errors associated with
$\sigma_i$ and $b_i$.  Because $\sigma_i$ and $b_i$ are non-negative, evaluating
the right-hand side of equation~(\ref{eqn:arss}) using the corresponding mean
values should give a reasonable estimate of the mean value of $a_{\rm rss}$.  A
Taylor series expansion of the right-hand side of equation~(\ref{eqn:arss})
evaluated at the mean parameter values therefore yields the uncertainty
\begin{equation}
\label{eqn:errarss}
\begin{array}{l}
\!\Delta a_{\rm rss} =\nonumber \\
\;\frac{1}{\sqrt{2}a}\left[\sigma_1^2(\Delta\sigma_1)^2 
  + \sigma_2^2(\Delta\sigma_2)^2 + b_1^2(\Delta b_1)^2 + b_2^2(\Delta
  b_2)^2\right]^{1/2}, 
\end{array}
\end{equation}
where $(\Delta X)^2$ represents the variance in $X$, and where
\begin{displaymath}
a = \left\{ 
  \begin{array}{c l}
    a_{\rm rss} & \quad a_{\rm rss} > 0 \\
    \sqrt{b_1^2+b_2^2} & \quad a_{\rm rss} = 0.
  \end{array}
\right.
\end{displaymath}

\subsubsection{Combining Intrinsic Source Extent Estimates from Multiple Observations}

Measurements of the mean intrinsic source extent derived from multiple
independent observations, $a_{{\rm rss},i} \pm \Delta a_{{\rm rss},i}$, are
combined using the multivariate optimal weighting formalism,
equations~(\ref{eqn:mow}). The minimum variance estimator of the intrinsic source
size is the variance-weighted mean,
\begin{displaymath}
\overline{a_{\rm rss}} = {\rm Var}[\overline{a_{\rm rss}}]\sum_i{\rm Var}[a_{{\rm rss},i}]^{-1}a_{{\rm rss},i},
\end{displaymath}
where ${\rm Var}[a_{{\rm rss},i}] = (\Delta a_{{\rm rss},i})^2$.  The variance
in $\overline{a_{\rm rss}}$ is
\begin{displaymath}
{\rm Var}[a_{\rm rss}] = \left[\sum_i{\rm Var}[a_{{\rm rss},i}]^{-1}\right]^{-1}.
\end{displaymath}

\subsection{Aperture Photometry\label{sec:apphot}}

Net source counts, count rates, and photon and energy fluxes for point sources
are computed from counts and exposure data accumulated in independent source and
background apertures, $R_s$ and $R_b$.  Typically, these apertures are simple
elliptical regions and surrounding elliptical annuli, but arbitrary areas from
either aperture may be excluded to avoid contamination from nearby sources, or
missing data due to detector edges.  The net aperture areas, $A_s$ and $A_b$,
and the fractions, $\alpha$ and $\beta$, of source counts expected in both
apertures are determined from
\begin{eqnarray*}
A_s=\int_{R_s}dx dy &;\quad& A_b =\int_{R_b}dx\,dy
\end{eqnarray*}
and
\begin{eqnarray*}
\alpha = \int_{R_s}dx\,dy\,\mathit{PSF}(x,y) &;\quad& \beta = \int_{R_b}dx\,dy\,\mathit{PSF}(x,y).
\end{eqnarray*}
We use the PSFs described in \S~3.6.1 to estimate $\alpha$ and $\beta$.
Although the finite number of PSF counts leads to some uncertainty in the 
estimate for $\beta$, the effect of this uncertainty on the derived net counts, 
count rates, and fluxes is small (typically $\ll 1\%$).

If a uniform background over the scale of $R_s$ and $R_b$ is assumed, then the
net source counts with aperture corrections applied, $S$, can be determined by 
solving the simultaneous set of linear equations
\begin{eqnarray}
\label{eqn:getC}
C = \alpha S + b &;\quad& B = \beta S + r b,
\end{eqnarray}
where $C$ and $B$ are the total counts in $R_s$ and $R_b$, respectively, $b$
represents the background in $R_s$, and $r=A_b/A_s$.  The solution is
\begin{eqnarray*}
S = (rC-B)/(r\alpha-\beta).
\end{eqnarray*} 

In general, selecting a background aperture so that $\beta \rightarrow 0$ is
difficult, since the inner radii of such annuli could range from $\sim\!25''$
to $>\!1000''$, depending on $\theta$ and energy.  Such large background
apertures would be subject to errors due to intrinsic background variations,
diffuse source emission, and background contributions from multiple detector
chips.  We choose rather to use smaller apertures, containing $\sim\!5$--10\%
of the source flux (see Fig.~14), whose effects can be modeled
more accurately.

To apply a consistent statistical approach in determining confidence bounds for
all photometric quantities (see below), we assume that the generic photometric 
quantity $S$ (whether counts, count rate, photon flux, or energy flux) can be 
converted to counts by multiplying by appropriate generic conversion factors 
$f$ and $g$ defined below, averaged over $R_s$, $R_b$, and generalize 
equations~(\ref{eqn:getC}) to include these terms,
\begin{eqnarray}
\label{eqn:Cgeneral}
C = f S + b &;\quad& B = g S + r b.
\end{eqnarray}

For example, if $S$ represents a count rate, $f=\alpha\canon{T_s}$ and
$g=\beta\canon{T_b}$, where $\canon{T_s}$ and $\canon{T_b}$ represent average
exposure times in $R_s$ and $R_b$, respectively.  The corresponding definitions
for photon flux are $f=\alpha\canon{E_s}$ and $g=\beta\canon{E_b}$, where
$\canon{E_s}$ and $\canon{E_b}$ are the average exposure map values (in $\rm
cm^2\,s$), computed at the monochromatic effective energy of the band, in $R_s$
and $R_b$, respectively.  For energy flux, $f=\alpha/\canon{F_s}$ and
$g=\beta/\canon{F_b}$, where $\canon{F_s}$ and $\canon{F_b}$ represent average
fluxes (in $\rm erg\,cm^{-2}\,s^{-1}$) in $R_s$ and $R_b$, respectively.  The
values $\canon{F_s}$ and $\canon{F_b}$ are determined by applying quantum
efficiency and effective area corrections to individual event energies in $R_s$
and $R_b$, and computing the averages of the resulting quantities.

Finally, we relax the assumption of uniform background over $R_s$ and $R_b$ by
defining $r=A_b\canon{T_b}/A_s\canon{T_s}$ for source rate,
$r=A_b\canon{E_b}/A_s\canon{E_s}$ for photon flux, and
$r=A_b\canon{F_s}/A_s\canon{F_b}$ for energy flux. With these definitions, the
general solution for $S$ may be written as
\begin{eqnarray}
\label{eqn:Sgeneral}
S = (rC-B)/(rf-g).
\end{eqnarray} 

To determine confidence bounds for $S$, the background marginalized posterior
probability density is computed first, with the assumption that $C$ and $B$ are
Poisson-distributed random variables whose means are $\theta=fS+b$ and
$\phi=gS+rb$, respectively. The posterior probability density for $S$ may then
be written
\begin{displaymath}
p(S|CB) = \int_0^\infty db\,p(Sb|CB).
\end{displaymath}
To determine $p(Sb|CB)$, we use Bayes' Theorem to write the joint posterior
probability density for $p(\theta\phi|CB)$, taking advantage of the fact that
$R_s$ and $R_b$ are independent:
\begin{displaymath}
p(\theta\phi|CB) = \frac{p(\theta)p(C|\theta)p(\phi)p(B|\phi)}{p(CB)}.
\end{displaymath}
The likelihoods are simple Poisson probabilities,
\begin{eqnarray*}
p(C|\theta) = \frac{\theta^Ce^{-\theta}}{\Gamma(C+1)} &;\quad& p(B|\phi)
= \frac{\phi^Be^{-\phi }}{\Gamma(B+1)},
\end{eqnarray*}
and we use generalized $\gamma$-priors for $p(\theta)$ and $p(\phi)$:
\begin{eqnarray*}
p(\theta) =
\frac{\rho_S^{\pi_S}\theta^{\pi_S-1}e^{-\rho_S\theta}}{\Gamma(\pi_S)}
&;\quad& p(\phi) = \frac{\rho_B^{\pi_B}\phi^{\pi_B-1}e^{-\rho_B\phi}}{\Gamma(\pi_B)}, 
\end{eqnarray*}
where the parameters $\pi_S$, $\rho_S$, $\pi_B$, and $\rho_B$ define the
function shapes, and $P(CB)$ is determined through normalization of
$p(\theta\phi|CB)$.  Once $p(\theta\phi|CB)$ is known, $p(Sb|CB)$ may be found
from simple substitution of variables,
\begin{eqnarray*}
p(\theta\phi|CB)d\theta d\phi & = & p(\theta(S,b)\phi(S,b) |CB)\left 
  |\frac{\partial(\theta,\phi)}{\partial(S,b)}\right |dS db\\
& = & p(Sb|CB)(rf-g)dSdb.
\end{eqnarray*}

Details of the derivation may be found in \citetkas, but here we merely 
cite the final results under the additional assumption of non-informative priors 
$\pi_S=\pi_B=1$ and $\rho_S=\rho_B=0$:
\begin{eqnarray*}
p(S&&\!\!|CB)dS =dS(rf-g)\sum_{k=0}^C\sum_{j=0}^B\frac{(fS)^ke^{-fS}}{\Gamma(k+1)}\frac{(gS)^je^{-gS }}{\Gamma(j+1)}\\
&&{}\times e^{(B-j)ln(r)+ln(\Gamma(C+B-k-j+1))} \nonumber \\
&&{}\times e^{-ln(\Gamma(C-k+1))-ln(\Gamma(B-j+1))-(C+B-k-j+1)ln(1+r)}\nonumber.
\end{eqnarray*}
Because of the computationally intensive nature of this expression, $p(S|CB)dS$
is approximated with an equivalent Gaussian distribution when $C+B > 50$ counts.
The validity of this approximation is verified through simulations.  Examples
of $p(S|CB)$ for three CSC sources are shown in Figure~20.

By using the different definitions for $f$ and $g$ as described above,
probability densities for net counts, rates, and photon and energy fluxes can
then be computed.

\begin{figure*}
\epsscale{1.0}
\plotone{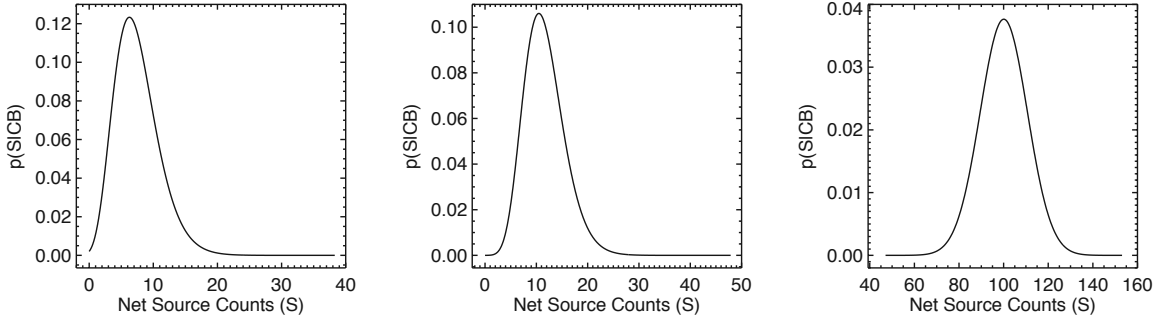}
\caption{\label{fig:pscb}
Probability distributions $p(S|CB)$ for the number of net source counts $S$ 
in the ACIS broad ({\it b\/}) band for three different CSC source observations.  
$C$ and $B$ are the total counts in the source and background region apertures, 
respectively; $\alpha$ and $\beta$ are the fractions of the source counts 
expected in each aperture; and $r$ is the ratio of the background to source 
region aperture areas. $p(S|CB)$ is computed as described in the text.  {\it 
Left:\/} $C=7$, $B=41$, $\alpha=0.83731$, $\beta=0.13587$, $r=23.41$ ({\it b\/} 
band flux significance = 1.94). {\it Center:\/} $C=11$, $B=26$, $\alpha=0.89741$, 
$\beta=0.08607$, $r=16.36$ ({\it b\/} band flux significance = 3.04). {\it 
Right:\/} $C=90$, $B=10$, $\alpha=0.89832$, $\beta=0.07743$, $r=24.02$ ({\it 
b\/} band flux significance = 9.4).}
\end{figure*}
 
Confidence bounds are determined by numerically integrating $p(S|CB)$ in
alternating steps above and below its mode until the desired confidence level is
achieved.  The values of $S$ at these points then determine the confidence
bounds. If the value of $S=0$ is reached before summation is complete, or if the
mode itself is 0, integration continues for $S$ above the mode, and the
resulting bound is considered an upper limit.

We note that in our approach, photon flux and energy flux are determined
somewhat differently.  While average exposure map values are used when computing
photon flux, in the case of energy flux the values $\canon{F_s}$ and
$\canon{F_b}$ are determined by applying quantum efficiency and effective area
corrections to individual event energies.  For sources with few counts in either
the source or background region, a single photon detected at an
uncharacteristically low or high energy (where the {\it Chandra\/} effective
area is small) can make a dominant contribution to the estimated energy flux.
In such cases, the true uncertainty will be significantly larger than our
estimated errors.  A {\it post facto\/} comparison of energy flux estimates
computed in this manner with energy fluxes calculated using an assumed canonical
power-law spectral model (see \S~3.10, below), indicates that fewer
than 1\% of ACIS broad energy band fluxes are affected by this problem.  A more
detailed analysis of the statistical accuracy of the energy flux determinations
is provided by \citetfap.

\subsubsection{Determining Flux Significance\label{sec:fluxsig}}

Significances for all aperture photometry quantities are determined directly
from the probability densities $p(S|CB)$.  Our goal is to provide a simple
statistic that is robust to calculate, easily interpretable by non-expert users,
and consistent with the classical SNR definition for high count sources.  To
this end, we compute the $\mbox{\rm FWHM}$ of $p(S|CB)$, since the latter has a
well-defined width even for low-significance sources in the catalog, as shown in
Figure~20.  If $S=0$ is reached before the half-maximum point below
the mode is found, the $\mbox{\rm HWHM}$ is computed from values above the mode
and $\mbox{\rm FWHM}$ is set equal to $2\times\mbox{\rm HWHM}$.  The $\mbox{\rm
FWHM}$ is then used to compute the ``equivalent $\sigma$'' for a Gaussian
probability density,
\begin{displaymath}
\sigma_e = \frac{\mbox{\rm FWHM}}{2\sqrt{2\ln 2}}.
\end{displaymath}
The flux significance value that is reported in the catalog for a source is
defined to be $\mbox{\rm SNR} = S/\sigma_e$, where $S$ is determined from
equation~(\ref{eqn:Sgeneral}).  This value must be at least 3.0 in at least one
energy band for an observation of a source to be included in the first release
of the catalog.

The flux significance threshold that we use imposes a conservative limit on
sources included in the CSC, which we deem necessary to reduce the number of
spurious sources at low count levels to an acceptable value.  Comparing our
results to those of other large {\it Chandra\/} surveys whose source lists are
derived from {\tt wavdetect}, but whose detection procedures differ, is useful.
In Figure~21, we compare the distribution of net counts for CSC
sources detected in the ACIS broad ($0.5$--$7.0\rm\,keV$) energy band with
distributions of similar quantities for four other {\it Chandra\/} catalogs
derived from a range of ACIS exposures comparable to those in the CSC: AEGIS-X
\citep[][$0.5$--$7.0\rm\,keV$]{lai09}, the Galactic Center catalog
\citep[][$0.5$--$8.0\rm\,keV$]{mun09}, C-COSMOS
\citep[][$0.5$--$7.0\rm\,keV$]{elv09}, and ChaMP
\citep[][$0.5$--$8.0\rm\,keV$]{kim07}.  We note that while these other catalogs
do include sources with fewer net counts than the CSC, the additions are in
general not large, comprising $\sim\!5\%$, $\sim\!9\%$, $\sim\!9\%$, and
$\sim\!24\%$, for AEGIS-X, the Galactic Center catalog, C-COSMOS, and ChaMP,
respectively.  We attribute the larger percentage in ChaMP to the restricted
fields-of-view and the careful manual screening of source detections used when
constructing that catalog.  The CSC appears to fare worse in comparison to the
XBootes survey \citep[][$0.5$--$7.0\rm\,keV$]{ken05}, most of whose sources have
fewer than 10 net counts.  However, the XBootes survey is composed of many
$5\rm\,ks$ non-overlapping observations for which the very low ACIS background
allows a lower count threshold. In contrast, the CSC is constructed from
observations comprising a wide range of exposures, $\sim\!70\%$ of which are
greater than $5\rm\,ks$ and $\sim\!10\%$ of which are greater than $50\rm\,ks$.
Finally, as mentioned in \S~3.4, $\sim\!1/3$ of all sources
detected by {\tt wavdetect} in the ACIS broad energy band fall below the flux
significance threshold.  However, we expect that a substantial fraction of these
sources are spurious.

\vskip 40pt

\subsubsection{Combining Aperture Photometry from Multiple Observations}

\begin{figure}
\epsscale{1.0}
\plotone{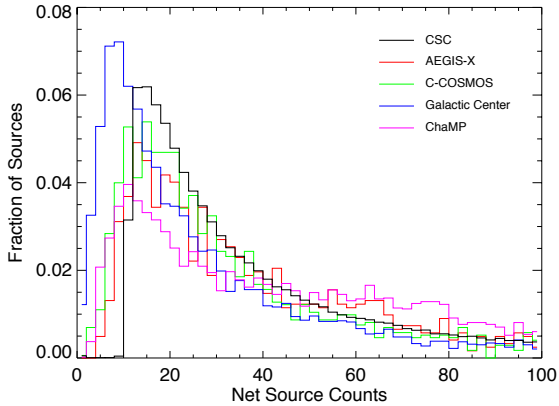}
\caption{\label{fig:srccnts}
Comparison of the distribution of net counts for CSC sources detected in the 
ACIS broad energy band with distributions of similar quantities for four other
{\it Chandra\/} catalogs derived from a range of ACIS exposures comparable to
those in the CSC.}
\end{figure}
 
Ideally, one should compute aperture photometry quantities for combined
observations by computing a joint probability density, using $p(S|CB)$ from one
observation as the prior for the next.  However, this approach is difficult to
implement and computationally expensive, especially when probability densities
from individual observations do not overlap significantly.  We have therefore
chosen simply to combine aperture data from various observations and compute a
single $p(S|CB)$ from those data.  We compute
\begin{eqnarray*}
\sum_i C_i = S \sum_i f_i + \sum_i b_i &;& \sum_i B_i = S \sum_i g_i + \sum_i r_i b_i.
\end{eqnarray*}
To cast these in the same form as equations~(\ref{eqn:Cgeneral}), we define
$r^\prime=\sum_i r_i b_i/ \sum_i b_i$, where the $b_i$ are determined from the
solutions to equations~(\ref{eqn:Cgeneral}) for individual observations.  One can
then write
\begin{eqnarray*}
\sum_i C_i = S \sum_i f_i + \sum_i b_i &;& \sum_i B_i = S \sum_i g_i + r^\prime \sum_i b_i.
\end{eqnarray*}
which are identical in form to equations~(\ref{eqn:Cgeneral}).  The combined
aperture photometry quantities and bounds can then be determined as described
earlier.

\subsection{Computing Limiting Sensitivity}

For the purposes of the catalog, ``limiting sensitivity'' is defined to be the
flux of a point source that meets but does not exceed the flux significance
threshold for inclusion in the catalog. Limiting sensitivity is a function of
source position, background, and the algorithm used to calculate flux and flux
significance. At any point within the field of view of an observation, the
limiting sensitivity can be used as a simple X-ray flux limit for individual
sources detected at other wavelengths. For the catalog, a full-field sensitivity
map is provided for each observation and energy band as a file-based data
product (see Table~3). These data are also required to
calculate sky coverage histograms (solid angle surveyed as a function of
limiting flux), which are themselves needed to calculate luminosity functions
and source surface brightness versus number density relationship.

As described above, the flux significance of a catalog source is defined to be
the ratio of the source flux to the equivalent $\sigma$ determined from the
width of the flux's posterior probability density.  There is no equivalent
quantity for sensitivity, and for simplicity and ease of computation, we use a
technique similar to that developed by \citet{mac82} for the {\it Einstein\/}
Observatory Medium Sensitivity Survey, namely, we approximate significance using
the aperture photometry relations of \S~3.7, under the
assumption of Gaussian statistics, and use the model background maps, randomized
to provide statistics appropriate to the observation in question, to determine
aperture counts.

Recall from equation~(\ref{eqn:Sgeneral}) that the flux may be written as 
\begin{equation}
\label{eqn:flux}
S = (rC-B)/(rf-g).
\end{equation}
Since $C$ and $B$ are independent random variables, the variance on $S$ may be
written
\begin{eqnarray*}
\sigma_S^2 &=& \frac{r^2\sigma_C^2 + \sigma_B^2}{(rf-g)^2} =  \frac{r^2C + B}{(rf-g)^2},
\end{eqnarray*}
assuming Gaussian statistics.  The significance, $S/\sigma_S$ may then be
written as
\begin{equation}
\label{eqn:snr}
S/\sigma_S = \frac{(rC-B)}{\sqrt{r^2C + B}}.
\end{equation}

The limiting sensitivity is found by determining the minimum number of counts
$C_{\rm min}$ in the source aperture that yields the flux significance threshold
$\mathit{SNR}_{\rm min}$ in equation~(\ref{eqn:snr}),
\begin{displaymath}
\mathit{SNR}_{\rm min} = \frac{(rC_{\rm min}-B)}{\sqrt{r^2C_{\rm min} + B}},
\end{displaymath}
whose solution is 
\begin{displaymath}
rC_{\rm min} = B + \frac{r \mathit{SNR}_{\rm min}^2}{2}
  \left\{1+\sqrt{1+\frac{4B}{r \mathit{SNR}_{\rm
        min}^2}\left(1+\frac{1}{r}\right)}\right\}, 
\end{displaymath}
and the limiting sensitivity for that aperture is then given by
equation~(\ref{eqn:flux}),
\begin{eqnarray}
\!\!\!\!\!\!\!\!\!\!\!S_{min} &=& (rC_{min}-B)/(rf-g) \nonumber\\
&=& \frac{r \mathit{SNR}_{min}^2}{2} \left\{1+\sqrt{1+\frac{4B}{r
      \mathit{SNR}_{min}^2}\left(1+\frac{1}{r}\right)}\right\} \nonumber\\
&&{}\times(rf-g)^{-1} \nonumber\\
\label{eqn:ls}
&=& \frac{\mathit{SNR}_{min}^2}{2f} \left\{1+\sqrt{1+\frac{4B}{r
      \mathit{SNR}_{min}^2}\left(1+\frac{1}{r}\right)}\right\}, 
\end{eqnarray}
where we have approximated 
\begin{displaymath}
(rf-g)^{-1} \approx (rf)^{-1}\left\{1+\frac{g}{rf}\right\} \approx (rf)^{-1}.
\end{displaymath}

Since the limiting sensitivity maps are computed from background maps with no
real sources, information about real source apertures is unavailable.  Rather,
for each element in the map, circular source and annular background apertures
appropriate to the 90\% ECF source aperture at that location are constructed,
and used to determine $B$, $r$, and $f$ for use in equation~(\ref{eqn:ls}).

The assumption of Gaussian statistics, and the subsequent simplification in the
algorithm, is made of necessity, since limiting sensitivity must be computed not
for each source but for each pixel in each of 5 energy band images.  We have,
however, verified the performance of the algorithm by comparing detected source
fluxes with values of limiting sensitivity at the source locations, for
thousands of catalog sources in all energy bands \citepfap, and find good
agreement.

\subsection{Spectral Model Fits\label{sec:spmdl}}

For observations of sources with at least 150 net counts in the energy band
$0.5$--$7\rm\,keV$ obtained using the ACIS detector, we further characterize the
intrinsic source properties by attempting to fit the observed counts spectrum
with both an absorbed black-body spectral model and an absorbed power-law
spectral model.  These two models represent basic spectral shapes of thermal and
non-thermal X-ray emission.

The standard forward fitting method used in X-ray spectral analysis computes the
predicted counts produced by the spectral model with the observed counts in the
detector channel space, and iteratively refines the model parameters to improve
the quality of the fit.

Instrumental response functions \citep{jed01a} define the mappings between
physical (source) space and detector space. \citet{geo07} describe two of these
calibration files, the detector redistribution matrix file (RMF) and the
ancillary response file (ARF)\null. The former specifies the energy dispersion
relation $R(E^\prime, \hat{p}^\prime; E, \hat{p}, t)$ that defines the
probability that a photon of actual energy $E$, location $\hat{p}$, and arrival
time $t$ will be observed with an apparent energy $E^\prime$ and location
$\hat{p}^\prime$, while the instrumental effective area $A(\hat{p}^\prime; E,
\hat{p}, t)$ is recorded in the latter. The final dispersion relation is the
photon spatial dispersion $P(\hat{p}^\prime; E, \hat{p}, t)$ transfer function
due to the instrumental point spread function.

With these definitions, the model $M(E^\prime, \hat{p}^\prime, t)$ that
describes the expected distribution of counts arriving at the detector is then
\begin{eqnarray}
\label{eqn:spmod}
&&\!\!M(E^\prime, \hat{p}^\prime, t) = \\
&&\int dE \, d\hat{p} \, R(E^\prime; E,
\hat{p}, t) \, P(\hat{p}^\prime; E, \hat{p}, t) \, A(E, \hat{p}^\prime, t) \, 
S(E,\hat{p}, t),\nonumber
\end{eqnarray}
where $S(E, \hat{p}, t)$ is the physical model that defines the physical energy
spectrum, spatial morphology, and temporal variability of the source.

We follow standard practice by ignoring the dependency on photon arrival time,
and instead consider only the total number of photons that arrived during the
observation in the forward fitting process.  The source position and shape are
taken as known, and we assume that the source photons are collected from the
detector area containing an entire source region of interest.  The latter
assumption is valid provided that sources are spatially separated on scales of
order the size of the PSF or larger.  In crowded fields, or for sources that
have a complex diffuse structure, the contribution from the other sources are
important.  With the assumptions listed above, equation~(\ref{eqn:spmod})
reduces to
\begin{displaymath}
M(E^\prime) = \int dE \, R(E^\prime; E) \, A(E) \, S(E),
\end{displaymath}
where the source emitted spectrum $S(E)$ depends on the source physics.  The
forward fitting procedure solves for the best fit parameters for $S(E)$,
assuming a pre-defined fit statistic.  Since spectral fitting is only performed
for sources with a minimum of 150 net counts, a $\chi^2$ fit statistic is used,
but note that this assumes a Gaussian distribution for the source counts.

For {\it all\/} sources observed using the ACIS detector (i.e., not just those
with at least 150 net counts in the broad energy band), the catalog processing
pipelines extract the observed energy spectra of the photons included in the
source and background regions of each detected source and store these in a
standard format \citep[PHA file;][]{arn09}.  An appropriate associated ARF and
RMF are computed by weighting the instrumental responses based on the history of
how the source and background regions move over the surface of the detector due
to the spacecraft dither motion.  The extracted spectra, and associated ARF and
RMF are stored as file-based data products (see Table~3) and
can be retrieved by the user for further analysis such as low-count spectral
fitting or spectral stacking.

To fit the background subtracted data, each PHA spectrum is grouped to a minimum
of 16 counts per channel bin, and the source model parameters are varied to
minimize the $\chi^2$ statistic (assuming data variance, $\sigma_i^2 = N_{i,S} +
(A_S/A_B)^2 N_{i,B}$).  Two models are applied to the data in order to evaluate
source properties: (1)~an absorbed blackbody model $f(E) = \exp^{-N_{\rm H}
  \sigma_E}\, A (E^2/(\exp^{E/kT}-1))$; and (2)~an absorbed power law model
$f(E) = \exp^{-N_{\rm H}\sigma_E}\, A\, E^{-\Gamma}$.  In these models, $N_{\rm
  H}$ is the equivalent Hydrogen column density, $\sigma_E$ is the
photo-electric cross-section based on \cite{bal92} and metal abundances from
\cite{and89}, $A$ is the model normalization at $E=1\,\rm keV$, $kT$ is the
blackbody temperature, and $\Gamma$ is the power-law photon index.  Forward
fitting is performed using the {\it Sherpa\/} fitting engine \citep{fre01,
  doe07}.  {\it Sherpa\/} finds the best fit model parameters and calculates
two-sided confidence intervals for each significant parameter.  The model flux
for the best fit parameters over the energy range $0.5$--$7\,\rm keV$ is also
calculated.

The $68\% \, (1\,\sigma)$\, confidence limits for each parameter are calculated
using the ``projection'' method in {\it Sherpa\/}.  This method finds the
two-sided confidence bounds independently for each parameter.  The algorithm
assumes that the current model has been fitted, and that all of the parameters
are at the values corresponding to a best fit which is at the minimum of the fit
statistic ($\chi^2_{\rm min}$).  For each parameter of interest, the search for
the lower or upper bound starts at the best fit value, which is then varied
along the parameter axis.  At each new value, the parameter of interest is
frozen and a new best fit model is determined by minimizing $\chi^2$ over the
remaining thawed parameters.  The new $\chi^2$ statistic, $\chi^2_{\rm new}$, is
determined and the difference between the new and the minimum statistics,
$\Delta \chi^2 = \chi^2_{\rm new} - \chi^2_{\rm min}$, is calculated.  A change
in $\Delta \chi^2$ equal to 1 corresponds to a $68\%$ confidence bound
\citep{avn76}, so the parameter of interest is varied until this value of
$\Delta \chi^2$ is obtained.

We note here that energy-dependent aperture corrections are not applied when
performing the spectral model fits.  Since the {\it Chandra\/} PSF is somewhat
more extended at higher energies, the lack of correction has the effect of
slightly softening the calculated spectral slope.  The correction in $\Gamma$ is
approximately 0.03--0.05 for power-law spectra with a wide range of spectral
indices.  For sources included in release~1 of the CSC for which spectral fits
have been performed, the error in $\Gamma$ introduced by not applying
energy-dependent aperture corrections is roughly six times smaller than the
median computed $1\,\sigma$ confidence limits.  About 2.5\% of sources with
spectral fits have computed confidence limits $\lesssim 0.05$, and these cases
appropriate caution should be exercised when using the spectral fit properties.

For most source properties, values recorded in the Master Sources Table are
computed by combining the relevant data from the set of observations in which
the source is detected.  However, for simplicity, the spectral model fit
parameter values recorded in the Master Sources Table are taken directly from
the single observation of the source that has the highest significance,
equation~(\ref{eqn:snr}).  In this case, data from multiple observations are not
combined to compute the Master Source Table spectral fit properties.

\subsection{Spectral Model Energy Fluxes\label{sec:mef}}

Spectral model fits are not performed for sources with $< 150$ net counts.
However, for all sources we estimate energy fluxes using canonical absorbed
power-law and black-body spectral models.

For a canonical source model $S(E)$ whose integral over the energy band is
$S^\prime$, an corresponding band count rate, $C^\prime$ in $\rm
counts\,s^{-1}$, can be computed from the effective area calibration, $A(E)$,
and the RMF, $R(E^\prime, E)$.  The count rate is $\int dE \, R(E^\prime; E) \,
A(E) \, S(E)$, where the integral is performed over the energy band.  For HRC
observations, a diagonal RMF is assumed.  The actual flux of a source can be
estimated from $S^\prime$ by scaling the latter by the ratio of the measured and
modeled source aperture count rates in the energy band.  Since the only free
parameter in this case is the normalization of the model, the calculation can be
performed for sources with too few counts for a reliable spectral fit.

The canonical power-law spectral model has a fixed photon index $\Gamma = 1.7$,
which falls in the range of values ($\Gamma\sim\!1.5$--2.5) that are typical of
AGN spectra \citep{ish10}.  The value chosen matches the photon index used to
convert count rates to energy fluxes in the second XMM-Newton serendipitous
source catalog \citep[2XMM;][]{wat08}, to simplify comparison of CSC and 2XMM
source fluxes.  Since we anticipate that the majority of sources with spectra
that are best fit by a power-law model are AGN, we fix the total neutral
Hydrogen absorbing column $N_{\rm H}$ equal to the Galactic column, $N_{\rm
H}({\rm Gal})$, under the assumption that this represents a lower limit to 
the true column density.

The canonical black-body spectral model has a fixed temperature $kT = 1.0\rm\,
keV$ and total neutral Hydrogen column density $N_{\rm H} = 3\times 10^{20}\rm\,
cm^{-2}$.  The latter value matches the median column density identified by
\citet{sax03}, and also corresponds to the typical column density found within
$1\rm\,kpc$ of the Sun \citep{lis83}.  As discussed by \citet{mcc10}, the choice
of black-body temperature is a compromise between the possible ranges of values
for different classes of thermal X-ray emitters.  Sources for which a thermal
model best represents the data will likely lie in our galaxy, and so in this
case setting $N_{\rm H} = N_{\rm H}({\rm Gal})$ would overestimate the total
neutral Hydrogen absorbing column.

Similar to spectral model fits, master source spectral model energy fluxes are
taken directly from the single observation of the source that has the highest
significance.

\subsection{Spectral Hardness Ratios}

While the spectral model fits described in \S~3.9 provide detailed
information about a source's spectral properties, only about 10\% of the source
observations included in the CSC have sufficient net counts to perform the
fitting process.  As an aid to characterizing the spectral properties of the
remaining catalog sources, hardness ratios are computed between the hard,
medium, and soft energy bands for all sources observed with the ACIS detector.

The spectral hardness ratio for the pair of energy bands $x$ and $y$ is defined
as
\begin{equation}
\label{eqn:hr}
\mathcal{HR}_{xy} = \frac{F_x - F_y}{F_b},
\end{equation}
where $F_x$ and $F_y$ are the photon fluxes measured in the energy bands $x$ and
$y$ respectively ($x$ is always the higher-energy band of the pair), and $F_b$
is the photon flux in the ACIS broad energy band, $F_b = F_h+F_m+F_s$.

A catalog source may be readily detected in one or more energy bands, but
remain undetected or include very few total counts in other bands.  Since
hardness ratios are cross-band measures, a technique that applies rigorous
statistical methods in the Poisson regime is required to compute these values
and their associated confidence limits robustly.  The hardness ratios included
in the CSC are computed using a Bayesian approach developed by \citet{par06},
which should be consulted for a detailed description of the algorithm.

To ensure that the Poisson errors are propagated correctly, the conversion
between counts and photon flux for each energy band is modeled as a linear
process, with a scale factor that is determined from the effective area of the
telescope/instrument combination computed at the monochromatic effective energy
of the band.  This implies that the photon fluxes in equation~(\ref{eqn:hr}) may
not match exactly the aperture photometry fluxes derived in \S~3.7.

Specifically, we model the observed total and background counts, $C_x$ and
$B_x$, in the hard, medium, and soft ACIS energy bands as
\begin{eqnarray*}
C_x &\!\sim\!& {\rm Poisson}[e_x(\lambda_x+\xi_x)], \\
B_x &\!\sim\!& {\rm Poisson}[r e_x\xi_x],
\end{eqnarray*}
where $x$ represents the energy band (one of {\it h\/}, {\it m\/}, or {\it
  s\/}); $\lambda_x$ and $\xi_x$ are the expected source and background counts
intensities, respectively; $e_x$ are the conversion factors that scale counts to
photon fluxes; and $r$ is the ratio of the background aperture area to the
source aperture area.

With these definitions, the spectral hardness ratio for the pair of energy bands
$x$ and $y$ is determined by computing the expectation value
\begin{displaymath}
\mathcal{HR}_{xy} = \frac{\lambda_x-\lambda_y}{\lambda_h+\lambda_m+\lambda_s}.
\end{displaymath}

\begin{deluxetable*}{ccl}
\tablecolumns{3}
\tablewidth{0pt}
\tablecaption{Intra-Observation Variability Indices\label{tab:intravar}}
\tablehead{
\colhead{Variability} & \colhead{Condition\tablenotemark{a}} & \colhead{Meaning} \\
\colhead{Index}
}
\startdata
0	& $p_{\rm GL} \leq 0.5$							& Definitely not variable \\
1	& $0.5 < p_{\rm GL} < 0.667$ AND $f_3 > 0.997$ AND $f_5 = 1.0$		& Not considered variable \\
2	& $0.667 \leq p_{\rm GL} < 0.9$ AND $f_3 > 0.997$ AND $f_5 = 1.0$		& Probably not variable \\
3	& $0.5 \leq p_{\rm GL} < 0.6$ AND ($f_3 \leq 0.997$ OR $f_5 < 1.0$)	& May be variable \\
4	& $0.6 \leq p_{\rm GL} < 0.667$ AND ($f_3 \leq 0.997$ OR $f_5 < 1.0$)	& Likely to be variable \\
5	& $0.667 \leq p_{\rm GL} < 0.9$ AND ($f_3 \leq 0.997$ OR $f_5 < 1.0$)	& Considered variable \\
6	& $0.9 \leq p_{\rm GL}$ AND $O < 2.0$					& Definitely variable \\
7	& $2.0 \leq O < 4.0$							& Definitely variable \\
8	& $4.0 \leq O < 10.0$							& Definitely variable \\
9	& $10.0 \leq O < 30.0$							& Definitely variable \\
10	& $30.0 \leq O$								& Definitely variable \\
\enddata
\tablenotetext{a}{$p_{\rm GL}$ is the Gregory-Loredo variability probability,
equation~(22); $f_3$ and $f_5$ are the fractions of the light
curve that fall within $3\,\sigma$ and $5\,\sigma$ of the average rate,
respectively; and $O = \sum_{j=2}^{m_{\rm max}} O_j$ is the sum of the
odds-ratios, equation~(21), for two or more bins.} 
\end{deluxetable*}

Following the lead of \citet{par06}, the joint posterior probability
distribution can be written as
\begin{eqnarray*}
p(\lambda_s, \lambda_m, \lambda_h&&\!\!|C_s, C_m, C_h, B_s, B_m, B_h) = \\
&&p(\lambda_s|C_s, B_s) p(\lambda_m|C_m, B_m) p(\lambda_h|C_h, B_h),
\end{eqnarray*}
where we have made use of the fact that the $\lambda_x$ are independent.
Marginalizing over nuisance variables yields the posterior distribution for the
hardness ratios \citep[equivalent to equation~\hbox{[14]} of][]{par06}:
\begin{eqnarray*}
p(\mathcal{HR}_{xy}&&\!\!|C_s, C_m, C_h, B_s, B_m, B_h)\, d\mathcal{HR}_{xy} = \\
&&d\mathcal{HR}_{xy}\,\int_{\psi,
  \omega}\left[d\psi\,d\omega\,\left(\frac{2}{\omega}\right)\right.\\
&&\left.{}\times p(\mathcal{HR}_{xy}, \psi,
\omega |C_s, C_m, C_h, B_s, B_m, B_h)\vphantom{\left(\frac{2}{\omega}\right)}\right],
\end{eqnarray*}
where $\psi=\lambda_x+\lambda_y$ and 
$\omega=\lambda_s+\lambda_m+\lambda_h$.

The spectral hardness ratios that are included in the CSC are determined
separately for each observation in which a source is detected, and also from the
ensemble of all observations of the source.  The former quantities are recorded
in the Source Observations Table, while the latter are recorded in the Master
Sources Table.  The prior probability distributions used to derive the Bayesian
posterior probabilities are computed differently in these two cases.

For a single observation, non-informative conjugate $\gamma$-prior distributions
\citep{van01} are used for the source and background intensities.  These
distributions ensure that the posterior probabilities conjugate to the expected
Poisson distributions of counts with no other prior information.  When multiple
observations are combined, the ensemble hardness ratios are computed by stepping
through all of the observations of a source in order of increasing net
broad-band source counts.  The posterior probability distribution computed from
each observation is used as the prior probability distribution for the
subsequent step.  If the propagated prior probability distribution is not
consistent with the observed counts in any step, then a conjugate $\gamma$-prior
is used instead, and a catalog flag is set to indicate that the source spectrum
is variable.

\subsection{Estimating Source Variability}

The CSC includes estimates of the probability that the flux from a source is
temporally variable both within a single observation and between two or more
observations in which the source was detected.  These estimates are
distinguished not only by the fact that they measure variability on different
time scales, but also because their definitions differ fundamentally.  Within an
observation, we measure the probability that the source flux is not consistent
with a constant level during a (largely) continuous observation, which is
equivalent to estimating the probability that the source is variable, and is a
positive statement with respect to variability.  The inter-observation
variability estimates measure the probability that the average flux levels
during the different observations are consistent with a uniform source
intensity.  This provides only a lower limit to the probability of the source
being variable, since we have no information about the source's behavior during
the gaps between the observations, which are often widely separated.

The intent of the various variability measures included in the CSC is to provide
users a means to easily select {\it potentially\/} variable sources.  The
individual source light curves or event lists should be assessed to reveal the
true nature of the source's temporal characteristics.  Moderately intense
background flares that are not rejected as part of the enhanced background event
screening (see \S~3.2) may cause sources to be incorrectly
identified as variable.  This possibility can be evaluated by comparing the
structure of the source and background light curves.

\subsubsection{Intra-Observation Variability\label{sec:intravar}}

The probability that a source is variable is estimated separately in each energy
band using the Gregory-Loredo and Kolmogorov-Smirnov (K-S) algorithms, and
Kuiper's variation on the latter.  A brief description of each of the three
algorithms is provided below.  Gregory-Loredo probabilities are used to
construct intra-observation variability indices that provide a shorthand measure
of variability.

All three algorithms directly use the photon event arrival times to compute the
variability probabilities, and apply corrections for variations of the geometric
areas of the source and background region apertures due to the spacecraft
dither-induced motion during the observation.  The latter corrections are
necessary since a source region that is moving across the edge of the detector
or over a bad detector region might otherwise be erroneously classified as
variable.  Optimal resolution light curves are generated as by-products of the
Gregory-Loredo test, and their power spectra are evaluated for the presence of
the fundamental spacecraft pitch and yaw dither frequencies or associated beat
frequencies.  If there is a peak in the power spectrum at one of these
frequencies that is at least $5\times$ the RMS value, then a catalog warning
flag is set for the source observation to indicate that the intra-observation
variability properties are unreliable.

The K-S test \citep{mas51} is a familiar and well established robust test for
comparing two distributions that are a function of a single variable. In the
simplest case we compare the cumulative sum of photon events, as a function of
time, against a linearly increasing function that represents a constant flux.
This null-hypothesis function is modified as necessary to account for data gaps
and variations in effective area.

For an observation with $N$ events, let $S_N(t)$ be the cumulative sum of
detected events as a function of time $t$, and $P(t)$ the cumulative function
that represents a constant flux.  The K-S statistic $D_N$ is defined as
\begin{displaymath}
D_N = \sup_t{|S_N(t) - P(t)|}.
\end{displaymath}
The K-S derived probability that the two distributions $S_N(t)$ and $P(t)$ do
not belong to the same population, and therefore that the source is variable, is
given by
\begin{equation}
\label{eqn:pvar}
p_{\rm var} = Q_{\rm KS}(\sqrt{N} D_N)
\end{equation}
where
\begin{displaymath}
Q_{\rm KS}(\lambda) = 2 \sum_{j=1}^\infty{(-1)^{j-1}\, e^{-2j^2\lambda^2}}.
\end{displaymath}
Equation~(\ref{eqn:pvar}) is strictly valid only in the asymptotic limit as $N
\rightarrow \infty$.  In practice $N \gtrsim 20$ is ``large enough,''
especially if conservative significance levels $\lesssim 0.01$ are required
\citep[e.g.,][]{numrec}. 

\citet{kui62} proposed a variation on the K-S test that involves replacing the
expression for $D_N$ by the difference between the largest positive and negative
deviations,
\begin{displaymath}
D_N = \sup_t[S_N(t) - P(t)] - \inf_t[S_N(t) - P(t)].
\end{displaymath}
Folding this expression into equation~(\ref{eqn:pvar}) yields the Kuiper derived
probability that the source is variable.  While the K-S test is primarily
sensitive to differences between the median values of the cumulative
distribution functions, the Kuiper test statistic is as sensitive to differences
in the tails of the distributions.  In many cases, this makes the Kuiper test a
more robust variation of the traditional K-S test for evaluating the probability
that a source is variable.

The Gregory-Loredo test \citep{gre92} is based on a Bayesian approach to
detecting variability.  The method works very well on photon event data and is
capable of dealing with data gaps.  We have incorporated the capability to
include temporal variations in effective area.  Although the algorithm was
developed for detecting periodic signals, it is a perfectly suitable method for
detecting random variability by forcing the period to equal the length of the
observation.

Briefly, the Gregory-Loredo algorithm bins the $N$ observed photon events into a
series of histograms containing $m$ bins, where $m$ runs from 2 to $m_{\rm
  max}$.  If the observed distribution of events across the $m$ histogram bins
is $n_1, n_2, \ldots, n_m$, then the probability that this distribution came
about by chance can be determined from the ratio of the multiplicity of the
distribution, $N!/(n_1! \cdot n_2! \cdots n_m!)$, to the total number, $m^N$, of
possible distributions.  The inverse of this ratio is a measure of the
significance of the distribution.  Following \citet{gre92}, we calculate an odds
ratio $O_m$ for $m$ bins versus a flat light curve as
\begin{equation}
\label{eqn:glodds}
O_m = T\; \frac{N!\; (m-1)!}{ (N+m-1)!}\ \frac{S_m\; m^N}{W_m},
\end{equation}
where we have rewritten the multiplicity of the distribution, $W_m$, as  
\begin{displaymath}
W_m = \frac{N!}{\prod_{j=1}^{m}{n_j!}}.
\end{displaymath}
Data gaps are accounted for through the binning factor, $S_m$, which is
\citep[Appendix~B of][]{gre92}
\begin{displaymath}
S_m = \prod_{j=1}^{m}{s_j}^{-n_j},
\end{displaymath}
where
\begin{displaymath}
s_j = \frac{t_j}{T/m},
\end{displaymath}
$t_j$ is the amount of good exposure time in bin $j$, and $T$ is the total good
exposure time for the observation.  The odds are summed over all values of $m
\ge 2$ to determine the odds that the source is time-variable.  $m_{\rm max}$ is
chosen for each case in such a way that the odds ratios corresponding to higher
values of $m$ contribute negligibly to the total.  The probability, $p_m$, of a
particular binning, $m$, is simply
\begin{displaymath}
p_m = O_m / \sum_{j=1}^{m_{\rm max}} O_j.
\end{displaymath}
Summing over bins $m \ge 2$ corresponding to a non-constant source flux 
yields the Gregory-Loredo variability probability
\begin{eqnarray}
\label{eqn:glprob}
p_{\rm GL} &=& \sum_{j=2}^{m_{\rm max}} p_j \\
&=& \frac{O}{1+O},
\end{eqnarray}
where $O = \sum_{j=2}^{m_{\rm max}} O_j$, and we have made use of the fact that
$O_1 = 1$.

\begin{deluxetable}{ccccc}
\tablecolumns{5}
\tablewidth{0pt}
\tablecaption{Inter-Observation Variability Indices\label{tab:intervar}}
\tablehead{
\colhead{Variability} & \multicolumn{4}{c}{Reduced $\chi^2$} \\
\colhead{Index} & \multicolumn{2}{c}{2 Observations} & \multicolumn{2}{c}{$>2$ Observations}\\
}
\startdata
0 &             &  $<0.4$ &              & $<0.8$  \\
3 &  $\geq 0.4$ &  $<0.7$ & $\geq  0.8$  & $<1.0$  \\
4 &  $\geq 0.7$ &  $<1.0$ & $\geq  1.0$  & $<1.15$ \\
5 &  $\geq 1.0$ &  $<2.7$ & $\geq  1.15$ & $<2.1$  \\
6 &  $\geq 2.7$ &  $<7.0$ & $\geq  2.1$  & $<3.8$  \\
7 &  $\geq 7.0$ & $<12.0$ & $\geq  3.8$  & $<5.5$  \\
8 & $\geq 12.0$ &         & $\geq  5.5$  &         \\
\enddata
\end{deluxetable}

The Gregory-Loredo algorithm bins the events into a series of light curves of
varying resolution, corresponding to the number of bins, $m$, in the range 2 to
$m_{\rm max}$.  Using the definitions above, the bins that comprise the
normalized light curve, $h_m$, associated with a specific value of $m$ are
\begin{displaymath}
h_{j,m} = \frac{n_j}{s_j N},
\end{displaymath}
and the corresponding standard deviations derived from the posterior
distribution are
\begin{displaymath}
\sigma_{j,m} = \frac{1}{s_j} \sqrt{\frac{s_j h_{j,m} (1 - s_j h_{j,m})}{N+m+1}}.
\end{displaymath}

As described by \citet{gre92}, an optimal resolution, light curve, $h$, can be
obtained by combining the individual light curves, $h_m$, weighted by the
probabilities, $p_m$:
\begin{displaymath}
h = (1 - p_{\rm GL})h_1 + \sum_{j=2}^{m_{\rm max}}p_mh_m.
\end{displaymath}
The optimal resolution light curve computed from the events included in the
source region aperture for each source is recorded as a file-based data product
(see Table~3) that is accessible through the catalog.  As well
as the light curve, $h$, this data product includes the uncertainty, $\sigma$,
and upper and lower confidence intervals, $h - 3\,\sigma$, and $h + 3\,\sigma$,
respectively.  To allow the users to verify the significance of features that
may be present in the light curve, the file also includes the corresponding
quantities derived from the events extracted from the background region
aperture, using the same binning.

Careful judgement should be applied when assessing the reliability of source
variability indicators using the source and background light curves.  Since the
background region aperture may contain up to $\sim\!10\%$ of the source flux
(see \S~3.4.1), the background and source light curves may appear
similar for very bright sources.  The PSF wings of unrelated but nearby strongly
variable sources may contaminate both the source and background region apertures
of the source being investigated.  An observation may have experienced
background variations intense enough to be noticeable when compared to the
target source's count rate, but not strong enough to have been removed by
background screening during observation recalibration.  In the first case the
source is truly variable, but this is not necessarily so in the latter two
examples.  A helpful, though not definitive, test is to scale the amplitude of
the flux variations in the source and background region apertures by their
respective areas (recorded in the FITS keyword {\tt APERTURE}).  If the
variations of the two scaled amplitudes are similar, then there is a good chance
that a background problem is responsible.  If the source region scaled amplitude
is considerably larger than the background region scaled amplitude and the
source is strong, then one is likely to have a truly variable source.

The Gregory-Loredo test appears to provide a more uniform and reliable measure
of variability than either the K-S or Kuiper tests, although the Gregory-Loredo
algorithm is more ``conservative'' than the other tests.  In cases where the K-S
and/or Kuiper tests detect variability, but the Gregory-Loredo test does not,
close inspection of the light curve often, but not always, demonstrated that
the level of variability does not exceed the $3\,\sigma$ bounds on the light
curve.  In cases where there are considerable data gaps, Gregory-Loredo is not
always be able to detect variability on time scales comparable to those gaps.
 
To provide the user with a short-hand measure of variability that allows
selection of sources on different degrees of confidence, the CSC includes a set
of integer ``variability index'' values in the range $[0, 10]$.  These indices
are based on a combination of the Gregory-Loredo probability, $p_{\rm GL}$, the
logarithm of the odds ratio, $O$, and a secondary criterion that addresses the
overall deviation of the light curve from the mean value.  The latter criterion
is based on the parameters $f_3$ and $f_5$, which are the fractions of the light
curve that falls within $3\,\sigma$ and $5\,\sigma$ of the average rate,
respectively.  Table~6 defines the mapping of the test
parameters to variability index values.

\subsubsection{Inter-Observation Variability}

Inter-observation variability is based on comparison of source region aperture
photon fluxes, and their confidence intervals, from multiple observations in
which the source is detected.  The catalog provides a probability that the data
are not consistent with a constant-flux source, as well as an inter-observation
variability index that is similar to the index defined for intra-observation
variability.  These measures of variability are assessed for each spectral
energy band independently, and consequently no cross-instrument comparison is
performed.  In the first release of the CSC, observations that cover the same
region of the sky, but in which the source is not detected, are {\it not\/}
considered when computing inter-observation variability.  These observations
should enter into the variability assessment as flux upper limits, since they
could conceivably be inconsistent with a constant source flux.  A future release
of the catalog will address this limitation.

As mentioned above, the inter-observation variability probability must be
interpreted differently from the intra-observation variability probability.
Whereas the light curve can be used to declare a source to be variable or 
non-variable within the time range of a single observation, one can never
conclude that a source does not vary between multiple observations.  If
inter-observation variability is detected, then the source is definitely
variable; however the converse is not true.

\begin{deluxetable*}{ll}
\tablecolumns{2}
\tabletypesize{\small}
\tablecaption{Source Codes\label{tab:codes}}
\tablewidth{0pt}
\tablehead{
\colhead{Property} & \colhead{Bit Encoding\tablenotemark{a}}
}
\startdata
conf\_code &  \phn 0: Source is not confused \\
& \phn 1: Multiple source in source region \\
& \phn 2: Source region overlaps another source region \\
& \phn 4: Source region overlaps another background region \\
& \phn 8: Background region overlaps another source region \\
& 16: Background region overlaps another background region \\
\tableline
edge\_code & \phn 0: Source does not dither off detector boundary \\
& \phn 1: Source position dithers off detector boundary \\
& \phn 2: Source region dithers off detector boundary \\
& \phn 4: Background dithers off detector boundary \\
\tableline
multi\_chip\_code & \phn 0: Source does not dither between detector chips\tablenotemark{b} \\
& \phn 1: Source position dithers across 2 chips \\
& \phn 2: Source region dithers across 2 chips \\
& \phn 4: Background region dithers across 2 chips \\
& \phn 8: Source position dithers across $>2$ chips \\
& 16: Source region dithers across $>2$ chips \\
& 32: Background region dithers across $>2$ chips \\
\tableline
var\_code & \phn 0: Intra-observation source variability not detected in any band \\
& \phn 1: Intra-observation variability detected in the ACIS ultrasoft ({\it u\/}) energy band \\
& \phn 2: Intra-observation variability detected in the ACIS soft ({\it s\/}) energy band \\
& \phn 4: Intra-observation variability detected in the ACIS medium ({\it m\/}) energy band \\
& \phn 8: Intra-observation variability detected in the ACIS hard ({\it h\/}) energy band \\
& 16: Intra-observation variability detected in the ACIS broad ({\it b\/}) energy band \\
& 32: Intra-observation variability detected in the HRC wide ({\it w\/}) energy band \\
\enddata
\tablenotetext{a}{Non-zero bit encodings are additive, so that (e.g.) $\rm
  var\_code = 28$ would mean the intra-observation variability was detected in
  the ACIS medium, hard, and broad energy bands.}
\tablenotetext{b}{``Chip'' refers to either an ACIS CCD or a HRC micro-channel plate.}
\end{deluxetable*}

The inter-observation variability probability is simply based on the reduced
$\chi^2$ of the distribution of the source region aperture photon fluxes of the
individual observations and their confidence intervals.  For a source detected
in $n$ separate observations, we first use the source region aperture photon
flux, $S_i$, and the associated lower and upper $1\,\sigma$ confidence limits,
$S_i^-$ and $S_i^+$, respectively, to compute an initial estimate of the
variance-weighted mean source region aperture photon flux
\begin{displaymath}
S_0 = \sum_{i=1}^n\frac{S_i}{\sigma_{0,i}^2}/\sum_{i=1}^n\frac{1}{\sigma_{0,i}^2},
\end{displaymath}
where we take $\sigma_{0,i} = (S_i^+ - S_i^-)/2$.  Using this estimate of the
mean flux, we define the ``effective $\sigma$'' for the $i$th observation of
the source as 
\begin{displaymath}
\sigma_i = \left\{
  \begin{array}{c l}
    S_i - S_i^- & \quad S_i > S_0 \\
    S_i^+ - S_i & \quad S_i < S_0 \\
    (S_i^+ - S_i^-) / 2 & \quad S_i = S_0.
  \end{array}
\right.
\end{displaymath}
A refined estimate of the variance-weighted mean flux is then given by 
\begin{displaymath}
S = \sum_{i=1}^n\frac{S_i}{\sigma_i^2}/\sum_{i=1}^n\frac{1}{\sigma_i^2}.
\end{displaymath}
and the reduced $\chi^2$ is
\begin{displaymath}
\chi^2 = \frac{1}{n-1}\sum_{i=1}^n\frac{(S_i-S)^2}{\sigma_i^2}.
\end{displaymath}

The inter-observation variability index is assigned on the basis of the reduced
$\chi^2$, according to Table~7. Note that the values 1, 2, 9,
and 10 are not used.

\subsection{Source Codes and Flags}

Each entry in both the Master Sources Table and the Source Observations Table
includes several source-specific flags and codes that identify specific
circumstances that may be of relevance to the catalog user.  Some flags and
codes are used to encode source properties that are commonly searched for by
users, as an aid to simplify catalog queries.  However, in most cases flags
and codes are intended to warn the user of conditions that may degrade the
quality of measured source properties, or that may limit the usefulness of the
source detection for some investigations.

The codes and flags included in the Source Observations Table are defined in
Table~1.  Flags are Boolean quantities that describe ``yes/no'' or
``true/false'' properties, whereas codes are multi-bit data values that encode
several levels of information.  Translations of the bit-encodings can be found
in Table~8.

The extent and variability codes require additional explanation.  The former
encodes a conservative estimate of whether the intrinsic extent of a source,
$a_{\rm rss}$ [equation~(\ref{eqn:arss})], is inconsistent with the extent of
the local PSF in each energy band.  Specifically, a source is considered
extended in an energy band if $a_{\rm rss} > 5\,\Delta a_{\rm rss}$ in that
energy band, where $\Delta a_{\rm rss}$ is the uncertainty in $a_{\rm rss}$,
given by equation~(\ref{eqn:errarss}).  The variability code bit corresponding
to a specific energy band is set if the intra-observation variability index
(Table~6) $\geq 3$.  A zero code therefore implies that the
source is either definitely not variable, not considered variable, or probably
not variable, depending on the value of the variability index.  Similarly, a
non-zero code implies that the source either may be variable, is likely to be
variable, is considered variable, or is definitely variable.

\begin{figure}
\epsscale{1.0}
\plotone{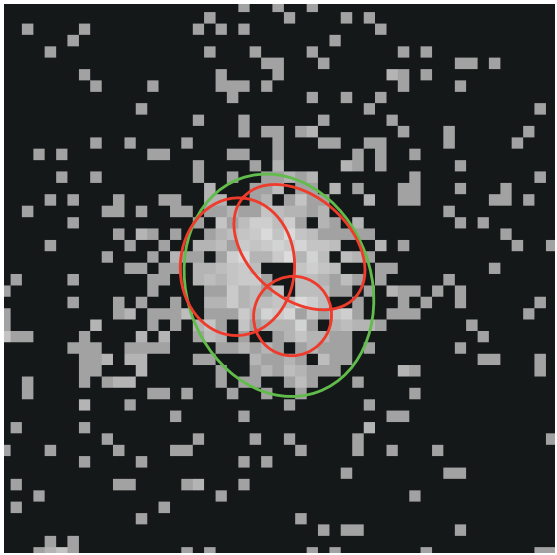}
\caption{\label{fig:crater}
Example of a highly piled-up source detected using ACIS.  The core of the image
has been eroded by photon pile-up, so that the source has a cratered appearance
(i.e., the photon density has an apparent minimum at the source location).
Bright spots on the ring are detected as distinct sources, shown in red.  These
source detections are manually adjusted to include only a single source centered
on the crater, shown in green.}
\end{figure}
 
The remaining codes and flags all warn of conditions that may affect derived
source properties to some extent.  The streak source flag, if set, indicates
that the source detection is located on an ACIS readout streak.  If the readout
streak is associated with a bright source, then there is a significant
probability that the source properties may be compromised.  This is particularly
true of aperture photometry values.  If the streak is especially intense, then
the source detection may not be real.  If the saturated source flag is set, then
the source is definitely real, but is so bright that photon pile-up has eroded
the core of the source image so that a single source has a ``cratered''
appearance (Fig.~22).  All source properties are compromised.

The Master Sources Table includes only flags, and these are defined in
Table~2.  In most cases, Master Sources Table flags summarize the
corresponding Source Observations Table flags and codes for all of the source
detections that have ``unique'' linkages to the master source.  The confusion
flag is an exception to this, in that it is set for a master source if the
confusion codes for any of the associated source detections indicate that
multiple sources are present in the source region or that the source region
overlaps another source region, {\it or\/} if there are any source detections
that have ``ambiguous'' linkages to the current master source.  

The master extent and variability flags are set if the corresponding codes for
{\it any\/} uniquely matched source detections indicate that the source is
extended or variable (as appropriate) in any energy band.  The remaining
master source flags are set only if the corresponding Source Observations Table
flags are set for {\it all\/} uniquely associated source detections, indicating
that the corresponding warning criteria are violated in all observations of the
source.

\subsection{Quality Assurance\label{sec:qa}}

\begin{deluxetable}{ccc}
\tablecolumns{3}
\tablewidth{0pt}
\tablecaption{ACIS Observation False Source Rate\label{tab:falsrc}}
\tablehead{
\colhead{Configuration} & \colhead{Livetime (ks)} & \colhead{False Source Rate}
}
\startdata
ACIS-012367 & \phn\phn 9 & 0.0\phn \\
ACIS-235678 & \phn 10 & 0.02 \\
ACIS-012367 & \phn 29 & 0.0\phn \\
ACIS-235678 & \phn 30 & 0.12\tablenotemark{a} \\
ACIS-235678 & \phn 51 & 0.21 \\
ACIS-012367 & \phn 68 & 0.22 \\
ACIS-235678 & 118 & 1.2\phn \\
ACIS-012367 & 125 & 1.28 \\
\enddata
\tablenotetext{a}{For this set of simulations, background data for CCD S4 (ACIS-8)
were unavailable; the false source rate was renormalized to account for the missing
chip data.}
\end{deluxetable}

The scientific integrity of the CSC is guaranteed through a set of quality
assurance steps that are performed as part of the catalog construction process
\citep{eva08}. Many of these analyses are executed automatically at the
completion of each stage of catalog pipeline processing, so that any issues can
be identified and corrected before they can affect downstream processing. These
mechanisms detect pipeline processing errors, and identify potential data
quality issues by comparing key diagnostic output products with predefined
standards. Each standard that is violated will either trigger a human review to
determine how to proceed, or will initiate one or more automated actions. The
latter typically result in termination of the processing thread for a subset of
the input data.

The vast majority of violations that occur because of data quality issues
address the reality of detected sources, and are typically resolved without
human intervention. Following the source detection step, detected source
regions that are either significantly smaller than the dimensions of the local
PSF or significantly larger than the maximum expected source size, or which
exceed a maximum ellipticity threshold, are deemed to be artifacts, and the
processing thread for the source region is terminated immediately to avoid
evaluating source properties unnecessarily. Sources that have too few counts, or
that have a detection significance that is too low to pass the catalog SNR
threshold are similarly discarded.  

The cores of sources observed with ACIS that are sufficiently bright can be
eroded by photon pile-up. The source detection algorithm incorrectly detects
bright spots on the ring surrounding the dark center of the image as distinct
sources.  Saturated sources are identified using a sliding matched filter
algorithm. The source detections are manually adjusted so that a single source
centered on the crater is included in the Source Observations Table for the
source, and the source properties are flagged as having been manually modified,
so that the user can exclude such sources if they so wish.  The tabulated source
position errors are unreliable for sources whose regions have been manually
modified.

\begin{figure*}
\epsscale{1.0}
\plotone{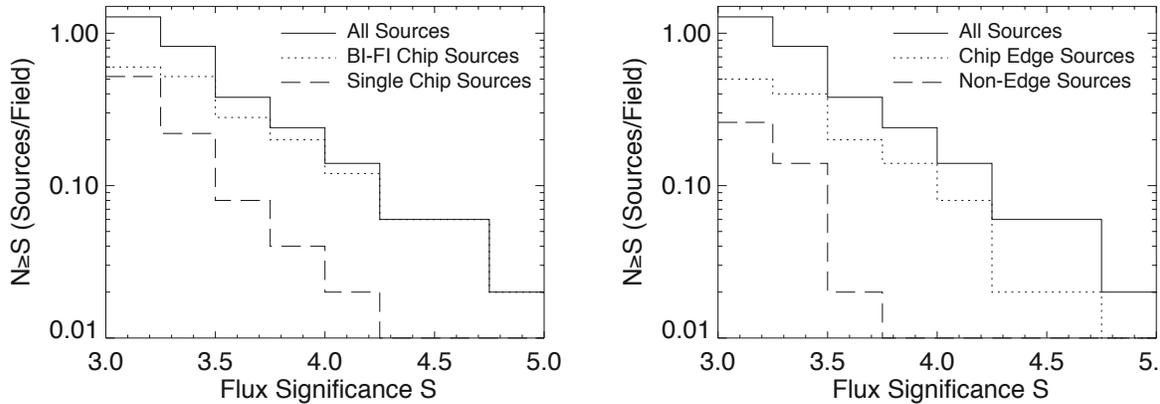}
\caption{\label{fig:falsrc}
False source rate as a function of flux significance for a simulated
$125\rm\,ks$ observation. The maximum flux significance across all science
energy bands is plotted.  {\it Left:\/} BI-FI Chip Sources are those whose
source regions dither across the CCD S2 (ACIS-6; back-illuminated)--S3
(ACIS-7; front-illuminated) boundary.  Single Chip Sources are those whose
source regions are completely contained on only a single chip.  {\it Right:\/}
Chip Edge Sources are those whose source regions dither off a chip edge during
the observation.  Non-Edge Sources are those whose source {\it and\/}
background regions do not dither off a chip edge.}
\end{figure*}

A more detailed comparison of the source dimensions with the local PSF is
performed after the source properties are computed, to identify sources that are
statistically smaller than the PSF in all energy bands. The fraction of the
local PSF that is included within the modified source region aperture (as
defined in \S~3.4.1) must be sufficiently large that the source
location and aperture photometry can be computed from the fraction of the
aperture that is not contaminated by overlapping sources.  Finally, the SNR of
the source is evaluated and compared with the minimum required for inclusion in
the catalog.

Human review is primarily required to address the occurrence of unexpected
pipeline warnings or errors. Although an automated process performs the
laborious task of scanning the log files associated with each processing
pipeline to identify problems, the wide diversity of possible error conditions
require human intelligence to assess the reason for the failure and determine
how to proceed.  The typical response is to terminate the current processing
thread, perform any needed repairs, and initiate reprocessing of the thread.

Other conditions that trigger a human review are precautionary in nature, and
include cases where the local spatial density of detected sources is too high,
or the total number of sources detected in the field of view exceeds a
predefined threshold. Although these conditions most likely arise because of
field crowding, they could indicate an error in the source detection process.
Errors that generate a large number of sources would require substantial
cleanup if processing was allowed to continue incorrectly.

\section{STATISTICAL PROPERTIES\label{sec:stat}}

A detailed characterization of the statistical properties of the CSC is beyond
the scope of this paper, but is the subject of a comprehensive discussion by
\citetfap.  Here, we merely present a summary of the principal statistical
properties of the catalog for reference.

Statistical characterization of catalog source properties is accomplished
primarily by using simulated and empty field (blank-sky) observations,
together with datasets consisting of empty fields that have simulated sources
with known properties added. These datasets are processed by the catalog
pipelines in the same manner as real observations.

\subsection{False Source Rate}

To estimate false source rates, a series of blank-sky simulations with exposure
times of $\sim\!10$, $30$, $60$, and $120\rm\, ks$ were constructed for typical
ACIS imaging CCD configurations.  For each simulation, a template background
event list for each active CCD was used to define the overall spatial variation
of the background, and the total number of background events was determined from
the nominal field background rates \citep{POG} and the simulated exposure time.
For all CCDs except chip S4 (ACIS-8) the template background event lists
recorded in the instrumental calibration database were used.  Chip S4 is
significantly affected by a variable pattern of linear streaks that appear to be
caused by a flaw in the serial readout which randomly deposits significant
amounts of charge along pixel rows as they are read out \citep{hou00}.  Because
of this issue, no adequate template is available for chip S4, and so one was
constructed by combining several CSC event lists that do not include bright
sources on that CCD\null.  Each simulated blank-sky event list was then
processed through the CSC pipeline source detection steps.  The false source 
rates derived from these simulations are reported in Table~9.
From these data we derive a simple linear relation for the number of false 
sources per field as a function of livetime, namely
\begin{displaymath}
\log(R_{\rm fs})= -3.345 + 1.6\times\log(t_{\rm live}),
\end{displaymath}
where $R_{\rm fs}$ is the false source rate, and $t_{\rm live}$ is the exposure 
livetime in units of ks.  Using this relation, we estimate that $\sim\!370$ 
sources ($\sim\!0.4\%$) included in the catalog are spurious.

As can be seen from the table, the false source rate is appreciable only for
exposures longer than $\sim\!50\rm\,ks$.  There is some evidence for a
clustering of false source detections near chip edges and at the boundaries
between the back- and front-illuminated CCDs.  This should not be surprising
since the low spatial frequency background is poorly constrained or changing
rapidly in these locations.  To investigate these effects further, the false
source rates near the chip edges and interfaces were examined separately for the
longest simulated exposures.  Figure~23 demonstrates that false
source rates are enhanced in these regions for the $125\rm\,ks$ simulation.

\subsection{Source Detection Efficiency}

\begin{figure*}
\epsscale{1.0}
\plotone{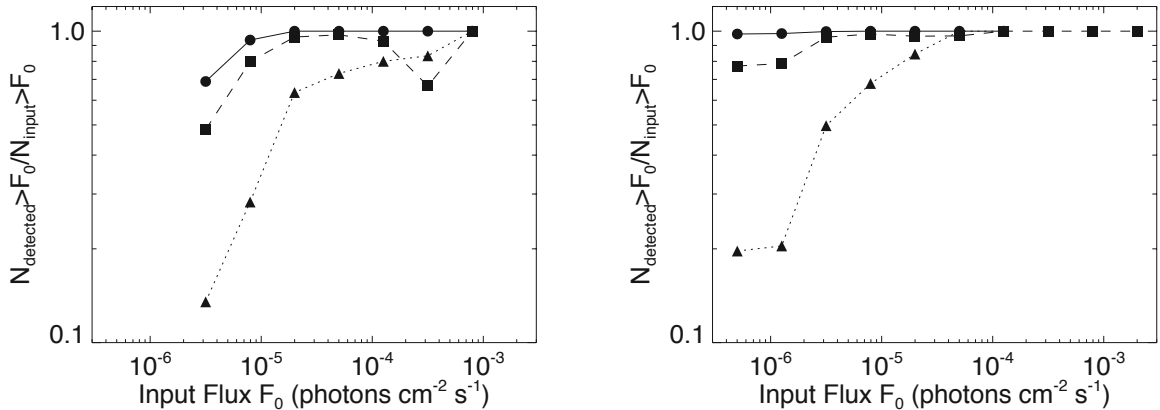}
\caption{\label{fig:deteff}
Cumulative detection efficiency estimates from ACIS-012367 simulations of
point sources with absorbed power-law spectral distributions ($\Gamma = 1.7$,
$N_{\rm H} = 3\times10^{20}\rm\,cm^{-2}$).  Simulated source fluxes were drawn
from a power-law $N > S$ distribution with index 1.5 and an overall
normalization adjusted to yield a few hundred detectable sources per
simulation.  Thirty simulations per exposure time were calculated.  Detection
efficiency is computed by comparing the measured and input $N > S$
distributions in the ACIS broad energy band.  The ratio of these two
distributions represents the fraction of input sources of a given incident
flux that are actually detected.  {\it Left:\/} Detection efficiency for a
$\sim\!9\rm\,ks$ exposure.  The solid, dashed, and dotted lines plot the
detection efficiencies for sources with $\theta < 5'$, $5' \leq \theta < 10'$,
and $10' \leq \theta < 15'$, respectively.  Because of the power-law $N > S$
distribution, relatively few bright sources were simulated and the plotted
detection efficiency does not smoothly approach 100\% for bright sources 
because of small number statistics.  {\it Right:\/} As for the left panel, 
except for a $\sim\!125\rm\,ks$ exposure time.}
\end{figure*}

Source detection efficiency is characterized using point source simulations. A
spatially random distribution of point sources is added to the blank sky
simulations described above using the MARX simulator \citep{MARX} to generate
the incident X-ray photons. Separate simulations were generated for non-thermal
sources with a power-law spectral distribution $F_E\propto\!E^{-1.7}$, and for
thermal black-body sources with temperature $kT=1.0\rm\,keV$, spectra. A neutral
Hydrogen absorbing column $N_{\rm H} = 3\times 10^{20}\rm\,cm^{-2}$ was assumed
for all sources. Source fluxes were drawn from a power-law $N > S$ distribution
with index 1.5. The overall normalization was adjusted to yield a few hundred
detectable sources per simulation, a compromise aimed at reducing source
confusion while limiting the total number of simulations required to obtain good
statistics. The effects of photon pile-up \citep{jed01b} and
observation-specific bad pixels were included by post-processing each simulation
with {\tt marxpileup} and {\tt acis\_process\_events}, respectively. The source
events from the MARX simulations were then merged with the appropriate simulated
blank-sky event lists, keeping only MARX-simulated source events that fell on
active CCDs for the observation. As with the blank-sky simulations, simulated
event lists were then processed through the CSC pipeline source detection and
source properties extraction steps, and the resulting sources that would have
been included in the catalog were tabulated. Finally, these sources were
cross-referenced with the input source lists to allow a source-by-source
comparison of input and derived properties.

Source detection efficiency is determined by comparing the measured $N > S$ and
input $N > S$ distributions.  The ratio of these two distributions represents
the fraction of input sources of a given incident flux that are actually
detected.  Results of the comparison for the ACIS broad energy band detections
from the shortest and longest ACIS-012367 power-law spectral distribution
simulation sets are shown in Figure~24.  The standard CSC
processing pipeline further combines ACIS source detections from the broad,
soft, medium, and hard energy bands to construct the final detected source list.
This step was not performed as part of these simulations.  However, since the
simulated source spectra are homogenous and well detected in the broad energy
band, the difference is not significant in this case.

\vskip 40pt

\subsection{Absolute Astrometric Accuracy\label{sec:aast}}

As mentioned in \S~3.5.1, the absolute astrometric accuracy of
release~1 of the CSC was evaluated {\it post facto\/} by cross-matching catalog
sources with their counterparts from the SDSS DR7 \citep{aba09}.  Like the CSC,
the SDSS DR7 is referenced to the International Celestial Reference System
\citep[ICRS;][]{ari95}, and has statistical positional uncertainties
$\sim\!45\rm\,milliarcseconds$ (mas) rms per coordinate for bright stars, with
systematic errors $<20\rm\,mas$ \citep{aba09}.  Only CSC-SDSS source pairs with
more than 90\% match probability, evaluated according to the Bayesian
probabilistic formalism described by \citet{bud08}, were evaluated, resulting in
6,310 source pairs associated with 9,476 sources detected in individual
observations.  Full details of the analysis and results are presented by
\citet{rot09} and \citetfap.  Here we summarize the main result.

For each matching CSC-SDSS source pair, the separation, $\rho$, and the total
$1\,\sigma$ positon error are computed, summing in quadrature the CSC and SDSS
errors (and remembering that CSC position errors are reported as 95\%
uncertainties).  We then examine the value of reduced $\chi^2 =
\sum(\rho/\sigma_{{\rm tot}})^2/(n-1)$ for bins in $\sigma_{{\rm tot}}$ covering
the range $\sim\!0.1$--$2''$\null.  The value of the reduced $\chi^2$ is
reasonably close to 1, except for $\sigma_{{\rm tot}} \lesssim 0.3''$
(indicating that the errors are underestimated in that range).  Adding a
systematic astrometric error component of $0\farcs16\pm0\farcs0.01$ to
$\sigma_{{\rm tot}}$ yields reduced $\chi^2$ near 1 for all values of
$\sigma_{tot}$.  We therefore adopt that value as the systematic astrometric
error present in release 1 of the CSC\null.

\begin{figure*}
\epsscale{1.0}
\plotone{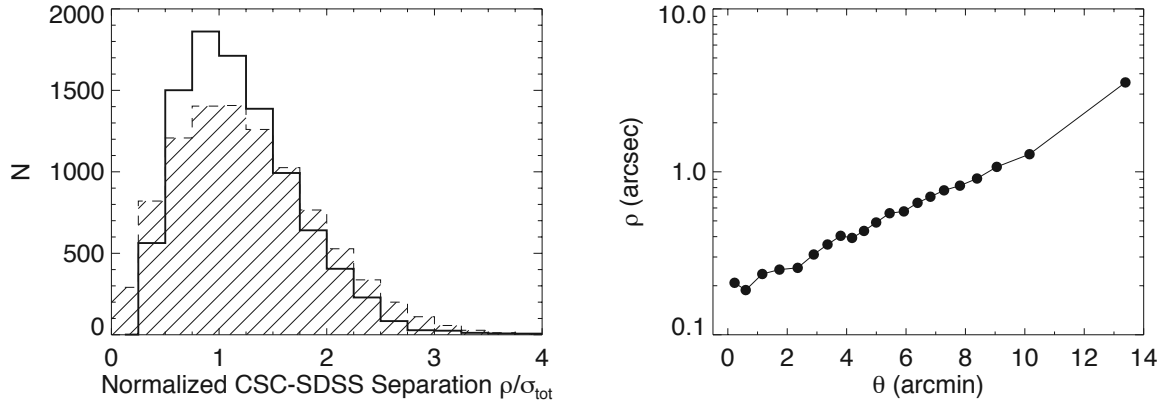}
\caption{\label{fig:aast}
{\it Left:\/} Distribution of the normalized separations for matching CSC-SDSS
source pairs (solid line).  The theoretical Rayleigh distribution for the same
number of sources is shown hatched.  {\it Right:\/} Average CSC-SDSS source
pair separation with $0\farcs16$ systematic astrometric error included, as a
function of off-axis angle, $\theta$.}
\end{figure*}

The distribution of the normalized separations for the CSC-SDSS source match
pairs is shown in the left panel of Figure~25, together with the
theoretical Rayleigh distribution for the same number of sources.  The overall
shape of the curve agrees with the Rayleigh distribution, although there is a
slight deficit at high values of normalized separation, suggesting that the
overall error may be overestimated for sources at large off-axis angles.  In the
right panel of Figure~25, we present the average CSC-SDSS
separation as a function of off-axis angle, with the systematic astrometric
error included. The average CSC $1\,\sigma$ positional error ranges from
$0\farcs2$ on-axis to $\sim\!3\farcs5$ at $\sim\!14'$ off-axis.

\subsection{Photometric Accuracy\label{sec:photacc}}

To assess the accuracy of CSC source fluxes, the measured source region aperture
photon fluxes are compared with the input photon fluxes of the simulated point
sources.  Figure~26 presents the comparison of ACIS broad energy band
photon fluxes for simulated sources with a power-law spectrum.  Inspection of
the figure reveals good agreement for sources with off-axis angles within $10'$
of the aimpoint.  For sources beyond $10'$, photon fluxes appear to be 
systematically overestimated by a factor of $\sim\!2$ for sources fainter than 
$\sim\!3 \times 10^{-6}\rm\,photons\,cm^{-2}\,s^{-1}$.

The systematic error in the faint flux bins is more prominent in the ACIS soft
energy band, in which the measured fluxes appear under-estimated for all
simulated input flux levels.  Further investigation of this effect will be
reported by \citetfap.  Preliminary analysis suggests that the effect results
from the use of a monochromatic exposure map (computed at the effective energy
of the band) when determining source fluxes.  Models based on this assumption
reproduce the general features of the apparent systematic errors, and for the
assumed model power-law spectrum the error is $\sim\!10\%$ in the broad, medium,
and hard energy bands, $\sim\!20$--$30\%$ in the soft energy band, and
$\sim\!30\%$ in the ultra-soft energy band.

\section{CONCLUSIONS}

The {\it Chandra\/} Source Catalog is a general purpose virtual X-ray
astrophysics facility that provides access to a carefully crafted set of
scientifically useful quantities for individual X-ray sources observed by the
{\it Chandra\/} X-ray Observatory.  The first release of the catalog was
published to the astronomical community in March 2009, and includes source
properties for 94,676 point and compact X-ray sources detected in a subset of
public ACIS imaging observations from roughly the first eight years of the {\it
  Chandra\/} mission.  This release of the catalog includes sources with
observed spatial extents are $\lesssim 30''$, and whose flux estimates are at
least 3 times their estimated $1\,\sigma$ uncertainties.  Observations that
include substantially extended sources are not included in the first release of
the catalog.  For each X-ray source, the catalog tabulates about 60 distinct
measured and derived source properties, generally with associated lower and
upper confidence limits, in several energy bands.  These properties are
generally derived from all of the observations in which a source is detected.
However, in the first catalog release, multiple observations are not {\it
  combined\/} prior to source detection, so the depth of the catalog is limited
by the duration of the longest single exposure of a field.  The catalog further
tabulates roughly 120 observation-specific properties for each observation of a
source, again with associated lower and upper confidence limits, and in several
energy bands.

Tabulated source properties include source position, spatial extent, multi-band
aperture fluxes computed in several different ways, X-ray hardness ratios and
spectral model fits, and intra- and inter-observation variability measures.  In
addition to these ``traditional'' catalog elements, for each source detection
the catalog includes an extensive set of FITS format file-based data products
that can be manipulated interactively by the user, including source images,
event lists, light curves, and spectra from each observation in which a source
is detected.

\begin{figure*}
\epsscale{1.0}
\plotone{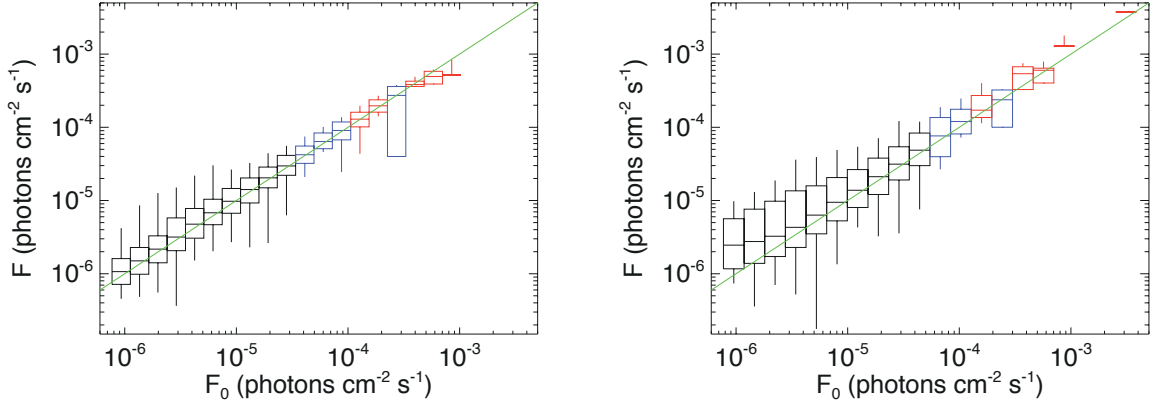}
\caption{\label{fig:phot}
Comparison of input ($\rm F_0$) and measured ($\rm F$) ACIS broad ({\it b\/}) 
band fluxes for simulated sources with power-law spectra and off-axis angles
$\theta\leq 10'$ ({\it left\/}) and $\theta > 10'$ ({\it right\/}).  For each 
bin, the horizontal line indicates the median measured flux value.  The boxes 
include 90\% of the measurements in each bin, and the vertical lines indicate 
the extreme values.  Bins colored red include fewer than 100 measurements; bins 
colored blue include 100--400 measurements; bins colored black include more 
than 400 measurements.  The green line has a slope of 1.\vglue 5pt}
\end{figure*}

Looking towards the future, release 1.1 of the catalog, scheduled for spring 2010,
will include data from public HRC-I imaging observations and newly public ACIS
imaging observations, but will otherwise retain the same limitations as release 1.
In release 2, we plan to co-add multiple observations of the same field that use 
the same or similar instrument configurations, and that have similar spacecraft
pointings (within $\sim\!30''$) prior to source detection, to achieve fainter 
limiting sensitivities in many fields.  We anticipate that new algorithms will 
allow this release to have a significantly fainter source detection threshold 
than release 1.  This release should also provide limited improvements in the
area of extended source handling (for example allowing for the inclusion of 
exposures containing moderately extended emission from galaxy cores up to 
$\sim\!60''$ spatial scale), as well as numerous algorithm enhancements that
will refine field and source property calculations.

\acknowledgments

The authors would like to thank the {\it Chandra\/} Source Catalog project
review visiting panel, who endorsed the catalog and proposed several key
recommendations that have guided the development of the catalog.  We would like
to acknowledge the support and guidance of the {\it Chandra\/} X-ray Center
director, Harvey Tananbaum, and manager, Roger Brissenden.

Former {\it Chandra\/} Source Catalog project team members who have contributed
significantly to the definition and development of the catalog include Martin
Elvis, St\'ephane Paltani, Adam Dobrzycki, Johnathan Slavin, Dan Harris, Peter
Freeman, and Michael Wise.  We would like to thank Taeyoung Park for his support
enhancing and testing the BEHR algorithm for use in the catalog processing
pipelines.  Lisa Paton and the CXC Systems Group provided extensive installation
and operational support for the catalog production Beowulf cluster.

The authors would also like to thank the anonymous referee, who performed a
very careful and comprehensive review of the first version of the manuscript, 
and whose suggestions materially improved the content and quality of the paper.

The development and operational construction of the catalog made extensive use
of the CIAO, ChIPS, and Sherpa software packages developed by the {\it
Chandra\/} X-ray Center, and the SAOImage DS9 imager developed by the
Smithsonian Astrophysical Observatory.

Support for development of the {\it Chandra\/} Source Catalog is provided by the
National Aeronautics and Space Administration through the {\it Chandra\/} X-ray
Center, which is operated by the Smithsonian Astrophysical Observatory for and
on behalf of the National Aeronautics and Space Administration under contract
NAS 8-03060.

\vskip 20pt

\appendix

\section{MASTER SOURCE MATCHING ALGORITHM\label{app:mrg}}

The procedure, referenced in \S~3.4.2, for matching source detections
from multiple observations that overlap the same region of the sky, is described
here.  These steps must be executed for each observation, but the outcome does
not depend on the order in which the observations are processed.

The algorithm defines the {\it overlap ellipse\/} of a source detection to be
the PSF 90\% ECF aperture in the energy band that has the highest number of
aperture source counts.  The following two assumptions are made: (a)~if a source
detection in one observation is resolved into multiple source detections in a
second observation, then the overlap ellipse corresponding to the former
detection will overlap all of the overlap ellipses corresponding to the latter
detections, and (b)~multiple source detections in a single observation
correspond to distinct sources on the sky, even if the overlap ellipses
intersect spatially.

For the set $\setS$ of source detections identified in the current observation,
the following 9 steps are performed.

(1)~Identify the sets $\setM_i, \setM_j, \ldots$ of {\it candidate matching
source detections\/} in observations $i, j, \ldots$ that overlap the current
observation.  Candidate matching source detections are those source detections
that have radial separations on the sky from any member of the set $\setS$ that
are smaller than some predefined radius, $r$.  For convenience, we designate the
union $\setM_i \cup \setM_j \cup \ldots$ as the set $\setM$.

(2)~Compute the overlap ellipses, defined above, for each member of the sets
$\setS$ and $\setM$.  

Two source detections $a$ and $b$ in observations $\1$ and $\2$, respectively,
are deemed to {\it overlap\/} {\it if and only if\/}
\begin{eqnarray*}
A[a \cap b]/A[a] > 0.15 &{\it or\/}& A[a \cap b]/A[b] > 0.15, 
\end{eqnarray*}
where $A[a]$ is the area of the overlap ellipse of source detection $a$, and
$A[a \cap b]$ is the area of the spatial intersection of the overlap ellipses of
$a$ and $b$.  This has the effect of dismissing very small overlaps.

\begin{figure}
\epsscale{0.35}
\plotone{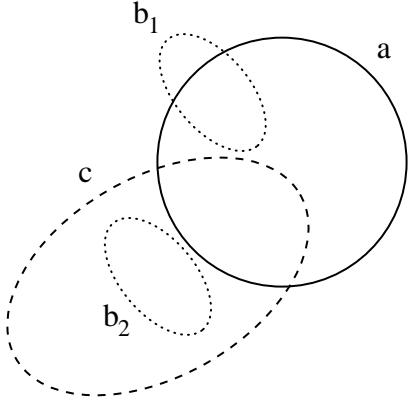}
\caption{\label{fig:confmatch}
Example ``confused match'' source detections.  Source detection $a$ from
observation $\1$, overlaps source detections $b_1$ and $c$ from observations
$\2$ and $\3$, respectively.  Source detection $c$ also comprises a partial
unambiguous match with source detection $b_2$ from observation $\2$.  The latter
connection implies that source detections $a$ and $c$ are confused matches with 
the pair of sources $b_1$ and $b_2$ from observation $\2$.}
\end{figure}

In addition, {\it if\/} a source detection $a$ in observation $\1$ overlaps
multiple source detections $b_1, b_2, \ldots, b_n$ ($n > \1$) in observation
$\2$, {\it and\/}
\begin{eqnarray*}
A[a \cap b_i]/A[a] > 0.67 &{\it and\/}& A[a \cap b_j]/A[a \cap b_i] < 0.33 
\end{eqnarray*}
for all $j \ne i$, {\it then\/} only the overlap between source detections $a$
and $b_i$ is recognized.  The remaining source detections $b_j$ included in
observation $\2$ are deemed {\it not\/} to overlap source detection $a$.  This
has the effect of recognizing only a single dominant overlap and ignoring
additional smaller overlaps from the same observation.

(3)~Compute the subset $\setN \subseteq \setS$ of source detections that do not
overlap any member of the set $\setM$ of candidate matching source detections.
{\it $\setN$ comprises the set of source detections that must be added to the
Master Sources Table as newly identified master sources.\/}

(4)~Compute the set $\setP$ of members of $\setS \cup \setM$ that comprise {\it
partial unambiguous matches\/}.  A source detection $a$ included in
observation $\1$ is a partial unambiguous match to a source detection $b$
included in a different observation $\2$ {\it if and only if\/} (a)~$a$
overlaps $b$, (b)~$a$ does not overlap any other source detection included in
observation $\2$, and (c)~$b$ does not overlap any other source detection
included in observation $\1$.  Sources included in $\setP$ are uniquely matched
between {\it pairs\/} of overlapping observations, but are not necessarily
uniquely matched between {\it all\/} overlapping observations.
  
(5)~Compute the subset $\setU \subseteq \setP$ of {\it unambiguous matches\/}.
An unambiguous match between source detections $a, b, c, \ldots$ included in
observations $\1, \2, \3, \ldots$ occurs when all pairs $([a, b], [a, c], [a,
\ldots], [b, c], [b, \ldots], \ldots)$ of source detections individually form
partial unambiguous matches.  {\it $\setU$ comprises the set of source
detections that are uniquely matched to existing master sources in the Master
Sources Table.\/}

Figure~15, {\it Left\/} is an example of an unambiguous match.

(6)~Compute the set $\setC$ of members of $\setS \cup \setM$ that comprise {\it
confused matches\/}.  A confused match results when a source detection $a$
included in observation $\1$ overlaps multiple source detections $b, c, \ldots$
that either (a)(i)~are included in a single observation $\2$ {\it and\/}
(ii)~overlap no other source detection in observation $\1$ than $a$, {\it or\/}
(b)~consist of partial unambiguous matches such that there is at least one
observation that is common amongst the partial unambiguous matches for all of
the sources $b, c, \ldots$.  {\it $\setC$ comprises the set of confused source
detections that must be flagged as confused, and linked ambiguously to the
corresponding master sources in the Master Sources Table.\/}

An example of case (a) above is shown in Figure~15, {\it
Center\/}. In the case (b) above, note that source detection $a$ is not required
to overlap all of the individual source detections that comprise each of the
partial unambiguous matches (e.g., Fig.~27).

(7)~Once the set $\setC$ of confused matches is determined, steps (2)--(5)
should be re-applied to the set $\{\setS \cup \setM\} \backslash \{\setN \cup
\setU \cup \setC\}$ to identify additional members of the sets $\setN$ and
$\setU$ that were previously missed because they were overlapped by one or more
confused source detections.

(8)~After all members of the sets $\setN$ and $\setU$ have been identified,
re-examine members of the set $\setC$ to verify that the overlaps of source
detections that caused each member to become assigned to set $\setC$ are
overlaps with members of $\setN$ and $\setU$.  {\it If\/} any source detection
which is not a member of $\setN$ or $\setU$ overlaps a member of $\setC$, {\it
and\/} that overlap was relied on to assign the member to $\setC$, {\it
then\/} remove the member from set $\setC$.  The removed member will revert to
an uncategorized source detection.

(9)~At this point, the set $\{\setS \cup \setM\} \backslash \{\setN \cup \setU
\cup \setC\}$ consists of source detections that cannot be merged or linked to
master sources using the above rules.  These source detections typically (but
not exclusively) overlap at least two other sources that were observed in
different observations and that do not overlap each other.  We designate members
of this set $\setH$ as {\it human-review matches\/}.  Manual review is required
to disambiguate the source matches.

Figure~15, {\it Right\/} is an example of a human-review match.

\section{COMBINING SOURCE POSITIONS\label{app:mow}}

As described in \S~3.5.2, a multivariate optimal weighting formalism
is used to improve the estimates of the position and positional uncertainty of
each X-ray source by combining the statistically independent source detections
included in the set of individual observations.  The source position error
uncertainties are expressed in the form of error ellipses centered upon the
estimated source positions.  Details of the derivations can be found in
\citet{jed07b}. 

The improved estimates of the source position, $X$, and associated covariance
matrix, $\sigma^2$, are
\begin{eqnarray}
\label{eqn:appmow}
X = \sigma^2 \sum_i \frac{X_i}{\sigma_i^2} &;& \sigma^2 = \left[ \sum_i
  \frac{1}{\sigma_i^2} \right]^{-1},
\end{eqnarray}
where $X_i$ represents the $i$th estimate of the mean of the two-dimensional
source position, and $\sigma_i^2$ denotes the $2 \times 2$ covariance matrix,
equation~(\ref{eqn:appcov}) below, associated with this estimate.

Before the covariance matrix $\sigma^2$ can be computed, the individual error
ellipses must be mapped from the celestial sphere onto a common tangent
plane.  The $i$th estimate of the source position is specified as a
confidence-ellipse centered upon the celestial coordinates $(\alpha_i,
\delta_i)$, with the major axis of the ellipse making an angle $\theta_i$ ($-\pi 
\leq \theta_i < \pi$) with respect to the local line of declination at the
center of the ellipse.  The celestial coordinates $(\alpha_i, \delta_i)$
correspond to a unit vector
\begin{displaymath}
\hat{p}_i = \hat{x} \cos \alpha_i \cos \delta_i + \hat{y} \sin \alpha_i \cos
\delta_i + \hat{z} \sin \delta_i 
\end{displaymath}
on the celestial sphere, where $(\hat{x}, \hat{y}, \hat{z})$ are orthonormal
basis vectors oriented such that $\hat{x}$ points to the origin of right
ascension on the celestial equator, $\hat{z}$ points to the North celestial
pole, and $\hat{y}$ completes the right-hand Cartesian system.

The common tangent plane is constructed on the celestial sphere at the position
$\hat{p}_0$, which is taken to be the arithmetic mean of the ellipse centers
$\hat{p}_i$:
\begin{equation}
\label{eqn:apptp0}
\hat{p}_0 = \sum_i \hat{p_i} / \left| \sum_i \hat{p_i} \right|.
\end{equation}
The tangent plane coordinates ($x_i$, $y_i$) corresponding to $(\alpha_i,
\delta_i)$ are 
\begin{eqnarray}
\label{eqn:apptpc}
x_i & = & (\hat{p}_i \cdot \hat{e}_x) / (\hat{p}_i \cdot \hat{p}_0) \nonumber \\
y_i & = & (\hat{p}_i \cdot \hat{e}_y) / (\hat{p}_i \cdot \hat{p}_0),
\end{eqnarray}
where $\hat{e}_x$ and $\hat{e}_y$ are orthonormal basis vectors parallel to the
local lines of right ascension and declination at $\hat{p}_0$, {\it i.e.\/}, 
\begin{eqnarray*}
\hat{e}_x & = & -\hat{x} \sin \alpha_0 + \hat{y} \cos \alpha_0 \\
\hat{e}_y & = & -\hat{x} \sin \delta_0 \cos \alpha_0 - \hat{y} \sin \delta_0
\sin \alpha_0 + \hat{z} \cos \delta_0,
\end{eqnarray*}
where $(\alpha_0, \delta_0)$ are the celestial coordinates that correspond to
$\hat{p}_0$. 

Similarly, the unit vectors on the celestial sphere corresponding to the
end-point positions of the semi-minor and semi-major axes of the ellipse are
given by 
\begin{eqnarray}
\label{eqn:apptpam}
\hat{p}_i^{\rm minor} & = & \hat{p}_i \cos \phi_i^{\rm minor} + \hat{\alpha}_i
\sin \phi_i^{\rm minor} \cos \theta_i - \hat{\delta}_i \sin \phi_i^{\rm minor}
\sin \theta_i \nonumber \\ 
\hat{p}_i^{\rm major} & = & \hat{p}_i \cos \phi_i^{\rm major} + \hat{\alpha}_i
\sin \phi_i^{\rm major} \sin \theta_i + \hat{\delta}_i \sin \phi_i^{\rm major}
\cos \theta_i, 
\end{eqnarray}
where $\phi_i^{\rm minor}$ and $\phi_i^{\rm major}$ are the arc-lengths of the
semi-minor and semi-major axes, respectively, and $\hat{\alpha}_i$ and
$\hat{\delta}_i$ are unit vectors that point along the directions of increasing
right ascension and declination, respectively, at the position $\hat{p}_i$.

The lengths of the semi-minor and semi-major axes on the tangent plane are given
by 
\begin{eqnarray}
\label{eqn:apptpl}
\sigma^\prime_{1,i} & = & \sqrt{(x_i^{\rm minor} - x_i)^2 + (y_i^{\rm minor} - y_i)^2} \nonumber \\
\sigma^\prime_{2,i} & = & \sqrt{(x_i^{\rm major} - x_i)^2 + (y_i^{\rm major} - y_i)^2},
\end{eqnarray}
respectively, where we have denoted the tangent plane coordinates of
$\hat{p}_i^{\rm minor}$ and $\hat{p}_i^{\rm major}$ as ($x_i^{\rm minor}$,
$y_i^{\rm minor}$) and ($x_i^{\rm major}$, $y_i^{\rm major}$), respectively.
The angle that the semi-major axis makes with respect to the local 
line of declination is 
\begin{equation}
\label{eqn:apptpt}
\vartheta^\prime_i = \tan^{-1} \left(\frac{x_i^{\rm major} - x_i}{y_i^{\rm
      major} - y_i}\right). 
\end{equation}

Armed with the projections of the individual error ellipses projected on the
common tangent plane, equations (\ref{eqn:apptpc})--(\ref{eqn:apptpt}), covariance
matrices can be computed as follows.

The three parameters that specify the geometry of each projected error ellipse
are the lengths of the semi-major and semi-minor axes, and the position angle
$\vartheta$ that the major axis of the ellipse makes with respect to the tangent
plane $y$ axis. The semi-major and semi-minor axis lengths correspond to the
$1\,\sigma$ confidence intervals along these axes. In a basis whose origin is at
the center of the ellipse, and whose $y$ axis is along the major axis of the
ellipse, the covariance matrix is
\begin{displaymath}
\sigma_i^{\prime 2} = \left(
\begin{array}{cc}
\sigma^{\prime 2}_{1,i} & 0 \\
0 & \sigma^{\prime 2}_{2,i}
\end{array}
\right),
\end{displaymath}
where $\sigma^\prime_{1,i}$ and $\sigma^\prime_{2,i}$ are the $1\sigma$
confidence values along the minor axis and major axis of the ellipse,
respectively ($\sigma^\prime_{2,i} \geq \sigma^\prime_{1,i}$). The form of the
covariance matrix in the unrotated system is
\begin{equation}
\label{eqn:appcov}
\sigma_i^2 = \left(
\begin{array}{cc}
\sigma^{\prime 2}_{1,i}\cos^2 \vartheta_i + \sigma^{\prime 2}_{2,i} \sin^2
\vartheta_i & (\sigma^{\prime 2}_{2,i} - \sigma^{\prime 2}_{1,i}) \cos
\vartheta_i \sin \vartheta_i \\ 
(\sigma^{\prime 2}_{2,i} - \sigma^{\prime 2}_{1,i}) \cos \vartheta_i \sin
\vartheta_i & \sigma^{\prime 2}_{1,i}\sin^2 \vartheta_i + \sigma^{\prime
  2}_{2,i} \cos^2 \vartheta_i 
\end{array}
\right),
\end{equation}
where $\vartheta_i$ is the angle that the major axis of the ellipse makes with
respect to the tangent plane $y$ axis.

At this point, equation~(\ref{eqn:appcov}) can be used to compute the covariance
matrices from the lengths of the semi-minor and semi-major axes of the source
position error ellipses in the tangent plane, equations~(\ref{eqn:apptpl}).  The
error ellipses for the individual source observations are then combined using
equation~(\ref{eqn:appmow}) to compute the optimally weighted source position and
position error ellipses {\it on the tangent plane\/} for the combined set of
observations.  The mapping of the optimally weighted error ellipse from the
tangent plane to the celestial sphere can be performed using the inverse
relations of equations (\ref{eqn:apptpc}), (\ref{eqn:apptpam}), (\ref{eqn:apptpl}),
and~(\ref{eqn:apptpt}).


\end{document}